\documentclass[12pt,a4paper]{article}
\pdfoutput=1

\usepackage{amssymb}
\usepackage{amsmath}
\usepackage{amsfonts}
\usepackage{authblk}
\usepackage{setspace}
\usepackage[pdftex]{graphicx} 

\numberwithin{equation}{section}

\newcommand{\ii}{{\mathrm{i}}}
\newcommand{\emath}{{\mathrm{e}}}

\newcommand{\Mhalfspace}{\widetilde{\mathcal{M}}}

\newcommand{\BbbR}{\mathbb{R}}

\DeclareMathOperator{\Realpart}{Re}

\DeclareMathOperator{\Imagpart}{Im}

\DeclareMathOperator{\sgn}{sgn}

\DeclareMathOperator{\si}{si}

\DeclareMathOperator{\Ci}{Ci}

\DeclareMathOperator{\arctanh}{arctanh}

\title{Onset and decay of the 1+1 Hawking-Unruh effect:
what the derivative-coupling detector saw}
\author{Benito A. Ju\'arez-Aubry\thanks{pmxbaju@nottingham.ac.uk} }
\author{Jorma Louko\thanks{jorma.louko@nottingham.ac.uk}}
\affil{School of Mathematical Sciences\\ 
University of Nottingham\\ 
Nottingham NG7 2RD\\ 
UK}

\date{{\small June 2014, revised August 2014\thanks{This 
is an author-created, un-copyedited version of an 
article published in 
Class.\ Quantum Grav.\ {\bf 31}, 245007 (2014). 
IOP Publishing Ltd is not responsible for any errors 
or omissions in this version of the manuscript or any 
version derived from it. 
The Version of Record is available online at 
doi:10.1088/0264-9381/31/24/245007.}}}

\begin{document}

\maketitle
\begin{abstract}
We study an Unruh-DeWitt particle detector that is coupled to 
the proper time derivative of a real scalar field in 1+1 spacetime 
dimensions. Working within first-order perturbation theory, 
we cast the transition probability into a regulator-free form, 
and we show that the transition rate remains well defined 
in the limit of sharp switching. 
The detector is insensitive to the infrared ambiguity 
when the field becomes massless, 
and we verify explicitly the regularity of the massless limit 
for a static detector in Minkowski half-space. 
We then consider a massless field for two scenarios of interest 
for the Hawking-Unruh effect: 
an inertial detector in Minkowski spacetime with 
an exponentially receding mirror, 
and an inertial detector in $(1+1)$-dimensional Schwarzschild spacetime, 
in the Hartle-Hawking-Israel and Unruh vacua. 
In the mirror spacetime the transition rate traces 
the onset of an energy flux from the mirror, 
with the expected Planckian late time asymptotics. 
In the Schwarzschild spacetime the transition rate of a detector that 
falls in from infinity gradually loses thermality, 
diverging near the singularity proportionally to~$r^{-3/2}$. 
\end{abstract}


\newpage

\section{Introduction\label{sec:intro}}

For quantum fields living on a pseudo-Riemannian manifold, the
experiences of observers coupled to the field depend both on the
quantum state of the field and on the worldline of the
observer~\cite{birrell-davies,wald-smallbook}.  A~celebrated example
is the Unruh effect~\cite{unruh}, in which uniformly accelerated
observers in Minkowski spacetime experience a thermal bath in the
quantum state that inertial observers perceive as void of particles.
Other well-studied examples arise with black holes that emit Hawking
radiation \cite{hawking} and with observers in spacetimes of high
symmetry~\cite{gibb-haw:dS}.

A useful tool for analysing the experiences of an observer 
is a model particle detector that follows 
the observer's worldline and has internal states that 
couple to the quantum field.  
Such detectors are known as Unruh-DeWitt (UDW) detectors~\cite{unruh,dewitt}. 
While much of the early literature on the UDW detectors focused 
on stationary situations, 
including the Unruh effect~\cite{Takagi:1986kn,Crispino:2007eb}, 
the detectors remain well defined also in time-dependent situations, 
where they can be analysed within first-order perturbation theory 
\cite{Davies:2002bg,schlicht,Langlois:2005nf,langlois-thesis,louko-satz-spatial,satz,louko-satz-curved,hodgkinson-louko,hodgkinson-louko-btz,Barbado:2012fy,Hodgkinson:2013tsa,Hodgkinson:2014,ng:schw-ads-stationary} 
as well as by nonperturbative techniques~\cite{Raval:1995mb,Lin:2006jw,Ostapchuk:2011ud,Brown:2012pw,Bruschi:2012rx}. A~review with further references can be found in~\cite{Hu:2012jr}. 

In this paper we consider a detector coupled to a quantised scalar
field in 1+1 spacetime dimensions.  A scalar field in 1+1 dimensions
has local propagating degrees of freedom, and it exhibits the Hawking
and Unruh effects just like a scalar field in higher dimensions.
However, the dynamics of the field in 1+1 dimensions is significantly
simpler than in higher dimensions, especially for a massless minimally
coupled field, for which the field equation is conformally invariant
and conformal techniques are available.  In particular, the evolution
of a massles minimally coupled field on a (1+1)-dimensional collapsing
star spacetime reduces essentially to that of a massless field on
(1+1)-dimensional Minkowski spacetime in the presence of a receding
mirror, and the system is explicitly
solvable~\cite{Davies:1976hi,Davies:1977yv}.

The simplifications in the dynamics in $1+1$ dimensions 
come however at a cost: the
Wightman function of a minimally coupled field in 1+1 dimensions is
infrared divergent in the massless limit.  While Hadamard states can
still can be defined in terms of the short distance expansion of the
Wightman function~\cite{decanini-folacci}, the Hadamard expansion
contains an additive constant that is not determined by the quantum
state.  While this undetermined additive constant does not contribute
to stress-energy expectation
values~\cite{birrell-davies,fulling-ruijsenaars,Kay:2000fi}, it does
contribute to the transition probability of an UDW detector that
couples to the value of the field at the detector's location. In
stationary situations the ambiguous contribution to transition
probablities can be argued to vanish, under suitable assumptions about
the switch-on and
switch-off~\cite{Takagi:1986kn,Langlois:2005nf,langlois-thesis}, but
in nonstationary situations the ambiguity is more severe and has been
found to lead to physically undesirable predictions in examples that
include a receding mirror spacetime~\cite{Hodgkinson:2013tsa}.

In this paper we analyse a detector that is insensitive to the
infrared ambiguity of the $(1+1)$-dimensional Wightman function in the
massless minimaly coupled limit: we couple the detector linearly to
the \emph{derivative\/} of the field with respect to the detector's
proper
time~\cite{Davies:2002bg,Raval:1995mb,Raine:1991kc,Wang:2013lex},
rather than to the value of the field.  Working in first-order
perturbation theory, the detector's transition probability involves
then the double derivative of the Wightman function, rather than the
Wightman function itself.  We show that the response of the
$(1+1)$-dimensional derivative-coupling detector is closely similar to
that of the $(3+1)$-dimensional detector with a non-derivative
coupling~\cite{satz,louko-satz-curved}.  In particular, the transition
probablity can be written as an integral formula that involves no
short-distance regulator at the coincidence limit but contains instead
an additive term that depends only on the switching function that
controls the switch-on and switch-off.  A consequence is that in the
limit of sharp switching the transition probability diverges but the
transition rate remains finite.

We carry out three checks on the physical reasonableness of the 
derivative-coupling detector in stationary situations in which  
the switch-on is pushed to asymptotically early times. 
First, we verify that the transition rate is continuous 
in the limit of vanishing mass for an inertial detector in 
Minkowski space, with the field in Minkowski vacuum, 
and we show that the same holds also for a static detector 
in Minkowski half-space with Dirichlet and Neumann boundary conditions. 
This is evidence that the derivative coupling manages 
to remove the infrared ambiguity of the massless field  
without producing unexpected discontinuities in the massless limit. 
Second, we show that the transition rate of a 
uniformly accelerated detector in Minkowski space, 
coupled to a massless field in Minkowski vacuum, coincides 
with that of an inertial detector at rest with a thermal bath, 
being in particular thermal in the sense of the Kubo-Martin-Schwinger 
(KMS) property~\cite{Kubo:1957mj,Martin:1959jp}. 
This shows that the derivative-coupling 
detector sees the usual Unruh effect for a massless field. 
Third, we show that in a thermal bath of a massless field in Minkowski space, 
the transition rate of an inertial detector 
that is moving with respect to the bath is a sum of two terms that 
are individually thermal but at different temperatures, 
related to the temperature of the bath by Doppler shifts to the red and to the blue. 
As these terms stem respectively from the right-moving and left-moving 
components of the field, the temperature shifts are exactly as one would expect.  

After these checks, we focus on the massless minimally coupled field in 
two nonstationary situations, each of interest for the Hawking-Unruh effect. 

We first consider a Minkowski spacetime with a mirror whose exponentially 
receding motion makes the field mimic the late time behaviour of a 
field in a collapsing star spacetime~\cite{birrell-davies,Davies:1976hi,Davies:1977yv}. 
We show that at late times the detector's transition rate is the sum of a 
Planckian contribution, corresponding to the field modes propagating away from the mirror, 
and a vacuum contribution, corresponding to the field modes propagating towards the mirror. While 
the detector couples to the sum of the two parts, 
the two parts can nevertheless be unambiguously identified 
by considering their dependence on the detector's velocity with 
respect to the rest frame in which the mirror was static in the asymptotic past. 
We also show numerical results on how the transition rate evolves 
from the asymptotic early time form to the asymptotic late time form. 
These properties of the transition rate are in full agreement with 
the energy flux emitted by the mirror~\cite{birrell-davies,Davies:1976hi,Davies:1977yv}. 

We then consider a detector falling inertially into the 
$(1+1)$-dimensional Schwarzschild black hole, 
with the field in the Hartle-Hawking-Israel (HHI) and Unruh vacua. 
Starting the infall at the asymptotic infinity, 
we verify that the early time transition rate 
in the HHI vacuum is as in a thermal state, 
in the usual Hawking temperature, 
while in the Unruh vacuum it is as in a thermal state for the outgoing 
field modes and in the Boulware vacuum for the ingoing field modes. 
The outgoing and ingoing contributions can again be unambiguously 
identified by considering their dependence on the detector's velocity in the asymptotic past. 
The transition rate remains manifestly nonsingular on horizon-crossing, and we present 
numerical evidence of how the thermal properties are gradually lost during the infall. 
Near the black hole singularity the transition rate diverges 
proportionally to $r^{-3/2}$ where $r$ is the Schwarzschild radial coordinate. 
These results are in full 
agreement with the known properties of the HHI and Unruh vacua, 
including their thermality, 
their invariance under Schwarzschild time translations and their 
regularity across the future horizon. 

We anticipate that the derivative-coupling detector will be a useful tool 
for probing a quantum field in other situations where 
the infrared properties raise technical difficulties for the conventional UDW detector. 
One such instance is when the field has zero modes, which typically occur 
in spacetimes with compact spatial sections~\cite{Martin-Martinez:2014qda}. 
Other instances may arise in analogue spacetime systems 
\cite{Anderson:2014jua} or in 
spacetimes where the back-reaction due to 
Hawking evaporation is strong 
(for a small selection of references see 
\cite{braunstein-et-al,Mathur:2009hf,Almheiri:2012rt,Hotta:2013clt,Almheiri:2013wka}). 

The structure of the paper is as follows. 
In Section \ref{sec:detector}
we introduce the derivative-coupling detector, 
write the transition probability in a regulator-free form
and provide the formula for the transition rate in the sharp switching limit. 
Consistency checks in three stationary situations are carried out in Section~\ref{sec:tests}. 
Sections
\ref{sec:recmir}
and 
\ref{sec:bh}
address respectively the receding mirror spacetime and the Schwarzschild spacetime. 
The results are summarised and discussed in 
Section~\ref{sec:conc}. Details of technical calculations are deferred to four appendices. 

Our metric signature is $(-+)$, 
so that the norm squared of a timelike vector is negative. 
We use units in which $c=\hbar= k_B = 1$, so that frequencies, 
energies and temperatures have dimension inverse length. 
Spacetime points are denoted by Sans Serif
characters ($\mathsf{x}$) and spacetime indices are denoted by $a,b,\ldots$. 
Complex conjugation is denoted by overline. 
$O(x)$~denotes a quantity for
which $O(x)/x$ is bounded as $x\to0$, 
$o(x)$~denotes a quantity for
which $o(x)/x \to0$ as $x\to0$, 
$O(1)$ denotes a quantity that is bounded in the limit under consideration, 
and $o(1)$ denotes a quantity that goes to 
zero in the limit under consideration.

\section{Derivative-coupling detector\label{sec:detector}}

In this section we introduce an UDW 
detector that couples linearly to the proper 
time derivative of 
a scalar field \cite{Davies:2002bg,Raval:1995mb,Raine:1991kc,Wang:2013lex}. 
We show, within first-order perturbation theory, that in 
$(1+1)$ spacetime dimensions the detector's 
transition probability and transition 
rate are closely similar to those of a $(3+1)$-dimensional UDW detector 
with a non-derivative coupling~\cite{satz,louko-satz-curved}. 

\subsection{Derivative-coupling detector in $d\ge2$ dimensions}

Our detector 
is a spatially point-like quantum system with two distinct energy eigenstates. 
We denote the normalised energy eigenstates by 
$|0\rangle_D$ and $|\omega\rangle_D$, with the respective energy eigenvalues
$0$ and~$\omega$, where $\omega\ne0$. 

The detector moves in a spacetime of dimension $d\ge2$ along the smooth 
timelike worldline~$\mathsf{x}(\tau)$, where $\tau$ 
is the proper time, and it couples to a real scalar field $\phi$ 
via the interaction Hamiltonian 
\begin{align}
H_{\text{int}}
&=
c\chi(\tau)\mu(\tau) \, \frac{d}{d\tau} \phi\bigl(\mathsf{x}(\tau)\bigr) 
\ , 
\label{eq:Hint-derivative}
\end{align}
where $c$ is a coupling constant, $\mu$ is 
the detector's monopole moment operator 
and the switching function $\chi$ specifies how the interaction is 
switched on an off. 
We assume that $\chi$ is real-valued, 
non-negative and smooth with compact support. 

Where $H_{\text{int}}$ \eqref{eq:Hint-derivative} differs 
from the usual UDW detector \cite{unruh,dewitt} 
is that the interaction is linear in the proper 
time derivative of the field, rather than in the field itself. 
An alternative expression is $H_{\text{int}} =
c\chi(\tau)\mu(\tau) \, \dot{\mathsf{x}}^a \nabla_a \phi\bigl(\mathsf{x}(\tau)\bigr)$, 
where the overdot denotes $d/d\tau$. 

We take the detector to be initially in the state $|0\rangle_D$ 
and the field to be in a state~$|\psi\rangle$, which we assume to 
be regular in the sense of the Hadamard property~\cite{decanini-folacci}. 
After the interaction has been turned on and off, 
we are interested in the probability for the detector to 
have made a transition to the state~$|\omega\rangle_D$, 
regardless the final state of the field. 
Working in first-order perturbation theory in~$c$, 
we may adapt the analysis of the usual UDW detector 
\cite{birrell-davies,wald-smallbook} to show that this probability 
factorises as 
\begin{align}
\label{eq:prob}
P(\omega)=c^2{|_D\langle0|\mu(0)|\omega\rangle_D|}^2\mathcal{F}(\omega)
\ , 
\end{align}
where ${|_D\langle0|\mu(0)|\omega\rangle_D|}^2$ depends only on the internal structure of the detector but neither on $|\psi\rangle$, the trajectory or the switching, 
while the dependence on $|\psi\rangle$, 
the trajectory and the switching is encoded in the response function~$\mathcal{F}$. 
With our $H_{\text{int}}$ \eqref{eq:Hint-derivative}, 
the response function is given by 
\begin{align}
\label{eq:respfunc-def}
\mathcal{F}(\omega)
=\int^{\infty}_{-\infty}\,d\tau'\,\int^{\infty}_{-\infty}\,d\tau''\, 
\emath^{-\ii \omega(\tau'-\tau'')} \,\chi(\tau')\chi(\tau'') \, 
\partial_{\tau'}
\partial_{\tau''}
\mathcal{W}(\tau',\tau'')
\ , 
\end{align}
where the correlation function 
$\mathcal{W}(\tau',\tau'') 
\doteq \langle\psi|\phi\bigl(\mathsf{x}(\tau')\bigr)\phi\bigl(\mathsf{x}(\tau'')\bigr)|\psi\rangle$ 
is the pull-back of the Wightman function 
$\langle\psi|\phi(\mathsf{x}')\phi(\mathsf{x}'')|\psi\rangle$
to the detector's world line. 

From now on we drop the constant prefactors in \eqref{eq:prob} 
and refer to the response function as the transition probability. 

As $|\psi\rangle$ is Hadamard and the detector's worldline is smooth and timelike, 
the correlation function $\mathcal{W}$ is a well-defined distribution on 
$\BbbR\times\BbbR$ \cite{Fewster:1999gj,junker,hormander-vol1,hormander-paper1}. 
As $\chi$ is smooth with compact support, it follows that
$\mathcal{F}$ \eqref{eq:respfunc-def} 
is well defined: 
given a family of functions $\mathcal{W}_\epsilon$
that converges to the distribution $\mathcal{W}$ 
as $\epsilon\to0_+$, 
$\mathcal{F}$ is evaluated by first making in 
\eqref{eq:respfunc-def} the replacement 
$\mathcal{W} \to \mathcal{W}_\epsilon$, then performing the integrals, 
and finally taking the limit $\epsilon\to0_+$. 
The limit $\epsilon \to 0_+$ 
may however not necessarily be taken under the integrals. 
For the usual UDW detector, for which 
the response function is given as 
in \eqref{eq:respfunc-def} but without the derivatives, 
this issue is known to become subtle if one wishes
to define an instantaneous transition rate by passing to the limit of 
sharp switching: the subtleties start in 
three spacetime dimensions and increase 
as the spacetime dimension increases and the 
correlation function becomes more singular at the coincidence limit 
\cite{satz,louko-satz-curved,hodgkinson-louko,hodgkinson-louko-btz}. 
We may expect similar subtleties for the derivative-coupling detector, 
and since the derivatives in \eqref{eq:respfunc-def} increase the 
singularity of the integrand at the coincidence limit, 
we may expect the subtleties to start in  
lower spacetime dimensions than for the usual UDW detector. 

We confirm these expectations 
in subsections 
\ref{subsec:detector:1+1resp}, 
\ref{subsec:detector:1+1sharp}
and 
\ref{subsec:stationary}
by analysing 
the response function \eqref{eq:respfunc-def} 
and the transition rate in $(1+1)$ spacetime dimensions.


\subsection{$(1+1)$ response function: isolating the 
coincidence limit\label{subsec:detector:1+1resp}}

We now specialise to $(1+1)$ spacetime dimensions. 
In this subsection we write the response function 
\eqref{eq:respfunc-def} 
in a way where the contribution from the singularity 
of the integrand at the coincidence limit has been isolated. 

We start from \eqref{eq:respfunc-def} and write 
$\mathcal{W} =  \left(\mathcal{W} - \mathcal{W}_\text{sing}\right) 
+ \mathcal{W}_\text{sing}$, 
where $\mathcal{W}_\text{sing}$ is the locally integrable function 
\begin{align}
\mathcal{W}_\text{sing}(\tau',\tau'')
\doteq 
\begin{cases}
{\displaystyle -\frac{\ii \sgn(\tau'-\tau'')}{4}  - \frac{\ln |\tau'-\tau''|}{2 \pi}}
&
\text{for $\tau'\ne\tau''\,$,}
\\[1ex]
0 
&
\text{for $\tau'=\tau''\,$,}
\end{cases}
\label{eq:Wsing-def}
\end{align}
and $\sgn$ denotes the signum function, 
\begin{align}
\sgn x 
\doteq 
\begin{cases}
1&
\text{for $x>0\,$,}
\\
0 
&
\text{for $x=0\,$,}
\\
-1 
&
\text{for $x<0\,$.}
\end{cases}
\label{eq:sgn-def}
\end{align}
We obtain 
\begin{subequations} 
\begin{align}
\mathcal{F}(\omega) &= \mathcal{F}_\text{reg}(\omega)+ \mathcal{F}_\text{sing}(\omega)
\ , 
\\[2ex]
\mathcal{F}_\text{reg}(\omega)
&=\int^{\infty}_{-\infty}\,d\tau'\,\int^{\infty}_{-\infty}\,d\tau''\, 
\emath^{-\ii \omega(\tau'-\tau'')} \,\chi(\tau')\chi(\tau'') \, 
\partial_{\tau'}
\partial_{\tau''}
\bigl[ \mathcal{W}(\tau',\tau'') - \mathcal{W}_\text{sing}(\tau',\tau'') \bigr] 
\ , 
\label{eq:Freg1}
\\
\mathcal{F}_\text{sing}(\omega)
&=
\int^{\infty}_{-\infty}\,d\tau'\,\int^{\infty}_{-\infty}\,d\tau''\, 
\emath^{-\ii \omega(\tau'-\tau'')} \,\chi(\tau')\chi(\tau'') \, 
\partial_{\tau'}
\partial_{\tau''}
\, 
\mathcal{W}_\text{sing}(\tau',\tau'')
\ ,  
\label{eq:Fsing0}
\end{align}
\end{subequations}
where the derivatives are understood in the distributional sense. 

Consider first $\mathcal{F}_\text{reg}$~\eqref{eq:Freg1}.  The
Hadamard short distance form of the Wightman function
\cite{decanini-folacci} implies that both $\mathcal{W}(\tau',\tau'')$
and $\partial_{\tau'}
\partial_{\tau''} \bigl[ \mathcal{W}(\tau',\tau'') -
\mathcal{W}_\text{sing}(\tau',\tau'') \bigr] $ are represented in a
neighbourhood of $\tau'=\tau''$ by locally integrable functions.  It
follows that the integral in \eqref{eq:Freg1} receives no
distributional contributions from $\tau'=\tau''$, and the integral can
hence be decomposed into integrals over the subdomains $\tau'>\tau''$
and $\tau'<\tau''$.  In the subdomain $\tau'>\tau''$ we write
$\tau'=u$ and $\tau'' = u-s$, where $u \in \BbbR$ and $0<s<\infty$,
and in the subdomain $\tau'<\tau''$ we write $\tau''=u$ and $\tau' =
u-s$, where again $u \in \BbbR$ and $0<s<\infty$. Using the property
$\mathcal{W}(\tau',\tau'') = \overline{\mathcal{W}(\tau'',\tau')}$ and
the explicit form of
$\mathcal{W}_\text{sing}(\tau',\tau'')$~\eqref{eq:Wsing-def}, we
obtain
\begin{align}
\mathcal{F}_\text{reg}(\omega)
=
2  \int_{-\infty}^{\infty} \! d u \, \int_{0}^{\infty} 
\! ds \,  
\chi(u) \chi(u-s) \Realpart \left[ \emath^{-\ii \omega s}
\left(\mathcal{A}(u,u-s) + \frac{1}{2\pi s^2}\right) \right] 
\ , 
\label{eq:Freg2}
\end{align}
where 
\begin{align}
\mathcal{A}(\tau',\tau'') 
\doteq 
\partial_{\tau'}
\partial_{\tau''}
\mathcal{W}(\tau',\tau'')
\ . 
\label{eq:A-distr}
\end{align}
Note that the integrand in \eqref{eq:Freg2} 
is still a distribution, but it is 
represented by a locally integrable 
function in a neighbourhood of $s=0$, 
and any distributional singularities
are hence isolated from $s=0$. 

We evaluate $\mathcal{F}_\text{sing}$ \eqref{eq:Fsing0} in  
Appendix~\ref{app:Fsing}. Combining \eqref{eq:Freg2} 
and~\eqref{eq:Fsing-clean1-appendix}, we find 
\begin{align}
\mathcal{F}(\omega)
&=
-\omega \Theta(-\omega) 
\int_{-\infty}^{\infty} d u \, {[\chi(u)]}^2
+ 
\frac{1}{\pi}
\int^{\infty}_{0} 
ds \, 
\frac{\cos(\omega s)}{s^2} 
\int_{-\infty}^{\infty} d u \, 
\chi(u) [\chi(u) - \chi(u-s)] 
\notag
\\[1ex]
&\hspace{3ex}
+ 2  \int_{-\infty}^{\infty} \! d u \, \int_{0}^{\infty} 
\! ds \,  
\chi(u) \chi(u-s) \Realpart \left[ \emath^{-\ii \omega s}
\left(\mathcal{A}(u,u-s) + \frac{1}{2\pi s^2}\right) \right] 
\ , 
\label{eq:F-final-nice}
\end{align}
where
$\mathcal{A}$ is given by \eqref{eq:A-distr}
and $\Theta$ is the Heaviside function, 
\begin{align}
\Theta (x) 
\doteq 
\begin{cases}
1&
\text{for $x>0\,$,}
\\
0 
&
\text{for $x\le0\,$.}
\end{cases}
\label{eq:heaviside-def}
\end{align}
An equivalent form, using for $\mathcal{F}_\text{sing}$ the alternative expression 
\eqref{eq:Fsing-clean2-appendix} given in
Appendix~\ref{app:Fsing}, is 
\begin{align}
\mathcal{F}(\omega)
&=
- \frac{\omega}{2} 
\int_{-\infty}^{\infty} \! d u \, {[\chi(u)]}^2
+
\frac{1}{\pi} \int_{0}^{\infty} \! \frac{ds}{s^2} \,
\int_{-\infty}^{\infty} \! du \,  
\chi(u) \left[ \chi(u) - \chi(u-s) \right] 
\notag
\\[1ex]
&\hspace{3ex}
+ 2  \int_{-\infty}^{\infty} \! d u \, \int_{0}^{\infty} 
\! ds \,  
\chi(u) \chi(u-s) \Realpart \left[ \emath^{-\ii \omega s}
\mathcal{A}(u,u-s) + \frac{1}{2\pi s^2} \right] 
\ . 
\label{eq:F-final-alt}
\end{align}
The integral over $s$ 
in the second term in \eqref{eq:F-final-nice} and \eqref{eq:F-final-alt} 
is convergent at small $s$ since 
the integral over $u$ produces 
an even function of $s$ that vanishes at $s=0$. 

The expression \eqref{eq:F-final-alt} 
for the response function is closely similar to that 
obtained in \cite{satz,louko-satz-curved} 
for the usual, non-derivative UDW detector in $(3+1)$ 
dimensions. 
This happens because of the similarity between 
the coincidence limit singularities of the 
twice differentiated $(1+1)$-dimensional  
correlation function, 
appearing in~\eqref{eq:respfunc-def}, 
and the undifferentiated $(3+1)$-dimensional 
correlation function that appears in the similar expression for 
non-derivative coupling. 

We re-emphasise that the last term in \eqref{eq:F-final-nice} 
and \eqref{eq:F-final-alt}
may contain distributional contributions from $s>0$. 
Similar distributional contributions 
were not considered for the $(3+1)$-dimensional 
non-derivative UDW detector in~\cite{louko-satz-curved}, 
but they can occur also there, and the analysis in 
\cite{louko-satz-curved}  
can be 
amended to include these contributions by 
proceeding as in the present paper. 
Similar distributional contributions can 
arise in any spacetime dimension: 
in $(2+1)$ dimensions,
examples on the Ba\~ndados-Teitelboim-Zanelli 
black hole were encountered in~\cite{hodgkinson-louko-btz}.

\subsection{$(1+1)$ sharp switching limit: 
transition rate\label{subsec:detector:1+1sharp}}

In this subsection we consider the sharp switching 
limit of the derivative-coupling detector in 
$(1+1)$ dimensions. 

%

Following \cite{satz,louko-satz-curved}, we consider a 
family of switching functions given by 
\begin{align}
\chi(u) = h_1 \! \left(\frac{u - \tau_0 + \delta}{\delta}\right)
\times h_2 \! \left(\frac{-u + \tau + \delta}{\delta}\right),
\label{eq:chi-profile}
\end{align}
where $\tau$ and $\tau_0$ are pararameters satisfying $\tau>\tau_0$, 
$\delta$ is a positive parameter, and 
$h_1$ and $h_2$ are smooth non-negative functions such that 
$h_1(x) = h_2(x) = 0$ for $x \le 0$ and $h_1(x) = h_2(x) = 1$ for $x \ge 1$. 
In words, the detector is switched on over an interval  
of duration $\delta$ before time~$\tau_0$, 
stays on at constant coupling from time $\tau_0$ 
to time~$\tau$, and is finally switched off over an interval of duration $\delta$ 
after time~$\tau$. 
The profile of the switch-on is determined by $h_1$ 
and the profile of the switch-off is determined by~$h_2$. 

We are interested in the limit $\delta\to0$. 
To begin with, suppose that 
$\mathcal{A}(\tau',\tau'')$ 
\eqref{eq:A-distr}
is represented by a smooth function for $\tau'\ne\tau''$. 
Given the similarity between
\eqref{eq:F-final-alt} and the $(3+1)$-dimensional 
non-derivative response function given 
by equation (3.16) in~\cite{louko-satz-curved}, 
we may follow the analysis that led to equations (4.4) and (4.5)
in~\cite{louko-satz-curved}. 
For the response function, we find 
\begin{align}
\mathcal{F}(\omega,\tau) 
& = 
- \frac{\omega \Delta\tau}{2}
+ 2  \int_{\tau_0}^{\tau} \! du \, \int_{0}^{u - \tau_0} 
\! ds \, 
\Realpart \left[\emath^{-\ii \omega s} \mathcal{A}(u,u-s) 
+ \frac{1}{2\pi s^2}\right] 
\notag 
\\[1ex]
& \hspace{3ex}
+\frac{1}{\pi} \ln \! \left(\frac{\Delta \tau}{\delta}\right) 
+ C + O(\delta),
\label{eq:sharpF}
\end{align}
where $\Delta\tau \doteq \tau - \tau_0$, 
$C$ is a constant that depends only on $h_1$ and~$h_2$, and
we have included in $\mathcal{F}(\omega,\tau)$ the second argument $\tau$ 
to indicate explicitly the dependence on~$\tau$. 

The response function \eqref{eq:sharpF}
hence diverges logarithmically as $\delta\to0$, 
but the divergent contribution is a pure switching effect, independent of 
the quantum state and of the detector's trajectory. The transition rate, 
defined as $\dot{\mathcal{F}}(\omega,\tau) 
\doteq \frac{\partial}{\partial\tau} \mathcal{F}(\omega,\tau)$, 
remains finite as $\delta\to0$, 
and is in this limit given by 
\begin{align}
\dot{\mathcal{F}}(\omega,\tau) 
= 
-\frac{\omega}{2}
+ 
2  \int_{0}^{\Delta \tau} 
\! ds \Realpart \left[\emath^{-\ii \omega s} \mathcal{A}(\tau,\tau-s) 
+ \frac{1}{2\pi s^2}\right] +\frac{1}{\pi \Delta \tau} 
\ . 
\label{eq:Fdot-sharp}
\end{align}
An equivalent expression, 
obtained by writing $1 = \cos(\omega s) + [1 - \cos(\omega s)]$ 
under the integral in \eqref{eq:Fdot-sharp}, 
is 
\begin{align}
\dot{\mathcal{F}}(\omega,\tau) 
&= 
-\omega \Theta(-\omega)
+ \frac{1}{\pi}
\left[
\frac{\cos(\omega \Delta\tau)}{\Delta\tau}
+ 
|\omega| \si(|\omega| \Delta\tau)
\right]
\notag
\\[1ex]
& \hspace{3ex}
+ 
2 \int_{0}^{\Delta \tau} 
\! ds \Realpart \left[ \emath^{-\ii \omega s} 
\left(\mathcal{A}(\tau,\tau-s) 
+ \frac{1}{2\pi s^2}\right) \right]
\ , 
\label{eq:Fdot-sharp-alt}
\end{align}
where $\si$ is the sine integral in the notation of~\cite{dlmf}. 
When the switch-on is in the asymptotic past 
and the fall-off of $\mathcal{A}(\tau,\tau-s)$ 
is sufficiently fast as large~$s$, 
the $\Delta\tau \to \infty$ limit of 
\eqref{eq:Fdot-sharp-alt} gives 
\begin{align}
\dot{\mathcal{F}}(\omega,\tau) 
= 
-\omega \Theta(-\omega)
+ 
2  \int_{0}^{\infty} 
\! ds \Realpart \left[\emath^{-\ii \omega s} 
\left(\mathcal{A}(\tau,\tau-s) 
+ \frac{1}{2\pi s^2}\right)\right]
\ . 
\label{eq:Fdot-sharp-aspast}
\end{align}
The observational meaning of the transition rate relates to ensembles
of ensembles of detectors (see Section 5.3.1 of
\cite{langlois-thesis} or Section 2 of~\cite{louko-satz-curved}).

When $\mathcal{A}(\tau',\tau'')$ \eqref{eq:A-distr} is not represented
by a smooth function for $\tau'\ne\tau''$, the estimates leading to
\eqref{eq:sharpF} and \eqref{eq:Fdot-sharp} need not hold, and the
transition rate need not have a well-defined $\delta\to0$ limit for
all $\tau_0$ and~$\tau$. In particular, if the detector is switched on
at a finite time $\tau_0$ and $\mathcal{A}(\tau',\tau'')$ has a
distributional singularity at $(\tau',\tau'') = (\tau,\tau_0)$, the
integral expressions in \eqref{eq:Fdot-sharp} and
\eqref{eq:Fdot-sharp-alt} would not be well defined because the
singularity occurs at an end-point of the integration.  If the
switch-on is in the asymptotic past, however, the transition rate
formula \eqref{eq:Fdot-sharp-aspast} is well defined even when
$\mathcal{A}(\tau',\tau'')$ has distributional singularities for
$\tau'\ne\tau''$ provided these singularities are sufficiently
isolated.  We shall encounter examples of such singularities in
subsection \ref{massFdot} and Section~\ref{sec:recmir}.

Similar remarks about singularities of 
the correlation function at timelike-separated points 
apply also to the sharp switching limit of the 
non-derivative UDW detector in $(3+1)$ dimensions. 
The transition rate results given in \cite{louko-satz-curved} 
for a switch-on at a finite time 
hold when no such singularities are present.

\subsection{Stationary transition rate\label{subsec:stationary}}

Suppose that the Wightman function is stationary with respect to the 
detector's trajectory, 
in the sense that $\mathcal{W}(\tau',\tau'')$ depends on $\tau'$ and $\tau''$ 
only through the difference $\tau'-\tau''$. 
When the detector is switched on in the asymptotic past, the transition rate 
\eqref{eq:Fdot-sharp-aspast} reduces to 
\begin{align}
\dot{\mathcal{F}}(\omega)
& = 
-\omega \Theta(-\omega)
+ 
2  \int_{0}^{\infty} 
\! ds \Realpart  \left[ \emath^{-\ii \omega s} 
\left(\mathcal{A}(s,0) 
+ \frac{1}{2\pi s^2}\right)\right]  
\notag
\\[1ex]
& = 
-\omega \Theta(-\omega)
+ 
\int_{-\infty}^{\infty} 
\! ds \, \emath^{-\ii \omega s} 
\left[\mathcal{A}(s,0) 
+ \frac{1}{2\pi s^2}\right]
\notag
\\[1ex]
& = 
-\omega \Theta(-\omega)
+ 
\int_C
\! ds \, \emath^{-\ii \omega s} 
\left[\mathcal{A}(s,0) 
+ \frac{1}{2\pi s^2}\right]  
\notag
\\[1ex]
& = 
\int_{-\infty}^{\infty} 
\! ds \, \emath^{-\ii \omega s} 
\, \mathcal{A}(s,0) 
\ , 
\label{eq:Fdot-stationary-naive}
\end{align}
where we have dropped the second argument $\tau$ from $\dot{\mathcal{F}}$ 
as the transition rate is now independent of~$\tau$, 
and $\mathcal{A}(s,0)$ is understood as a distribution everywhere, 
including $s=0$. 
In \eqref{eq:Fdot-stationary-naive} we have first used the properties 
$\mathcal{A}(\tau',\tau'') = \mathcal{A}(\tau'-\tau'',0)$ and  
$\mathcal{A}(\tau',\tau'') = \overline{\mathcal{A}(\tau'',\tau')}$. 
Next, we have deformed the real $s$ axis into a contour $C$ in the complex $s$ 
plane, such that $C$ follows the real axis except that it dips into the 
lower half-plane near $s=0$; this deformation is justified by the Hadamard 
short separation form of the Wightman function~\cite{decanini-folacci}. 
In the contour integral over~$C$, 
we have then separated the two terms in the integrand, 
evaluated the integral of the second term by a standard contour technique, 
and noted that in the first term $C$ can be deformed back 
to the real $s$ axis provided the integrand is 
understood as a distribution 
for all~$s$, including $s=0$. 

The result \eqref{eq:Fdot-stationary-naive} coincides with the 
transition rate that one obtains from the response function \eqref{eq:respfunc-def} 
by the usual procedure of setting $\chi=1$ and 
formally factoring out the infinite total detection time~\cite{birrell-davies}.

\section{Stationary checks: 
massless limit, the Unruh effect, and inertial response in a thermal bath\label{sec:tests}}

In this section we perform reasonableness checks on the 
derivative-coupling detector in three stationary situations.   
First, we verify that the transition rate is continuous in the 
massless limit for a static detector in 
Minkowski space, and also in Minkowski half-space 
with Dirichlet and Neumann boundary conditions. 
Second, we verify that the detector sees the usual Unruh effect 
when the field is massless. 
Third, we verify that the transition rate of an inertial detector 
in the thermal bath of a massless field sees a Doppler shift 
when the detector has a nonvanishing velocity in the inertial frame of the bath.

\subsection{Static detector in Minkowski (half-)space\label{massFdot}}

Let $\mathcal{M}$ be $(1+1)$ Minkowski spacetime, with standard Minkowski coordinates 
$(t,x)$ in which the metric reads $ds^2 = - dt^2 + dx^2$, and let  
$\Mhalfspace$ be the submanifold of $\mathcal{M}$ 
in which $x>0$.  
We consider in $\mathcal{M}$ and $\Mhalfspace$ 
a scalar field of mass $m\ge0$, and in $\Mhalfspace$ 
we impose the Dirichlet or Neumann boundary condition 
that the field or its normal derivative vanish at $x=0$. 

We set the field in $\mathcal{M}$ in the 
Minkowski vacuum $|0\rangle$, 
and 
the field in $\Mhalfspace$ 
in the Minkowski-like vacuum $|\tilde0\rangle$ 
that is the no-particle state with respect 
to the timelike Killing vector~$\partial_t$. 

Now, consider in $\mathcal{M}$ and $\Mhalfspace$ 
a detector on the static worldline 
\begin{align}
\mathsf{x}(\tau) = (\tau, d)
\ , 
\label{eq:Mink-static-traj}
\end{align}
where $d$ is a positive constant. 
In $\mathcal{M}$ the value of $d$ has no geometric significance, 
but in $\Mhalfspace$ $d$ is the distance of the detector from the mirror at $x=0$. 
We take the detector to be switched on in the asymptotic past, 
so that the detector's transition rate is stationary 
and given by~\eqref{eq:Fdot-stationary-naive}. 
We shall show that the transition rate is continuous in the limit $m\to0$.

\subsubsection{$m>0$}

For $m>0$, the Wightman function in $\mathcal{M}$ 
is \cite{birrell-davies}
\begin{align}
\langle0|
\phi(\mathsf{x}) \phi(\mathsf{x}') |0\rangle
= \frac{1}{2 \pi} K_0 
\left[
m \sqrt{ (\Delta x)^2 - (\Delta t - \ii \epsilon)^2}
\right] 
\ , 
\label{eq:mpos-Wightman-Mink}
\end{align}
where $\Delta x = x-x'$, $\Delta t = t-t'$, 
$K_0$ is the modified Bessel function of the second kind~\cite{dlmf}, 
and the expression is understood as a distribution 
in the sense of $\epsilon\to0_+$. 
The square root is positive when $\mathsf{x}$ and $\mathsf{x}'$ 
are spacelike separated and $\epsilon\to0_+$, 
and the continuation to general $\mathsf{x}$ and $\mathsf{x}'$ 
is specified by the $\ii\epsilon$ prescription. 
By the method of images, the 
Wightman function in $\Mhalfspace$ is the sum of 
\eqref{eq:mpos-Wightman-Mink} 
and the image piece 
\begin{align}
\langle\tilde0|
\phi(\mathsf{x}) \phi(\mathsf{x}') |\tilde0\rangle
- 
\langle0|
\phi(\mathsf{x}) \phi(\mathsf{x}') |0\rangle
= 
\frac{\eta}{2 \pi} K_0 
\left[
m \sqrt{ (x+x')^2 - (\Delta t - \ii \epsilon)^2}
\right] 
\ , 
\label{eq:mpos-Wightman-image}
\end{align}
where $\eta=-1$ for Dirichlet and $\eta=1$ for Neumann. 

We evaluate the transition rate \eqref{eq:Fdot-stationary-naive}
in Appendix~\ref{app:mink-eval}. We obtain 
\begin{subequations}
\label{eq:mpos-Fdot-combined}
\begin{align}
\mathcal{M}: 
\hspace{4ex}
\dot{\mathcal{F}}(\omega)
&= \frac{\omega^2}{\sqrt{\omega^2-m^2}} 
\, \Theta(-\omega-m)
\ , 
\label{eq:mpos-Fdot-Minkowski}
\\[1ex]
\Mhalfspace: 
\hspace{4ex}
\dot{\mathcal{F}}(\omega)
&= \frac{\omega^2 
\bigl[ 1 + \eta \cos\bigl(2d\sqrt{\omega^2 - m^2}\,\bigr) \bigr]}{\sqrt{\omega^2-m^2}} 
\, \Theta(-\omega-m)
\ . 
\label{eq:mpos-Fdot-Mtilde}
\end{align}
\end{subequations}
The transition rate is non-negative, and it is 
nonvanishing only for $\omega< -m$, that is, 
for de-excitations exceeding the mass gap. 
These are properties that one would expect of a 
reasonable detector coupled to a massive field. 

\subsubsection{$m=0$}

On $\mathcal{M}$, the massive Wightman function 
\eqref{eq:mpos-Wightman-Mink} diverges as $m\to0$. 
However, 
the quantity 
$\langle0|
\phi(\mathsf{x}) \phi(\mathsf{x}') |0\rangle 
+ \frac{1}{2\pi}
\ln[m \emath^\gamma/(2\mu)]$, 
where $\gamma$ is Euler's constant and $\mu$ is 
a positive constant of dimension inverse length, 
has at $m\to0$ a finite limit, given by \cite{dlmf}
\begin{align}
\langle0|
\phi(\mathsf{x}) \phi(\mathsf{x}') |0\rangle
\doteq - \frac{1}{2 \pi} 
\ln 
\left[
\mu \sqrt{ (\Delta x)^2 - (\Delta t - \ii \epsilon)^2}
\right] 
\ . 
\label{eq:mzero-Wightman-Mink}
\end{align}
We take \eqref{eq:mzero-Wightman-Mink} 
as the definition of the Wightman function for $m=0$. 
The constant $\mu$ is required for dimensional consistency, 
and its arbitrariness 
means that 
$\langle0|
\phi(\mathsf{x}) \phi(\mathsf{x}') | 0\rangle$
\eqref{eq:mzero-Wightman-Mink}
is unique up to an additive constant. 
The massless Wightman function on $\Mhalfspace$ is the sum of 
\eqref{eq:mzero-Wightman-Mink}
and the image piece 
\begin{align}
\langle{\tilde0}|
\phi(\mathsf{x}) \phi(\mathsf{x}') |{\tilde0}\rangle
- 
\langle0|
\phi(\mathsf{x}) \phi(\mathsf{x}') |0\rangle
= - \frac{\eta}{2 \pi} 
\ln 
\left[
\mu \sqrt{ (x+x')^2 - (\Delta t - \ii \epsilon)^2}
\right] 
\ , 
\label{eq:mzero-Wightman-image}
\end{align}
where again $\eta=-1$ for Dirichlet and $\eta=1$ for Neumann. 
Note that for $\eta=-1$, the massless Wightman function on 
$\Mhalfspace$ is independent of $\mu$ 
and can be obtained as the $m\to0$ limit of the massive 
Wightman function on $\Mhalfspace$ without introducing a subtraction by hand. 

We show in Appendix \ref{app:mink-eval} that the transition 
rate is given by  
\begin{subequations}
\label{eq:mzero-Fdot-combined}
\begin{align}
\mathcal{M}: 
\hspace{4ex}
\dot{\mathcal{F}}(\omega)
&= -\omega
\, \Theta(-\omega)
\ , 
\label{eq:mzero-Fdot-Minkowski}
\\[1ex]
\Mhalfspace: 
\hspace{4ex}
\dot{\mathcal{F}}(\omega)
&= - \omega \, [ 1 + \eta \cos(2d\omega) ] 
\, \Theta(-\omega)
\ . 
\label{eq:mzero-Fdot-Mtilde}
\end{align}
\end{subequations}
The transition rate is non-negative, and it is nonvanishing only for de-excitations, 
as one would expect of a reasonable detector coupled to a massless field. 

We see from 
\eqref{eq:mpos-Fdot-combined}
and 
\eqref{eq:mzero-Fdot-combined}
that the massless transition rate 
is equal to the massless limit of the massive transition rate. 
This is the property that we wished to verify.

\subsection{Unruh effect\label{thermalFdot}}

Let again $\mathcal{M}$ be $(1+1)$ Minkowski spacetime,
and consider in $\mathcal{M}$ 
a massless field in the Minkowski vacuum. 
We consider a detector on the uniformly accelerated worldline 
\begin{align}
\mathsf{x}(\tau) = \bigl(a^{-1} \sinh(a \tau), a^{-1} \cosh (a \tau) \bigr) 
\ , 
\label{eq:Mink-rindler-traj}
\end{align}
where the positive constant $a$ is the magnitude of the proper acceleration. 
The trajectory is stationary with respect to the boost Killing vector 
$t\partial_x + x\partial_t$, and $|0\rangle$ is invariant under this Killing vector. 
With the detector switch-on pushed to the asymptotic past, 
the transition rate is independent of time and given by~\eqref{eq:Fdot-stationary-naive}. 

From \eqref{eq:A-distr}, 
\eqref{eq:mzero-Wightman-Mink}
and 
\eqref{eq:Mink-rindler-traj}, 
we find 
\begin{align}
\mathcal{A}(\tau', \tau'')= - \frac{a^2}{8 \pi \sinh^2\bigl(a(\tau'-\tau''-\ii\epsilon)/2\bigr)} 
\ . 
\label{eq:Rindler-Afunction}
\end{align}
Substituting \eqref{eq:Rindler-Afunction} in~\eqref{eq:Fdot-stationary-naive}, 
deforming the contour of $s$-integration to $s = -i\pi/a + r$ where $r\in\BbbR$, 
and using formula 3.985.1 in~\cite{grad-ryzh}, we find 
\begin{align}
\dot{\mathcal{F}}(\omega)
& = 
\frac{\omega}{\emath^{2 \pi \omega/a}-1}
\ . 
\label{eq:Rindler-transrate}
\end{align}
The transition rate \eqref{eq:Rindler-transrate}
satisfies the KMS relation~\cite{Kubo:1957mj,Martin:1959jp}, 
\begin{align}
\dot{\mathcal{F}}(\omega) = \emath^{-\omega / T} \dot{\mathcal{F}}(-\omega)
\ , 
\label{eq:KMS-gen}
\end{align}
with $T = a/(2\pi)$, and is hence thermal at temperature $a/(2\pi)$ in the KMS sense. 
We conclude that the detector does see the usual Unruh effect \cite{unruh,Takagi:1986kn}. 
The Planckian form of the transition rate \eqref{eq:Rindler-transrate} 
is identical to that of a non-derivative detector coupled to a massless 
field on a uniformly accelerated trajectory in $(3+1)$ dimensions~\cite{unruh,Takagi:1986kn}.

\subsection{Inertial detector in a thermal bath}

We consider again a massless field in $(1+1)$ Minkowski spacetime~$\mathcal{M}$, 
but now in the thermal state $|T\rangle$ of positive temperature~$T$. 
Working in Minkowski coordinates $(t,x)$ in which the thermal bath is at rest, 
the thermal Wightman function is obtained from the vacuum Wightman function 
by taking an image sum in $t$ with period $\ii/T$~\cite{birrell-davies}. 
With the vacuum Wightman function~\eqref{eq:mzero-Wightman-Mink}, 
the sum reads
\begin{align}
\langle T|
\phi(\mathsf{x}) \phi(\mathsf{x}') |T\rangle
= - \frac{1}{4 \pi} 
\sum_{n=-\infty}^\infty 
\ln 
\left\{
\mu^2  \left[ (\Delta x)^2 - (\Delta t - \ii \epsilon + \ii n /T)^2 \right] 
\right\}
\ , 
\label{eq:mzero-Wightman-thermal-formalsum}
\end{align}
and does not converge. 
However, differentiation of the sum in \eqref{eq:mzero-Wightman-thermal-formalsum} term by 
term with respect to $\Delta x$ gives a new sum that converges and can be summed by 
residues into an elementary function. 
We integrate the elementary function with respect to $\Delta x$
and fix the $\Delta t$-dependent integration constant by 
requiring that the massless Klein-Gordon equation is satisfied, 
and requiring evenness in $\Delta t$ for $(\Delta x)^2 - (\Delta t)^2 > 0$. 
The outcome is 
\begin{align}
\langle T|
\phi(\mathsf{x}) \phi(\mathsf{x}') |T\rangle
\doteq 
& \,
-\frac{1}{4 \pi}
\ln \! \left\{\sinh[\pi T (\Delta x + \Delta t - \ii\epsilon)]\right\}
\notag
\\[1ex]
& \,
-\frac{1}{4 \pi}
\ln \! \left\{\sinh[\pi T (\Delta x - \Delta t + \ii\epsilon)]
\right\}
\ , 
\label{eq:mzero-Wightman-thermal-summed}
\end{align}
uniquely up to an additive constant, 
and we take \eqref{eq:mzero-Wightman-thermal-summed} 
as the definition of the thermal Wightman function. 
Note that \eqref{eq:mzero-Wightman-thermal-summed} 
decomposes into the right-mover contribution that depends on 
$\Delta (x - t)$ and the left-mover contribution that depends on 
$\Delta (x + t)$. 


We consider the inertial detector worldline 
\begin{align}
\mathsf{x}(\tau) = 
(\tau\cosh\lambda, - \tau \sinh\lambda)
\ , 
\label{eq:Mink-inertial-traj}
\end{align}
where $\lambda\in\BbbR$ is the detector's 
rapidity parameter with respect to the 
rest frame of the bath. 
For later convenience, we have chosen the sign in
\eqref{eq:Mink-inertial-traj} so that a detector with positive
$\lambda$ is moving towards decreasing~$x$.  
From \eqref{eq:A-distr}, 
\eqref{eq:mzero-Wightman-thermal-summed}
and 
\eqref{eq:Mink-inertial-traj}
we find 
\begin{align}
\mathcal{A}(\tau',\tau'') = -\frac{1}{16 \pi} 
\left( \frac{{(2 \pi T_+)}^2}{\sinh^2[\pi T_+ (\tau'-\tau'' - \ii\epsilon)]} 
+\frac{{(2 \pi T_-)}^2}{\sinh^2[\pi T_- (\tau'-\tau'' - \ii\epsilon)]}  \right)
\ , 
\end{align}
where $T_\pm \doteq \emath^{\pm \lambda} T$. 
Taking the detector to be switched on in the asymptotic past, 
and proceeding as with \eqref{eq:Rindler-Afunction}, 
we find that the transition rate is given by 
\begin{align}
\dot{\mathcal{F}}(\omega) 
= \frac{\omega}{2} \left( \frac{1}{\emath^{\omega/T_+}-1}
+\frac{1}{\emath^{\omega/T_-}-1}\right)
\ , 
\label{eq:Mink-drift-transrate}
\end{align}
simplifying in the special case $\lambda=0$ to 
\begin{align}
\dot{\mathcal{F}}(\omega) 
= \frac{\omega}{\emath^{\omega/T}-1}
\ . 
\label{eq:Mink-zerodrift-transrate}
\end{align}

The $\lambda=0$ transition rate \eqref{eq:Mink-zerodrift-transrate} 
satisfies the KMS relation \eqref{eq:KMS-gen} in temperature~$T$, and it 
coincides with the transition rate 
\eqref{eq:Rindler-transrate} of a uniformly accelerated detector 
when $T = a/(2\pi)$. 
The $\lambda\ne0$ transition rate 
\eqref{eq:Mink-drift-transrate}
is a sum of the right-mover and left-mover contributions, each 
satisfying the KMS relation but in the respective 
Doppler-shifted temperatures~$T_\pm$. 
These are properties that one would expect of a reasonable detector. 

\subsection{Inertial detector with vacuum 
left-movers and thermal right-movers}

In preparation for the nonstationary situations that will be addressed
in Sections \ref{sec:recmir} and~\ref{sec:bh}, we consider here the
inertial detector \eqref{eq:Mink-inertial-traj} in the state in which
the left-movers are in the Minkowski vacuum but the right-movers are
in a thermal bath with temperature~$T$. As the left-movers and the
right-movers decouple, the results can be read off from those given
above in a straightforward way. Taking the switch-on to the asymptotic
past, we find
\begin{align}
\dot{\mathcal{F}}(\omega)  = 
-\frac{\omega}{2} 
\Theta (-\omega) 
+ \frac{\omega}{2 \, (\emath^{\omega/T_+}-1)}
\ . 
\label{eq:Mink-mixed-trrate}
\end{align}
The first term in \eqref{eq:Mink-mixed-trrate} 
is the left-mover contribution, equal to half of
the Minkowski transition rate. The second term is the right-mover
contribution, which is Planckian in the Doppler-shifted temperature
$T_+ = \emath^{\lambda} T$.

\section{The receding mirror spacetime\label{sec:recmir}}

In this section we consider a massless field in $(1+1)$-dimensional
Minkowski spacetime with a receding mirror that asymptotes at late
times to a null line, in a fashion that mimics the late time
redshift that occurs in a collapsing star
spacetime~\cite{birrell-davies,Davies:1976hi,Davies:1977yv}.
Focusing on a specific mirror trajectory that is asymptotically
inertial at early times, and choosing a vacuum with no incoming
radiation from infinity, we compute the transition rate of an
inertial, sharply-switched detector that is turned on in the
asymptotic past. We show that the
early time transition rate is Minkowskian and the late time transition
rate has the expected form of Planckian radiation emitted from the
mirror.

\subsection{Mirror spacetime and the in-vacuum}

\begin{figure}[t]
\centering
\includegraphics[width=\textwidth]{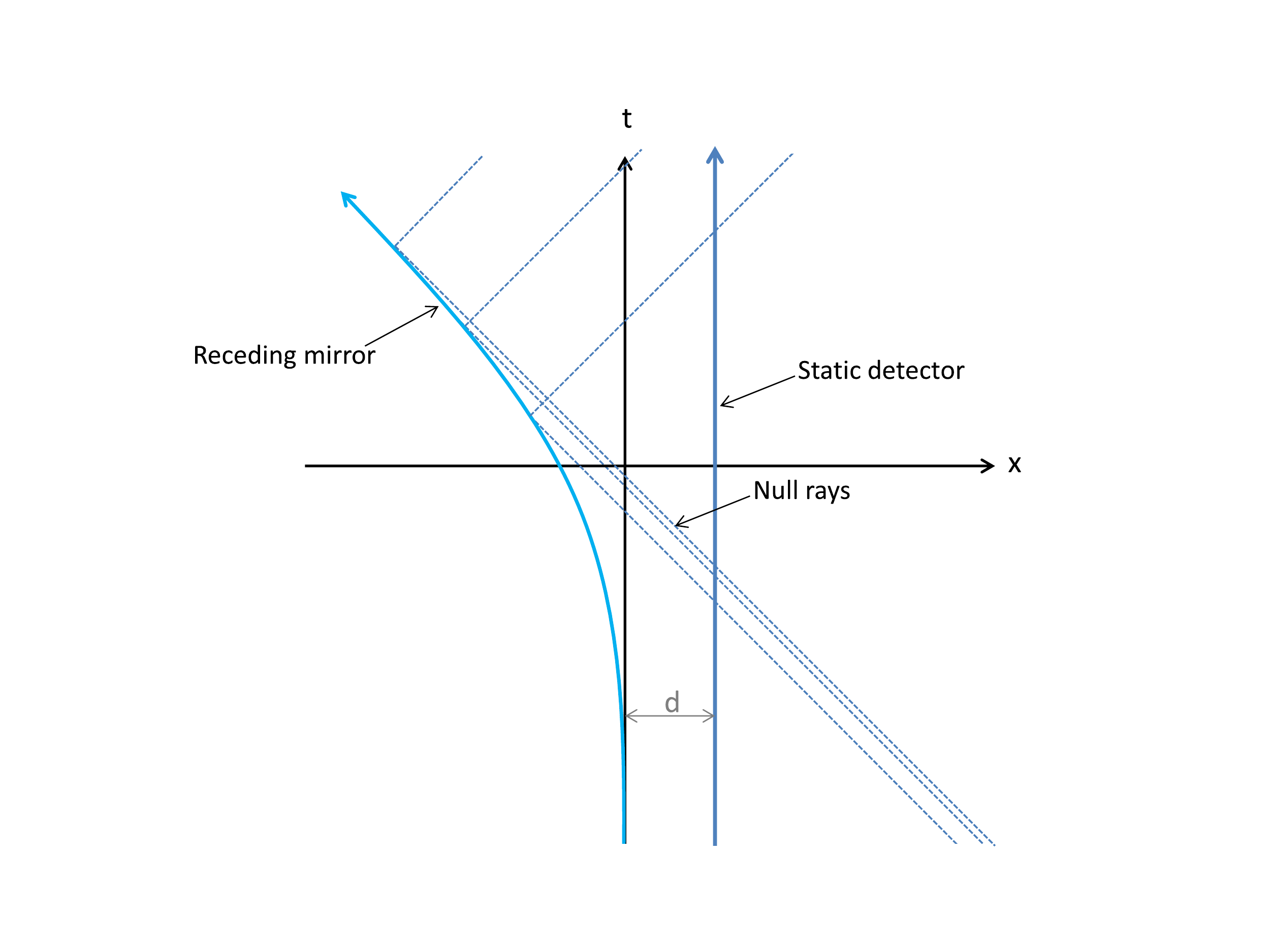}
\caption{Minkowski spacetime with the 
receding mirror \eqref{eq:mirror-trajectory}
and an inertial detector \eqref{eq:Mink-static-traj}
that is static with respect 
to the mirror in the asymptotic past. 
Dashed lines show a selection of null 
geodesics that bounce off the mirror.}
\label{fig:recmir}
\end{figure}

Denoting a standard set of Minkowski coordinates by $(t,x)$, 
we work in the double null coordinates 
\begin{subequations}
\label{eq:nullcoords-Mink}
\begin{align}
u &= t-x
\ , 
\\
v &= t+x
\ , 
\end{align}
\end{subequations}
in which $ds^2 = - du\,dv$. 
We take the mirror trajectory to be 
\begin{align}
v = -\frac{1}{\kappa} \ln(1+ \emath^{-\kappa u})
\ , 
\label{eq:mirror-trajectory}
\end{align}
where $\kappa$ is a positive constant. When parametrised in terms of
the proper time~$\tau$, the trajectory reads 
\begin{subequations}
\begin{align}
u = - \frac{2}{\kappa} \ln [\sinh(-\kappa\tau/2)]
\ , 
\\[1ex]
v = - \frac{2}{\kappa} \ln [\cosh(-\kappa\tau/2)]
\ , 
\end{align}
\end{subequations}
where $-\infty < \tau <0$. 
The velocity and acceleration are towards decreasing~$x$, 
and the proper acceleration has 
the magnitude $\kappa/\sinh(-\kappa\tau)$. 
At early times the trajectory is asymptotically inertial, 
asymptoting to $x=0$ from the left, with proper acceleration 
that vanishes exponentially in~$\tau$. 
At late times the trajectory asymptotes to the null line 
$v=0$ from below, receding to infinity as $\tau\to0_-$, 
and the proper acceleration diverges as~$-1/\tau$. 
A~spacetime diagram is shown in Figure~\ref{fig:recmir}. 

We consider the spacetime that is to the right of the mirror. 
The mirror is hence receding,
and the constant $\kappa$ is analogous to the surface
gravity in a collapsing star spacetime at late times 
\cite{birrell-davies,Davies:1976hi,Davies:1977yv}. 

We consider a massless scalar field $\phi$ with
Dirichlet boundary conditions at the mirror. As the 
positive frequency mode functions, 
we choose \cite{birrell-davies,Davies:1976hi,Davies:1977yv}
\begin{align}
u_k = \ii (4 \pi k)^{-1/2} 
\left[ 
\emath^{-\ii k v} -\emath^{-\ii k p(u)} 
\right]
\ ,
\label{eq:mirror-modes}
\end{align}
where $k>0$ and 
\begin{align}
p(u) =  -\frac{1}{\kappa} \ln(1+ \emath^{-\kappa u})
\ . 
\label{eq:pfunc-def}
\end{align}
These modes satisfy the massless Klein-Gordon equation, they
satisfy the Dirichlet boundary condition at the mirror, and they are
Dirac orthormal, 
$(u_k,u_{k'}) = - (\overline{u_k},\overline{u_{k'}}) 
= \delta(k-k')$ 
and 
$(\overline{u_k},u_{k'}) = (u_k,\overline{u_{k'}}) 
= 0$, 
where $(\,\cdot\, , \,\cdot\,)$ is the Klein-Gordon inner
product on a hypersurface of constant~$t$. 
In the distant past, the modes reduce 
to the usual Dirichlet boundary condition modes in the static half-space
$x>0$. We note that to verify the orthonormality, it suffices to
consider the static half-space limit on a 
constant $t$ hypersurface in the distant past: the inner product is
constant in $t$ due to the Klein-Gordon equation and the Dirichlet
boundary condition. 

We denote by $|0, \text{in}\rangle$  
the no-particle state with respect to the
modes~\eqref{eq:mirror-modes}. 
In the distant past, $|0, \text{in}\rangle$ coincides with 
the usual no-particle state in the half-space $x>0$, and we call it
the in-vacuum. Computing the Wightman as a mode sum from 
\eqref{eq:mirror-modes} gives~\cite{birrell-davies}
\begin{align}
\langle 0, \text{in}|
\phi(\mathsf{x}) \phi(\mathsf{x}') 
|0, \text{in}\rangle
= 
-\frac{1}{4 \pi} \ln \! \left[ 
\frac{\bigl(p(u)-p(u')-\ii \epsilon\bigr)
(v-v'-\ii \epsilon)} 
{\bigl(v-p(u')-\ii \epsilon\bigr)\bigl(p(u)-v'-\ii \epsilon\bigr)}
\right]
\ , 
\label{eq:rec-mirror-Wightman}
\end{align}
where $\ii \epsilon$ arises from the conditional 
ultraviolet convergence as usual. 
The mode sum is infrared convergent
because of the Dirichlet boundary condition.

\subsection{Inertial detector: 
static in the distant past\label{subsec:past-staticwithmirror}}

We consider a detector on the inertial worldline~\eqref{eq:Mink-static-traj}, 
where $d$ is again a positive constant. In the asymptotic past, 
the detector is hence at distance $d$ from a static mirror. 
We take the detector to be switched on in the asymptotic past, 
and we take the field to be in the in-vacuum $|0, \text{in}\rangle$. 

Using \eqref{eq:Mink-static-traj}, 
\eqref{eq:nullcoords-Mink} and~\eqref{eq:rec-mirror-Wightman}, we find 
\begin{align}
\mathcal{A}(\tau',\tau'') 
&=   - \frac{1}{4 \pi} \Biggl( \frac{p'(u') p'(u'')}{{[p(u') - p(u'') 
- \ii\epsilon]}^2} 
+ \frac{1}{{(v'-v''-\ii\epsilon)}^2} 
\notag
\\[1ex]
& \hspace{10ex}
- \frac{p'(u'')}{{[v'-p(u'')-\ii \epsilon]}^2} 
- \frac{p'(u')}{{[p(u')-v''-\ii \epsilon]}^2}\Biggr)
\ ,
\label{eq:Aepsilon-mirror-asstat}
\end{align}
where 
$u' = \tau' - d$, 
$v' = \tau' + d$, 
$u'' = \tau'' - d$
and 
$v'' = \tau'' + d$. 
The prime on $p$ denotes derivative with respect to the argument. 
From \eqref{eq:Fdot-sharp-aspast} we then have 
\begin{subequations}
\label{eq:mirror-asstat-Fdot}
\begin{align}
\dot{\mathcal{F}} (\omega,\tau) &= 
\dot{\mathcal{F}}_0 (\omega,\tau) + 
\dot{\mathcal{F}}_1 (\omega,\tau) + 
\dot{\mathcal{F}}_2 (\omega,\tau)
\ , 
\\[1ex]
\dot{\mathcal{F}}_0 (\omega,\tau)
&= 
- \omega \Theta(-\omega)
+ 
\frac{1}{2\pi} \int_0^\infty ds \, 
\cos(\omega s) \! 
\left( - \frac{p'(\tau - d) p'(\tau - d - s)}{{[p(\tau - d) - p(\tau - d - s)]}^2} 
+ \frac{1}{s^2} \right) 
\ , 
\label{eq:mirror-asstat-regfree0-Fdot}
\\
\dot{\mathcal{F}}_1 (\omega,\tau)
&=  
\frac{1}{2\pi} \int_0^\infty ds \, 
\frac{\cos(\omega s) \, p'(\tau -d - s)}{{[\tau + d -p(\tau -d -s)]}^2}
\ , 
\label{eq:mirror-asstat-regfree1-Fdot}
\\[1ex]
\dot{\mathcal{F}}_2 (\omega,\tau)
&=  
\frac{1}{2\pi} \int_0^\infty ds \, 
\Realpart 
\left( 
\frac{\emath^{-\ii \omega s} \, p'(\tau -d)}{[{p(\tau-d) -\tau -d +s-\ii \epsilon]}^2} 
\right)
\ . 
\label{eq:mirror-asstat-regulator-Fdot}
\end{align}
\end{subequations}
In \eqref{eq:mirror-asstat-regfree0-Fdot} and \eqref{eq:mirror-asstat-regfree1-Fdot} 
we have set $\epsilon=0$ 
as the integrand has no singularities. 
The $\epsilon$ in \eqref{eq:mirror-asstat-regulator-Fdot} 
needs to be kept as the integrand has a singularity, 
arising because the points 
$\tau-s$ and $\tau$ on the detector's trajectory are connected 
by a null ray that is reflected from the mirror, 
as shown in Figure~\ref{fig:recmir}. 
The integral is well defined despite this singularity since the switch-on is 
in the asymptotic past so that the range of $s$ cannot end at the singularity. 

We show in Appendix \ref{app:rec-mirror-calculation} 
that the early and late time forms of the 
transition rate \eqref{eq:mirror-asstat-Fdot} are
\begin{subequations}
\label{pastv0-futurev0}
\begin{align}
\dot{\mathcal{F}} (\omega,\tau)
&= 
-\omega \, [1-\cos (2 d \omega) ] \, \Theta (-\omega) 
+ O(\emath^{\kappa\tau}) 
\hspace{4ex}
\text{as $\tau \rightarrow -\infty$}
\ , 
\label{pastev0}
\\[1ex]
\dot{\mathcal{F}} (\omega,\tau)
&=
-\frac{\omega}{2} 
\Theta (-\omega) + \frac{\omega}{2 \, (\emath^{2\pi\omega/\kappa}-1)}
+ o(1) 
\hspace{4ex}
\text{as $\tau \rightarrow \infty$}
\ . 
\label{futurev0}
\end{align}
\end{subequations}

\subsection{Inertial detector: 
travelling towards the mirror 
in the distant past\label{subsec:past-towardsmirror}}

We next consider a detector on the inertial worldline 
\begin{align}
\mathsf{x}(\tau) = 
(\tau\cosh\lambda, - \tau \sinh\lambda)
\ , 
\label{eq:mirror-drifting-detector}
\end{align}
where $\lambda>0$. 
In the asymptotic past, where the mirror is static, 
the detector is moving towards the mirror with speed 
$\tanh\lambda$. 

Proceeding as above, we find 
\begin{subequations}
\label{eq:mirror-asdrift-Fdot}
\begin{align}
\dot{\mathcal{F}} (\omega,\tau) &= 
\dot{\mathcal{F}}_0 (\omega,\tau) + 
\dot{\mathcal{F}}_1 (\omega,\tau) + 
\dot{\mathcal{F}}_2 (\omega,\tau)
\ , 
\\[1ex]
\dot{\mathcal{F}}_0 (\omega,\tau)
&= 
- \omega \Theta(-\omega)
+ 
\frac{1}{2\pi} \int_0^\infty ds \, 
\cos(\omega s) \! 
\left( - \frac{p'(\emath^\lambda\tau) p'\bigl(\emath^\lambda(\tau - s)\bigr) \,  
\emath^{2\lambda}}{{\bigl[p(\emath^\lambda\tau) 
- p\bigl(\emath^\lambda(\tau - s)\bigr)\bigr]}^2} + \frac{1}{s^2} \right) 
\ , 
\label{eq:mirror-asdrift-regfree0-Fdot}
\\
\dot{\mathcal{F}}_1 (\omega,\tau)
&=  
\frac{1}{2\pi} \int_0^\infty ds \, 
\frac{\cos(\omega s) \, p'\bigl(\emath^\lambda(\tau -s)\bigr)}
{{\bigl[\emath^{-\lambda}\tau -p\bigl(\emath^\lambda(\tau-s)\bigr)\bigr]}^2}
\ , 
\label{eq:mirror-asdrift-regfree1-Fdot}
\\[1ex]
\dot{\mathcal{F}}_2 (\omega,\tau)
&=  
\frac{1}{2\pi} \int_0^\infty ds \, 
\Realpart 
\left( 
\frac{\emath^{-\ii \omega s} \, p'(\emath^\lambda\tau)}
{[{p(\emath^\lambda\tau) -\emath^{-\lambda}(\tau -s)-\ii \epsilon]}^2} 
\right)
\ , 
\label{eq:mirror-asdrift-regulator-Fdot}
\end{align}
\end{subequations}
and we show in Appendix \ref{app:rec-mirror-calculation} 
that the early and late time forms are 
\begin{subequations}
\label{drift:pastv0-futurev0}
\begin{align}
\dot{\mathcal{F}} (\omega,\tau)
&= 
-\omega \left[1- \emath^{2\lambda} 
\cos(2 \tau \sinh\lambda \,  \emath^\lambda \omega) \right] 
\Theta(-\omega) + O(\tau^{-1})
\hspace{4ex}
\text{as $\tau \rightarrow -\infty$}\,,
\label{drift:pastev0}
\\[1ex]
\dot{\mathcal{F}} (\omega,\tau)
&=
-\frac{\omega}{2} 
\Theta (-\omega) 
+ \frac{\omega}{2 \, (\emath^{2\pi\emath^{-\lambda}\omega/\kappa}-1)}
+ o(1) 
\hspace{4ex}
\text{as $\tau \rightarrow \infty$}\,.
\label{drift:futurev0}
\end{align}
\end{subequations}

\subsection{Onset of thermality\label{subsec:emergtherm}}

We are now ready to discuss the sense in which the transition rate 
exhibits the onset of thermality as the mirror continues to recede. 

Consider first the distant past. 
For the detector~\eqref{eq:Mink-static-traj}, 
static with respect to the mirror, 
the transition rate \eqref{pastev0} 
agrees with that \eqref{eq:mzero-Fdot-Mtilde} of the same detector 
in the static half-space~$\Mhalfspace$. 
For the detector~\eqref{eq:mirror-drifting-detector}, 
drifting towards the mirror, the transition rate 
\eqref{drift:pastev0}
can be verified to agree with that of the same detector 
in~$\Mhalfspace$. Compared with~\eqref{pastev0}, 
the non-Minkowski part of \eqref{drift:pastev0}
has the static distance $d$ replaced 
by the time-dependent distance~$-\tau \sinh\lambda$, 
$\omega$ replaced by the blueshifted frequency~$\emath^\lambda \omega$, 
and an additional blueshift factor~$\emath^\lambda$. 

Consider then the distant future. The distant future transition rates
\eqref{futurev0} and \eqref{drift:futurev0} agree with the transition
rate \eqref{eq:Mink-mixed-trrate} of an inertial detector in Minkowski
space when the left-movers are in the Minkowski vacuum and the
right-movers are thermal in temperature~$\kappa/(2\pi)$. Note that the
detector's velocity shows up by a Doppler blueshift in the right-mover
contribution.

The late time transition rates \eqref{futurev0} and \eqref{drift:futurev0} 
can hence be interpreted to consist of a 
contribution from the left-moving part of the field, 
undisturbed by the mirror, 
and and a contribution from the right-moving part of the field, 
excited by the mirror to induce a Planckian response. 
This interpretation is consistent with the fact that the 
stress-energy tensor of the field contains 
at late times an energy flux to the right
\cite{birrell-davies,Davies:1976hi,Davies:1977yv,Carlitz:1986nh,Good:2013lca}. 

This late time result is consistent with that quoted in 
\cite{birrell-davies} for a non-derivative UDW detector 
with a mirror trajectory 
with similar late time asymptotics, 
in the sense that 
the left-mover contribution was not 
explicitly written out in~\cite{birrell-davies}. 

\begin{figure}[!p]
\centering
\includegraphics[width=0.8\textwidth]{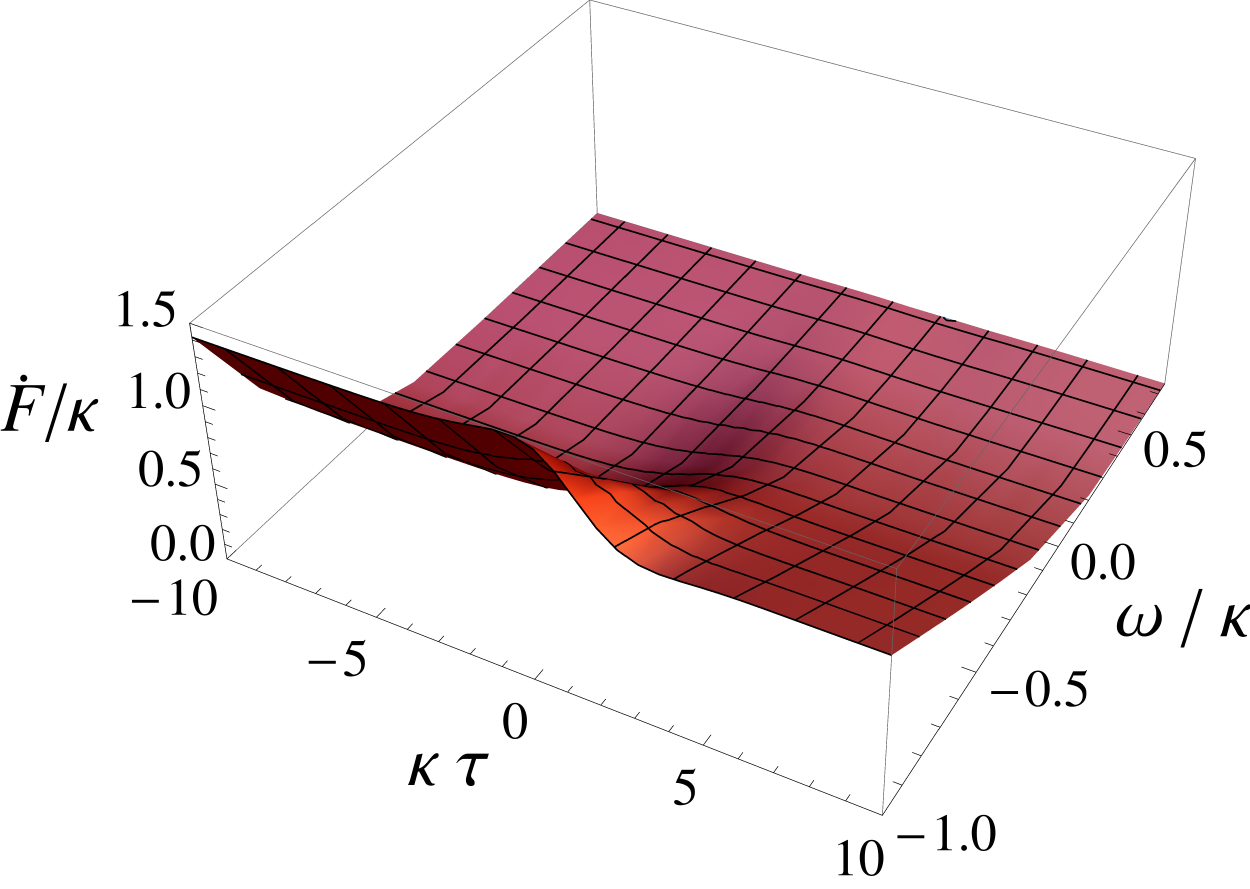}
\caption{The figure shows a perspective plot of the transition rate 
$\dot{\mathcal{F}} (\omega,\tau)$ 
\eqref{eq:mirror-asstat-Fdot}
for the detector \eqref{eq:Mink-static-traj} that is asymptotically static 
with respect to the mirror in the distant past, with $d = 1/\kappa$.}
\label{fig:mirror3d}
\end{figure}

\begin{figure}[!p]
\centering
\begin{tabular}{cc}
\includegraphics[width=0.48\textwidth]{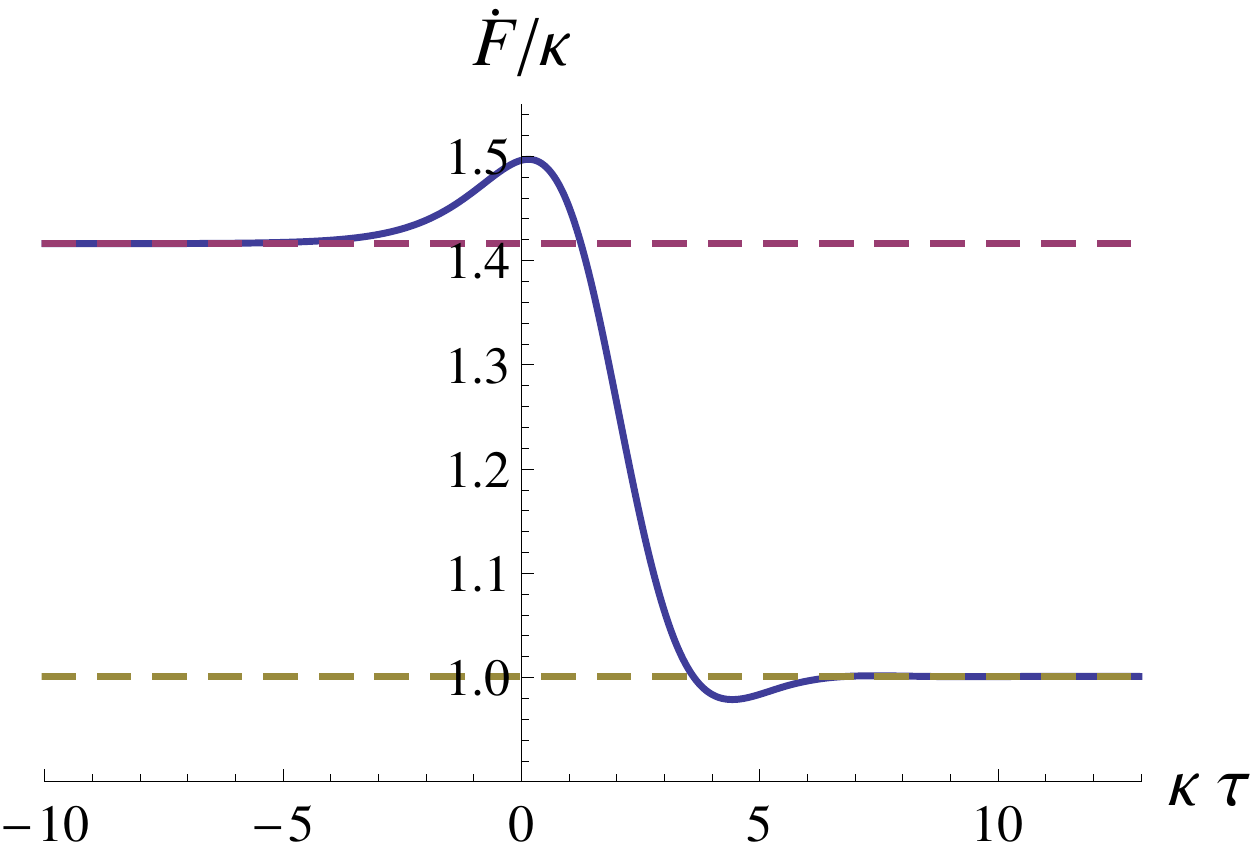}& 
\includegraphics[width=0.48\textwidth]{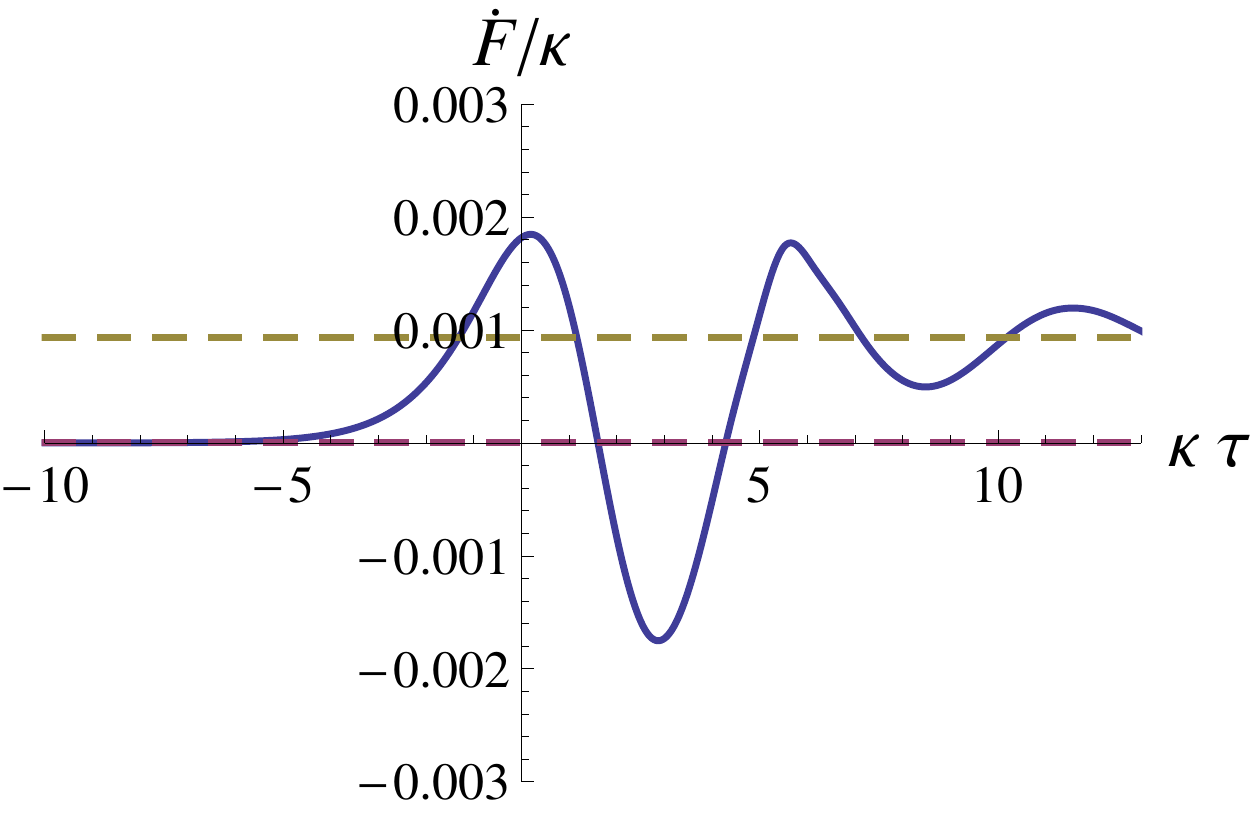}\\
(a) & (b)
\end{tabular}
\caption{Cross-sections of the plot in Figure \ref{fig:mirror3d}
at (a) $\omega = -\kappa$ and (b) $\omega = \kappa$. 
The dashed horizontal lines show the past and future 
asymptotic values~\eqref{pastv0-futurev0}.}
\label{fig:mirrorasympt}
\end{figure}

Figures \ref{fig:mirror3d} and \ref{fig:mirrorasympt}
show numerical plots for the evolution of the transition rate 
from early to late times, for the detector \eqref{eq:Mink-static-traj}
that is static with respect to the mirror in the distant past. 
The asymptotic late time value is reached 
via a ring-down of oscillations whose 
period equals $2\pi/\kappa$ 
within the range of the numerical experiments. 
We have not attempted to examine this 
oscillation analytically.

\section{$(1+1)$ Schwarzschild spacetime\label{sec:bh}}

In this section we consider a detector in the 
$(1+1)$-dimensional Schwarzschild spacetime, 
obtained by dropping the angular dimensions 
from the $(3+1)$-dimensional Schwarzschild metric~\cite{Misner:1974qy}. 
We first establish the notation, recall the definitions of the 
Boulware, HHI and Unruh vacua \cite{birrell-davies}, and discuss briefly  
the case of a static detector in the exterior. The main objective is to study a 
geodesically infalling detector.

\subsection{Spacetime and vacua}

We write the metric of the $(1+1)$-dimensional maximally 
extended Schwarzschild spacetime 
in the notation of \cite{birrell-davies} as 
\begin{align}
ds^2 = -\frac{2M \emath^{-r/(2M)}}{r} \, d\bar{u}\, d\bar{v}
\ ,
\label{eq:kruskalmetric}
\end{align}
where $M>0$ is the Schwarzschild mass parameter, the Kruskal null coordinates 
$\bar{u}$ and $\bar{v}$ increase towards the future and satisfy 
$\bar{u} \bar{v}< {(4M)}^2$, and $r \in \BbbR_+$ 
is the unique solution to 
\begin{align}
-\bar{u} \bar{v} = {(4M)}^2 \bigl[r/(2M)-1\bigr] e^{r/(2M)}
\ . 
\label{eq:schw-r-defeq}
\end{align}
The metric has the Killing vector 
$\xi = {(4M)}^{-1} (- \bar{u} \partial_{\bar{u}} + \bar{v} \partial_{\bar{v}})$, 
which is timelike for $\bar{u} \bar{v} < 0$ ($r>2M$), 
spacelike for $\bar{u} \bar{v} > 0$ ($r<2M$) 
and null at the Killing horizon $\bar{u} \bar{v}=0$ ($r=2M$). 
The right-going (respectively left-going) 
branch of the Killing horizon is $\bar{u}=0$ ($\bar{v}=0$). 
The Killing horizon divides the spacetime into four quadrants as 
summarised in Table~\ref{table:Schw-quadrants}. 


\begin{table}[b!]
\begin{center}
  \begin{tabular}{ | l | c | c | c | c |}
Quadrant & $\bar{u}$ & $\bar{v}$ & $\xi^a \xi_a$ & $r$ \\
\hline
\hbox to 7mm {I:} right-hand exterior
 & $-$ & $+$ & $-$ & $2M<r<\infty$
\\ \hline
\hbox to 7mm {II:} black hole interior & $+$ & $+$ & $+$ & $0<r<2M$
\\ \hline
\hbox to 7mm {III:} left-hand exterior
 & $+$ & $-$ & $-$ & $2M<r<\infty$
\\ \hline
\hbox to 7mm {IV:} white hole interior
 & $-$ & $-$ & $+$ & $0<r<2M$
\\ \hline
\end{tabular}
\end{center}
\caption{The four quadrants of the extended Schwarzschild spacetime.
The columns show the signs of the Kruskal coordinates $\bar{u}$ and $\bar{v}$, 
the norm squared of the Killing vector~$\xi$, and the range of the function~$r$. 
In the exteriors, where $\xi$ is timelike, it is future-pointing in 
Quadrant I and past-pointing in Quadrant~III\null. 
\label{table:Schw-quadrants}}
\end{table}

We denote by $u$ and $v$ the tortoise null coordinates defined by 
\begin{subequations}
\label{eq:tortoise-null}
\begin{align}
&\hbox to 25ex{$u = - 4M \ln [-\bar{u}/(4M)]$\hfill}
\hspace{3ex}
\text{for $\bar{u}<0$}
\ , 
\\[1ex]
&\hbox to 25ex{$v = 4M \ln [\bar{v}/(4M)]$\hfill}
\hspace{3ex}
\text{for $\bar{v}>0$} 
\ . 
\end{align}
\end{subequations}
In Quadrant I (right-hand exterior), where $r>2M$, we can hence introduce the usual
exterior Schwarzschild coordinates $(t,r)$ by 
\begin{subequations}
\label{eq:schw-vs-tortoisenull}
\begin{align}
u &= 
t 
- r 
- 2M 
\ln [r/(2M) -1 ] 
\ , 
\label{eq:EF-u-def}
\\
v &= 
t 
+ r 
+ 2M 
\ln [r/(2M) -1 ] 
\ , 
\label{eq:EF-v-def}
\end{align}
\end{subequations}
so that 
\begin{align}
t = 2M \ln (-\bar{v}/\bar{u})
\ , 
\label{eq:schw-ext-t-def}
\end{align}
the metric reads 
\begin{align}
ds^2 = - (1 - 2M/r) \, dt^2 + \frac{dr^2}{(1 - 2M/r)}
\ , 
\label{eq:Schw-metric}
\end{align}
and $\xi = \partial_t$. In Quadrant II, where $r<2M$, 
we can similarly introduce the Schwarzschild-like coordinates $(\tilde t, r)$ by 
\eqref{eq:schw-r-defeq} and 
\begin{align}
\tilde t = 2M \ln (\bar{v}/\bar{u})
\ , 
\label{eq:schw-int-ttilde-def}
\end{align}
so that the metric takes the form 
\begin{align}
ds^2 =  - \frac{dr^2}{[(2M/r) -1]} + [(2M/r) -1] \, d{\tilde{t}}^2 
\label{eq:Schw-int-metric}
\end{align}
and 
$\xi = \partial_{\tilde{t}}$. A~pair of coordinates that covers Quadrants I and II and 
the black hole horizon that separates them is $(\bar{u},v)$. 
We shall not need the explicit form of the metric in these coordinates. 

We consider a massless minimally coupled 
scalar field in three distinguished states.
First, in Quadrant I we consider the Boulware vacuum~$|0_B\rangle$,
defined by the positive and negative frequency decomposition with
respect to $\partial_t$
in~\eqref{eq:Schw-metric}~\cite{Boulware-scalar}.  At the
asymptotically flat infinity of Quadrant~I, $|0_B\rangle$ reduces to
the Minkowski vacuum.  Second, on the whole spacetime we consider the
HHI vacuum~$|0_H\rangle$, defined by the positive and negative
frequency decomposition with respect to $\partial_{\bar{u}}$ and
$\partial_{\bar{v}}$ on the Killing horizon~\cite{Hartle:1976tp,Israel:1976ur}.  In
Quadrant~I, $|0_H\rangle$ is a thermal equilibrium state with respect
to~$\partial_t$, at the local Hawking temperature 
\begin{align}
T_{\text{loc}} = \frac{1}{8\pi M \sqrt{1-2M/r}}
\ .
\label{eq:Tloc-Hawking}
\end{align}
Third, in Quadrants I and II
and on the black hole horizon that separates them, we consider the
Unruh vacuum~$|0_U\rangle$, defined by the positive and negative
frequency decomposition with respect to $\partial_{\bar{u}}$ and
$\partial_v$ in the coordinates~$(\bar{u},v)$~\cite{unruh}.
$|0_U\rangle$~mimics a state that results from the collapse of a star
at late times when there is initially no incoming radiation from
infinity, and it has the left-moving part of the field in a
Boulware-like state and the right-moving part of the field in a
HHI-like state. 

The Wightman functions for the three vacua are
\cite{birrell-davies}
\begin{subequations}
\label{eq:wightmans-in-schw}
\begin{align}
\langle 0_B| \phi(\mathsf{x})\phi(\mathsf{x}') |0_B \rangle
& = -\frac{1}{4\pi} \ln[(\epsilon + \ii \Delta u)(\epsilon + \ii \Delta v)] 
\ , 
\label{eq:B-wightman-in-schw}
\\
\noalign{\medskip}
\langle 0_H| \phi(\mathsf{x})\phi(\mathsf{x}') |0_H \rangle
& = -\frac{1}{4\pi} \ln[(\epsilon + \ii \Delta \bar{u})(\epsilon + \ii \Delta \bar{v})] 
\ ,
\label{eq:H-wightman-in-schw}
\\
\noalign{\medskip}
\langle 0_U| \phi(\mathsf{x})\phi(\mathsf{x}') |0_U \rangle
& = -\frac{1}{4\pi} \ln[(\epsilon + \ii \Delta \bar{u})(\epsilon + \ii \Delta v)] 
\ , 
\label{eq:U-wightman-in-schw}
\end{align}
\end{subequations}
where $\Delta u = u-u'$ and similarly for the other coordinates. 
Each of the Wightman functions is
unique up to a real-valued additive constant. The Boulware and HHI
vacuum Wightman functions are invariant under the isometries generated
by the Killing vector~$\xi$. The Unruh vacuum Wightman function
changes under these isometries by an additive constant; 
however, the Unruh vacuum may nevertheless be considered 
invariant under the isometries since the stress-energy tensor and 
other quantities built from derivatives of the Wightman function 
are invariant~\cite{fulling-ruijsenaars,Kay:2000fi}. 
The non-invariance of the Unruh vacuum Wightman function is due to the 
infrared properties of the $(1+1)$-dimensional conformal field 
and has no counterpart in higher dimensions~\cite{Dappiaggi:2009fx}.

\subsection{Static detector\label{subsubsec:static}}

We consider first a detector in Quadrant I on the static, noninertial
trajectory $r = R$, where $R>2M$ is a constant. Using
\eqref{eq:Fdot-stationary-naive} and~\eqref{eq:wightmans-in-schw}, the
calculations are closely similar to those in Section~\ref{sec:tests},
and we omit the detail. We find
\begin{subequations}
\begin{align}
\dot{\mathcal{F}}_B(\omega) 
& = -\omega \Theta (-\omega)
\ , 
\label{FRR}
\\[1ex]
\dot{\mathcal{F}}_H(\omega) 
& = \frac{\omega}{\emath^{\omega/T_{\text{loc}}}-1}
\ ,
\label{FHHR}
\\[1ex]
\dot{\mathcal{F}}_U(\omega) 
& = -\frac{\omega}{2} \Theta (-\omega) 
+ \frac{\omega}{2 \, (\emath^{\omega/T_{\text{loc}}}-1)}
\ , 
\label{FUR}
\end{align}
\end{subequations}
for respectively the Boulware, HHI and Unruh vacua,
where $T_{\text{loc}}$ is the local Hawking temperature 
\eqref{eq:Tloc-Hawking} evaluated at $r=R$. 

These results conform fully to expectations.  The Boulware vacuum
transition rate is that of an inertial detector in Minkowski space in
Minkowski vacuum, while the HHI vacuum transition rate is thermal in
the local Hawking temperature~\eqref{eq:Tloc-Hawking}.  The Unruh
vacuum transition rate is the average of the two, with the two pieces
arising respectively from the left-moving and right-moving parts of
the field.

These results are also consistent with what was reported for the
non-derivative UDW detector in~\cite{birrell-davies}, in the sense
that the left-mover contribution in \eqref{FUR} was not explicitly
written out in~\cite{birrell-davies}.  Finally, the similarity between
\eqref{FUR} and our receding mirror spacetime results \eqref{futurev0}
and \eqref{drift:futurev0} is an additional confirmation that the
Unruh vacuum mimics the late time properties of a state created in a
collapsing star spacetime~\cite{birrell-davies,unruh}.

\subsection{Interlude: geodesics\label{subsubsec:sch-geodesics}}

We next turn to inertial detectors. 
In this subsection we recall a convenient parametrisation for the geodesics. 
We give the full expressions in a form that applies only to Quadrant~I, 
where the equations of a timelike geodesic in the 
Schwarzschild coordinates \eqref{eq:Schw-metric} take the form 
\begin{subequations}
\label{eq:Schw-geod-eqs}
\begin{align}
\dot{t} &= \frac{E}{1-2M/r}
\ , 
\\[1ex]
\dot{r}^2 &= E^2-1 + 2M/r
\ ,  
\label{eq:Schw-rdot-geod-eq}
\end{align}
\end{subequations}
where $E$ is a positive constant and the overdot denotes derivative with 
respect to the proper time~$\tau$. The continuation beyond Quadrant 
I can be done by passing to the Kruskal coordinates $(\bar{u}, \bar{v})$. 

When $E>1$, the geodesic has at infinity the nonvanishing speed 
$\sqrt{1 - E^{-2}}$ with respect to the 
Killing vector~$\xi$. 
We consider a geodesic that is sent in from the infinity, so that 
$\dot{r}<0$. 
The geodesic can be parametrised as 
\begin{subequations}
\label{eq:schw-allcoords-as-of-chi}
\begin{align}
\tau & = \frac{M}{{(E^2-1)}^{3/2}} \, (\sinh \chi - \chi)
\ , 
\label{eq:schw-tau-as-of-chi}
\\[1ex]
r & = \frac{M}{(E^2-1)} \, (\cosh\chi -1)
\ , 
\label{eq:schw-r-as-of-chi}
\\[1ex]
t& = \frac{ME}{{(E^2-1)}^{3/2}}
\left[\sinh \chi + (2E^2 - 3)\chi \right]
+ 
2M 
\ln \Biggl(
\frac{-\tanh(\chi/2) + \sqrt{1 - E^{-2}}}{-\tanh(\chi/2) - \sqrt{1 - E^{-2}}}
\Biggr)
\ , 
\label{eq:schw-t-as-of-chi}
\end{align}
\end{subequations}
where the parameter $\chi$ takes values in $(-\infty, 0)$, 
so that the trajectory starts at the infinity in the asymptotic past at 
$\chi \to -\infty$ and hits the singularity at $\chi \to 0$. 
The additive constant in \eqref{eq:schw-tau-as-of-chi} is chosen so that $-\infty < \tau < 0$. 
The horizon-crossing occurs at 
$\chi =  \chi_h \doteq - 2\arctanh \bigl(\sqrt{1 - E^{-2}}\,\bigr)$. 
Equation \eqref{eq:schw-t-as-of-chi} applies
only in Quadrant~I, where $-\infty < \chi < \chi_h$. 

When $E=1$, the geodesic has at infinity a vanishing speed
with respect to~$\xi$. 
We consider again a geodesic that is sent in from 
the infinity. 
The geodesic takes the form 
\begin{subequations}
\label{eq:schw-allcoords-as-of-tau-E=1}
\begin{align}
r & = 2M 
{[-3\tau/(4M)]}^{2/3}
\ , 
\\[1ex]
t& = \tau 
- 4M {[-3\tau/(4M)]}^{1/3}
+ 
2M \ln \! \left(
\frac{{[-3\tau/(4M)]}^{1/3} +1}{{[-3\tau/(4M)]}^{1/3} -1}
\right)
\ , 
\label{eq:schw-t-as-of-tau-E=1}
\end{align}
\end{subequations}
where $-\infty < \tau < 0$. The 
horizon-crossing occurs at $\tau =  \tau_h \doteq - 4M/3$, 
and the singularity is reached as $\tau\to0$. 
Equation \eqref{eq:schw-t-as-of-tau-E=1}
applies only in Quadrant~I, where $-\infty < \tau < \tau_h$. 

When $0<E<1$, the geodesic has a maximum value of~$r$.
The geodesic can be parametrised as 
\begin{subequations}
\label{eq:schw-allcoords-as-of-phi}
\begin{align}
\tau & = \frac{M}{{(1-E^2)}^{3/2}} \, (\varphi + \sin\varphi)
\ , 
\label{eq:schw-tau-as-of-phi}
\\[1ex]
r & = \frac{M}{(1-E^2)} \, (1 + \cos\varphi)
\ , 
\label{eq:schw-r-as-of-phi}
\\[1ex]
t& = \frac{ME}{{(1-E^2)}^{3/2}}
\left[
\sin\varphi + (3- 2E^2)\varphi \right]
+ 2M 
\ln \Biggl(
\frac{1 + \sqrt{E^{-2}-1} \, 
\tan (\varphi/2)}{1 - \sqrt{E^{-2}-1} \, \tan (\varphi/2)}
\Biggr)
\ , 
\label{eq:schw-t-as-of-phi}
\end{align}
\end{subequations}
where the parameter $\varphi$ takes values in $(-\pi, \pi)$, 
so that the trajectory starts at the white hole singularity at $\varphi \to -\pi$ and ends at the black hole 
singularity at $\varphi \to \pi$. The additive constant in \eqref{eq:schw-tau-as-of-phi}
is chosen so that $\tau=0$ at the moment when $r$ reaches its maximum value, $2M/(1-E^2)$. 
The total proper time elapsed between the singularities is 
$2\pi M {(1-E^2)}^{-3/2}$. 
The horizon-crossings occur at $\varphi =  \mp \varphi_h$ where 
$\varphi_h \doteq 2 \arctan\bigl(\sqrt{E^{-2} -1}\,\bigr)$. 
Equation \eqref{eq:schw-t-as-of-phi} 
applies only in Quadrant~I, where $-\varphi_h < \varphi < \varphi_h$. 

Finally, there exist also timelike geodesics that pass from 
the white hole to the black hole through the 
horizon bifurcation point $\bar{u}=\bar{v}=0$, 
without entering Quadrant I (or Quadrant~III\null). 
These geodesics take the form  
\begin{align}
\bar{u} =
\bar{v} = 4 M \sin(\varphi/2) \exp\bigl[\tfrac12 \cos^2(\varphi/2) \bigr] 
\ , 
\label{eq:ubar-vbar-ito-varphi}
\end{align}
where the parameter $\varphi$ 
takes values in $(-\pi, \pi)$, 
and $\tau$ and $r$ are given by 
\eqref{eq:schw-tau-as-of-phi}
and 
\eqref{eq:schw-r-as-of-phi} 
with $E=0$. The isometry generated by $\xi$ has been used 
in \eqref{eq:ubar-vbar-ito-varphi}
to set $\bar{u} =
\bar{v}$ without loss of generality. 

\subsection{Inertial detector\label{subsubsec:BH-geod-detector}}

The transition rate of the inertial detector 
is obtained by inserting the Wightman functions 
\eqref{eq:wightmans-in-schw}
and the geodesic trajectories of subsection 
\ref{subsubsec:sch-geodesics}
into the integral formulas of subsection~\ref{subsec:detector:1+1sharp}. 
The transition rate is expressible as the integral of an elementary 
function for all values of~$E$; 
for 
$E>1$ (respectively $E<1$) this is accomplished by 
writing the differentiations and the integration
in terms of $\chi$~($\varphi$). 

We address the near-infinity 
and near-singularity limits analytically and the 
intermediate regime numerically. 

\subsubsection{Near the infinity}

We consider the $E>1$ trajectories \eqref{eq:schw-allcoords-as-of-chi}
and the $E=1$ trajectory \eqref{eq:schw-allcoords-as-of-tau-E=1}, 
all of which fall in from the infinity, and we 
push the switch-on to the infinite past. It is shown 
in Appendix \ref{app:schw} that at early times, $\tau \to -\infty$, 
we have 
\begin{subequations}
\label{eq:sch-transrate-aspast}
\begin{align}
\dot{\mathcal{F}}_B (\omega,\tau)
&= 
-\omega \Theta (-\omega)
+ o(1) 
\ , 
\\[1ex]
\dot{\mathcal{F}}_H (\omega,\tau)
&= 
\frac{\omega}{2 \, (\emath^{\omega/T_-}-1)} 
+ \frac{\omega}{2 \, (\emath^{\omega/T_+}-1)}
+ o(1) 
\ , 
\label{HHF}
\\[1ex]
\dot{\mathcal{F}}_U (\omega,\tau)
&= 
-\frac{\omega}{2} \Theta(-\omega) + \frac{\omega}{2 \, (\emath^{\omega/T_+}-1)}
+ o(1) 
\ , 
\end{align}
\end{subequations}
where $T_\pm \doteq \emath^{\pm\lambda}/(8\pi M)$ and 
$\lambda \doteq \arctanh \bigl(\sqrt{1 - E^{-2}} \, \bigr)$. 
For $E=1$, we have $T_+ = T_- = 1/(8\pi M)$,  
so that the two terms in \eqref{HHF} are equal and combine to the Planckian response. 

The asymptotic past results \eqref{eq:sch-transrate-aspast} conform fully to 
physical expectations. 
The Boulware vacuum transition rate is that in  
Minkowski vacuum~\eqref{eq:mzero-Fdot-Minkowski}, consistently with the  
interpretation of the Bouware vacuum as the no-particle state with respect to~$\xi$. 
The HHI vacuum transition rate
is that of an inertial detector in a thermal bath
in Minkowski space~\eqref{eq:Mink-drift-transrate}, 
with the temperature given by the Hawking temperature at the infinity, $1/(8\pi M)$, and with 
each of the two Planckian terms containing a Doppler shift factor 
that accounts for the detector's velocity at the infinity. 
The Unruh vacuum transition rate sees a Planckian term only 
in the outgoing part of the field, as confirmed by the 
Doppler shift to the blue in this term, while the term that 
corresponds to the ingoing part of the field is Minkowski-like.

\subsubsection{Near the singularity}

We consider the transition rate in the HHI and Unruh vacua in the
limit where the detector approaches the black black hole singularity.
We allow all values of the non-negative constant~$E$. We also allow
the switch-on moment to remain arbitrary, subject only to the
condition that for $0\le E <1$ the switch-on in the HHI vacuum takes
place after the trajectory emerges from the white hole singularity,
and the switch-on in the Unruh vacuum takes place after the trajectory
crosses the past horizon.

It is shown in Appendix \ref{app:schw} that in this near-singularity
limit we have 
\begin{align}
\dot{\mathcal{F}} (\omega,\tau)
&= 
\frac{1}{8 \pi M} \left[ \left(\frac{2M}{r(\tau)}\right)^{3/2}
+\frac{1+E^2}{2} \left(\frac{2M}{r(\tau)}\right)^{1/2}\right]
+ O(1) 
\ , 
\label{eq:sch-transrate-as-sing}
\end{align}
for both the HHI vacuum and the Unruh vacuum: 
the differences between the two vacua show up only in the $O(1)$
part. In terms of~$\tau$, 
the leading term in \eqref{eq:sch-transrate-as-sing} 
is $1/[6\pi(\tau_\text{sing} - \tau)]$, 
where $\tau_\text{sing}$ is the value of $\tau$ at the black
hole singularity.

\subsubsection{Intermediate regime: loss of thermality\label{subsubsec:losstherm}}

\begin{figure}[p]
\centering
\begin{tabular}{ccc}
\includegraphics[width=0.31\textwidth]{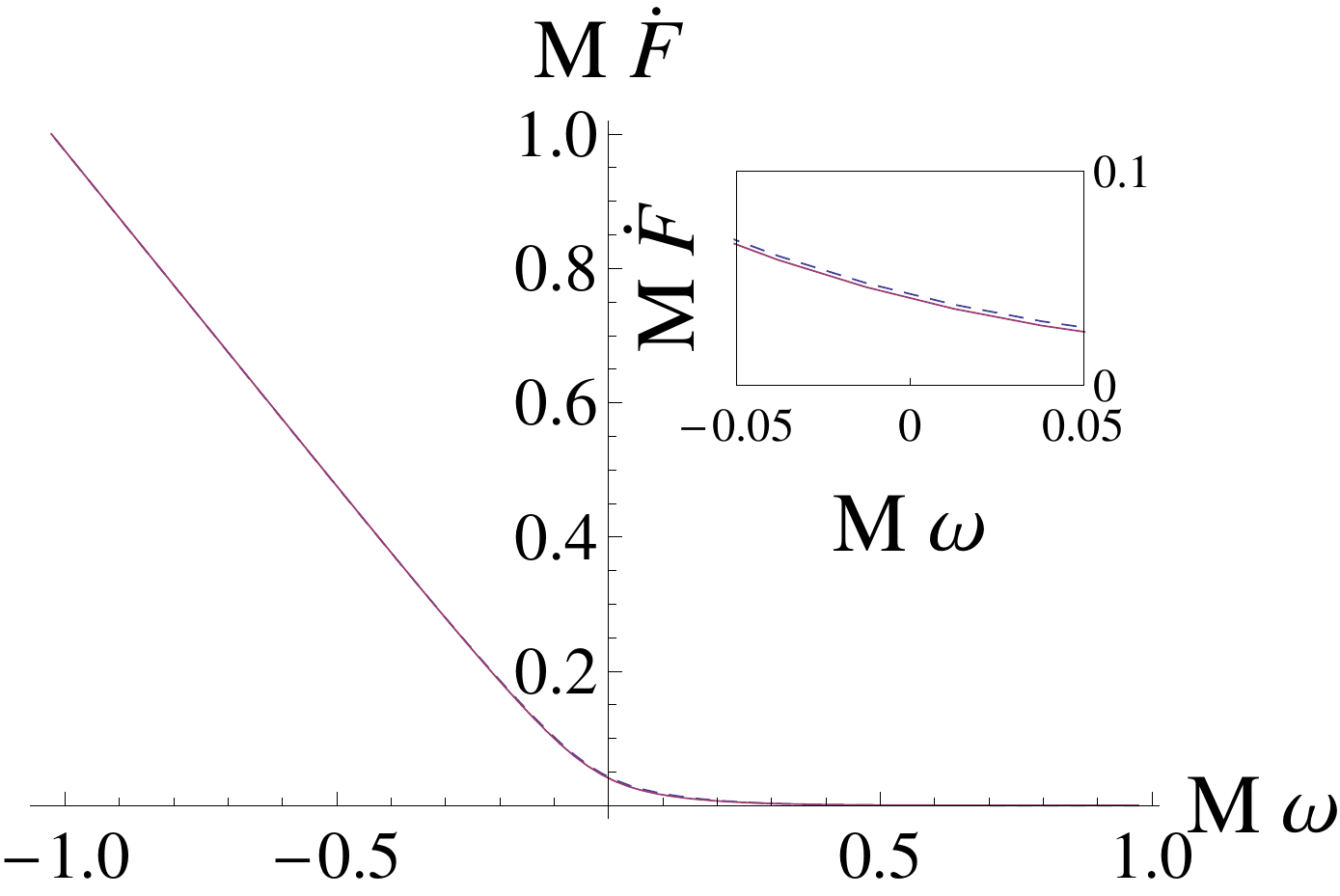}& 
\includegraphics[width=0.31\textwidth]{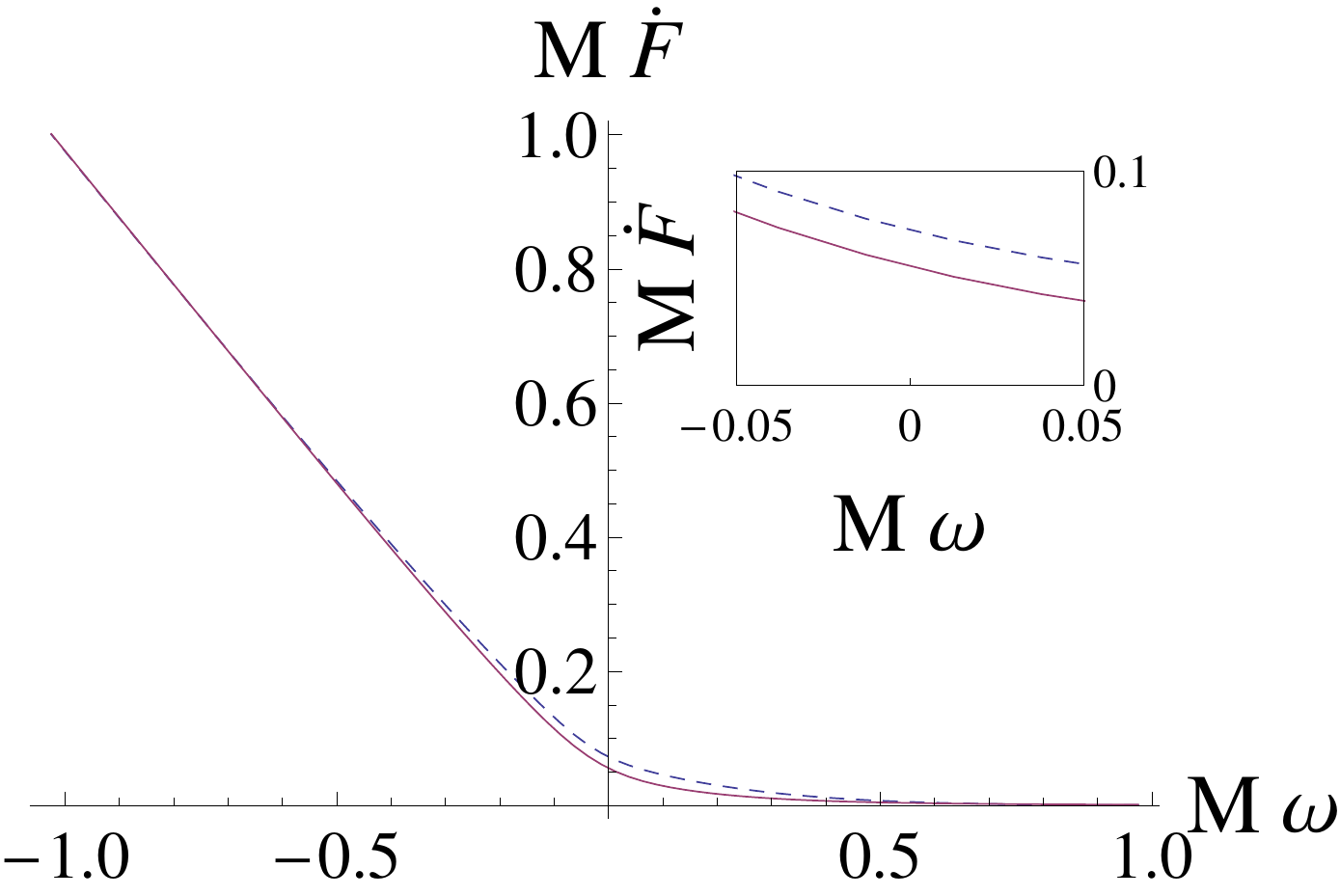} &
\includegraphics[width=0.31\textwidth]{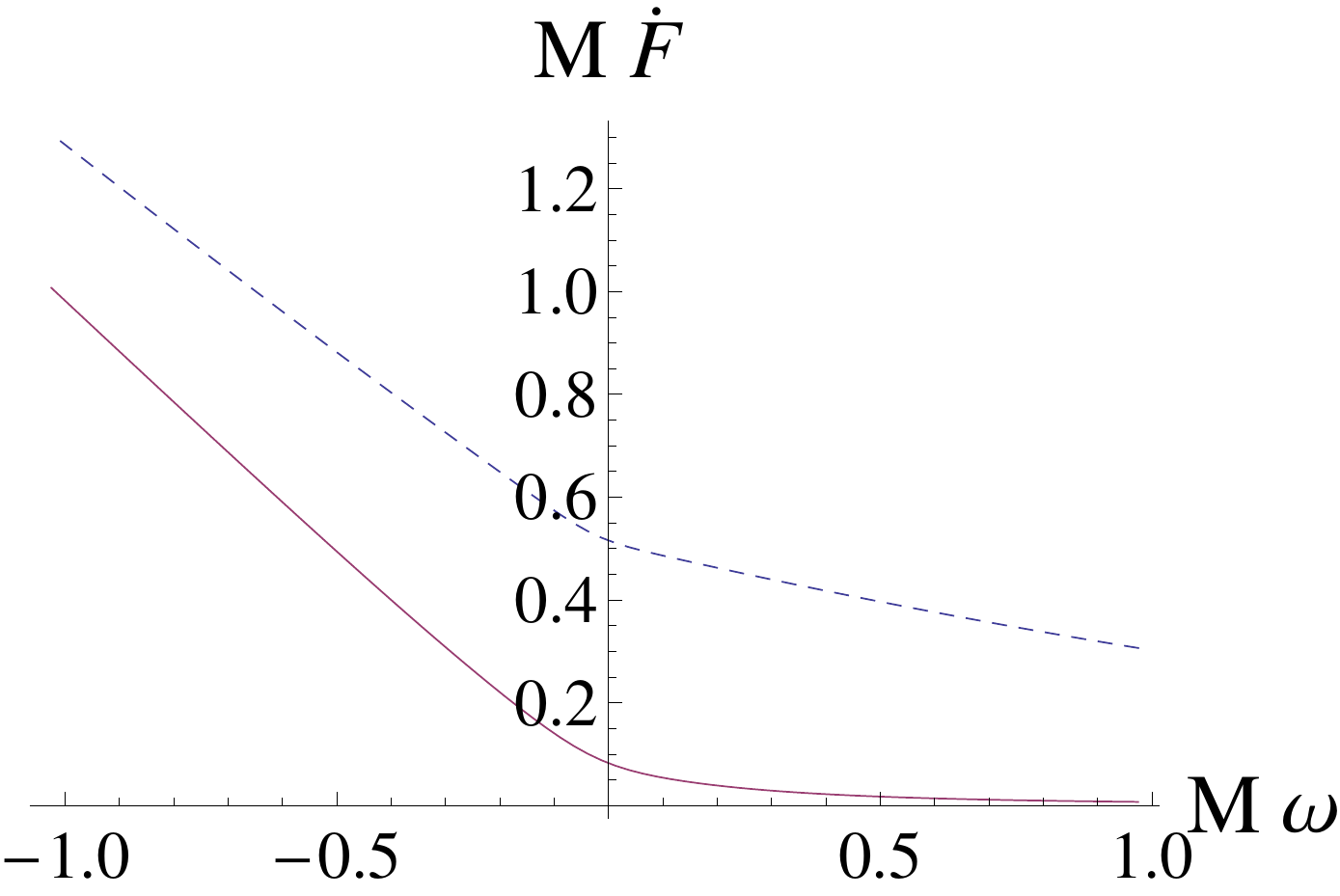}\\
(a) & (b) & (c)
\end{tabular}
\caption{The solid (red) curve shows 
$M \dot{\mathcal{F}}$ 
as a function of $M\omega$ for the $E=1$ trajectory in the 
HHI vacuum, at the times 
(a) $\tau = -10 M$, 
(b) $\tau = -3.5 M$ 
and 
(c) $\tau = -1.5 M$, all of them before the 
horizon-crossing, which occurs at $\tau = \tau_h = -(4/3) M$. 
The dashed (blue) curve shows $M$ times 
the Minkowski thermal bath response 
\eqref{eq:Mink-drift-transrate} at the local Hawking temperature 
$T_{\text{loc}}$ \eqref{eq:Tloc-Hawking} and 
with the Doppler shift factor 
$\lambda = \lambda_{\text{loc}} = \arctanh\bigl(\sqrt{2M/r}\,\bigr)$, 
as a function of $M\omega$. 
The discrepancy between the two curves shows that 
the Planckian character of the transition rate 
is lost as $\tau$ approaches~$\tau_h$, 
where the solid curve remains finite 
but the dashed curve disappears to $+\infty$.}
\label{fig:HHomega}
\end{figure}

\begin{figure}[p]
\centering
\begin{tabular}{ccc}
\includegraphics[width=0.31\textwidth]{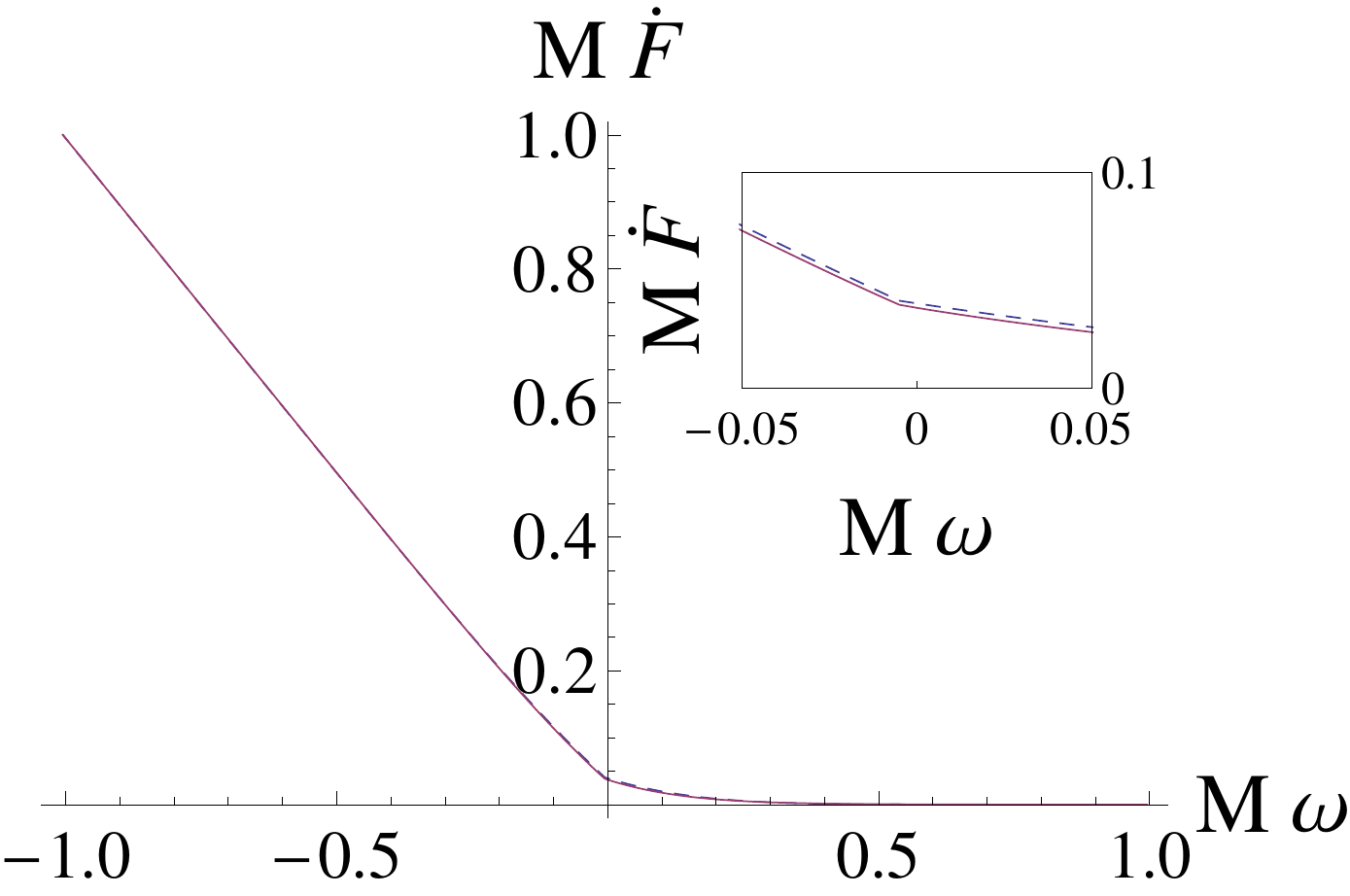}
& \includegraphics[width=0.31\textwidth]{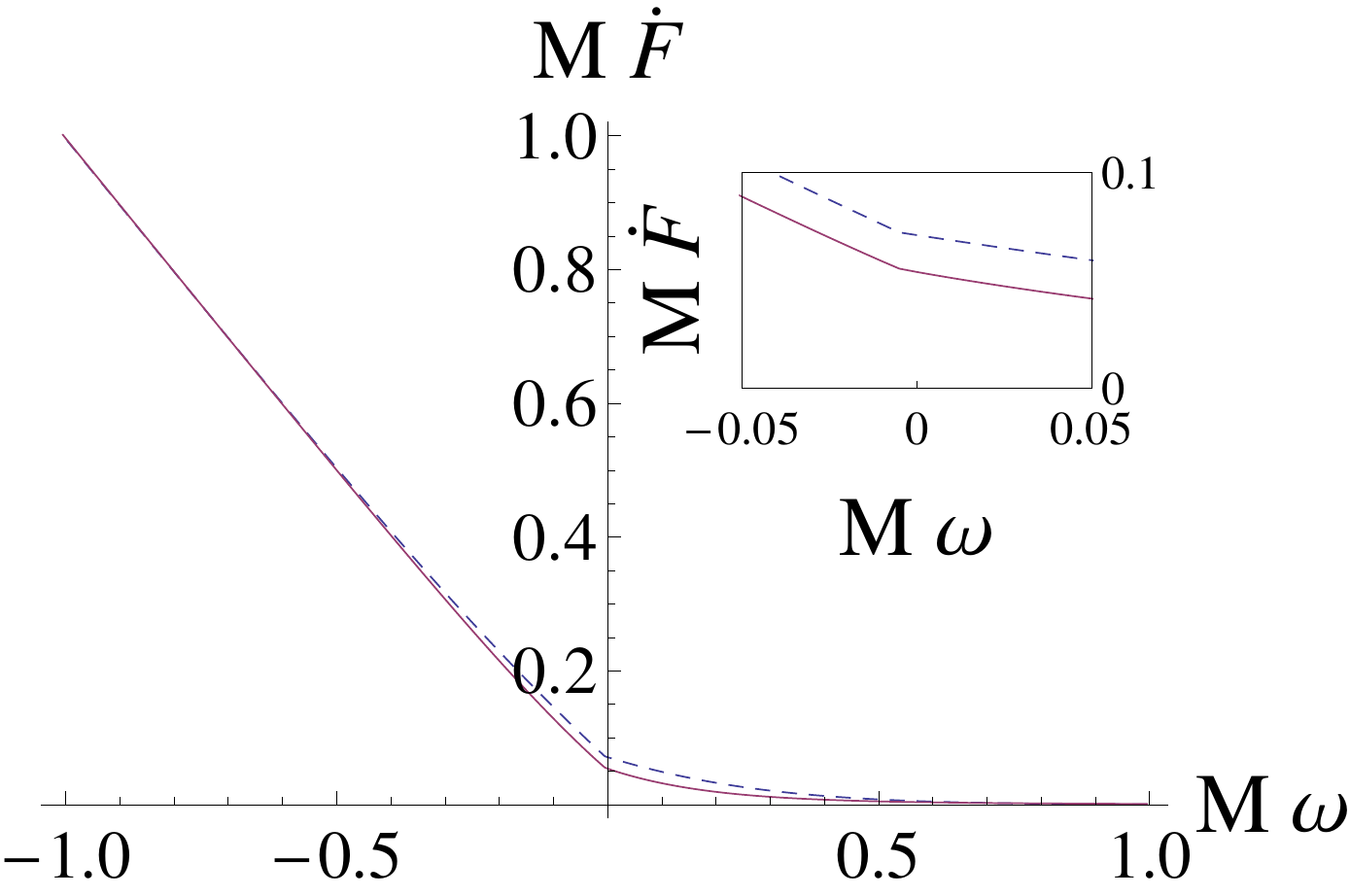} 
& \includegraphics[width=0.31\textwidth]{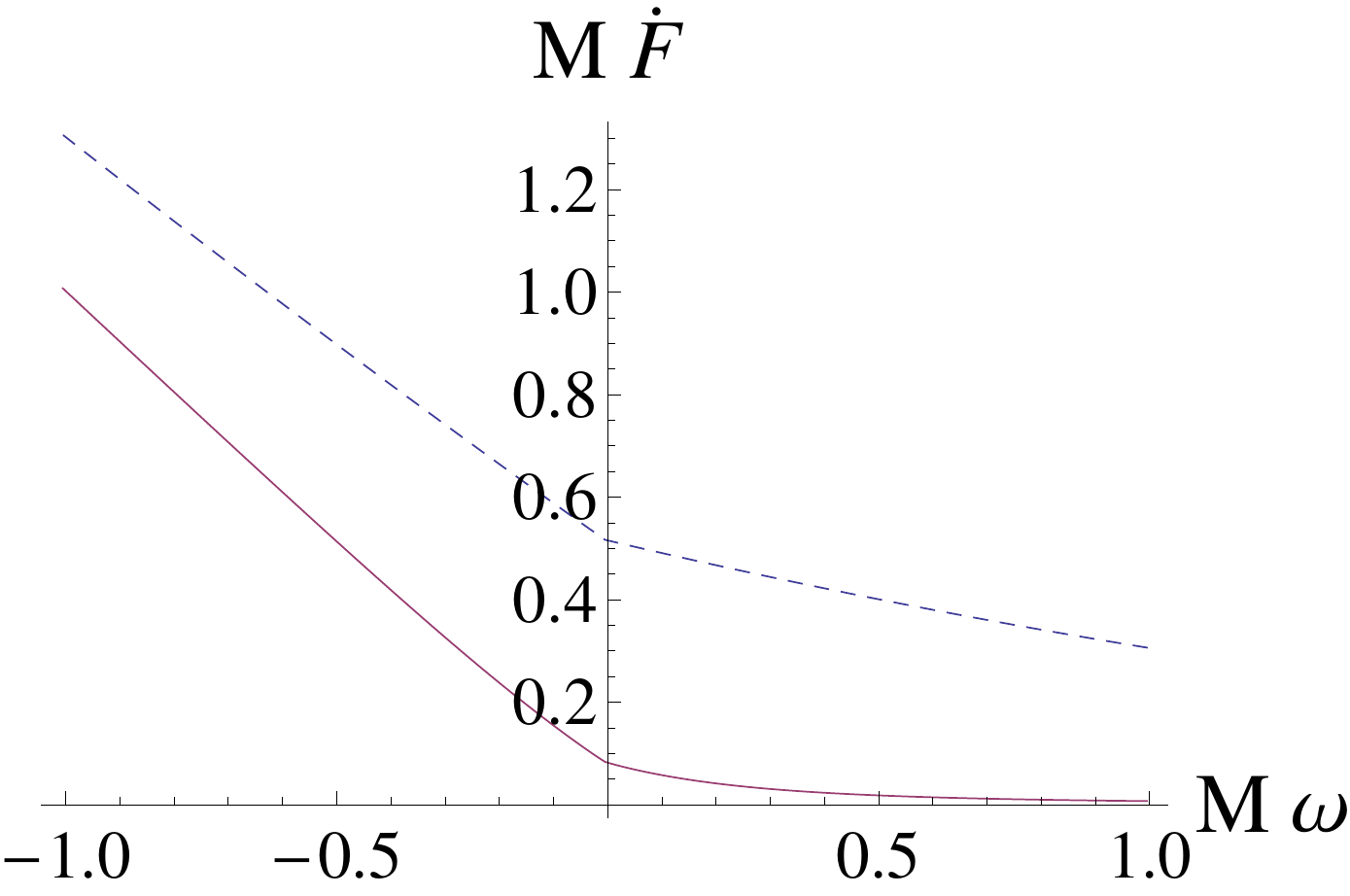}\\
(a) & (b) & (c)
\end{tabular}
\caption{The solid (red) curve is as in Figure \ref{fig:HHomega} 
but for the Unruh vacuum. 
The dashed (blue) curve shows $M$ times the Minkowski response 
\eqref{eq:Mink-mixed-trrate} in a state with vacuum left-movers and
thermal right-movers, at the local Hawking temperature  
$T_{\text{loc}}$ \eqref{eq:Tloc-Hawking} and with the Doppler shift
factor $\lambda = \lambda_{\text{loc}} = \arctanh\bigl(\sqrt{2M/r}\,\bigr)$. 
The discrepancy between the two curves again shows loss of the 
Planckian character as $\tau$ approaches~$\tau_h$, 
where the solid curve remains finite 
but the dashed curve disappears to $+\infty$. 
Note the discontinuous slope of the dashed curve at $\omega=0$.} 
\label{fig:Unruhomega}
\end{figure}

For a trajectory falling in from the infinity in the HHI and Unruh vacua, 
it is seen from the limits \eqref{eq:sch-transrate-aspast}
and 
\eqref{eq:sch-transrate-as-sing} that the thermal character 
of the transition rate 
is lost during the infall. 
Numerical evidence of how this loss takes place 
for the $E=1$ trajectory is shown in
Figures \ref{fig:HHomega}
and 
\ref{fig:Unruhomega}. The numerical evidence shows that the 
Planckian form of the transition rate is lost 
before the trajectory crosses the horizon. 

\begin{figure}[p]
\centering
\begin{tabular}{cc}
\includegraphics[width=0.45\textwidth]{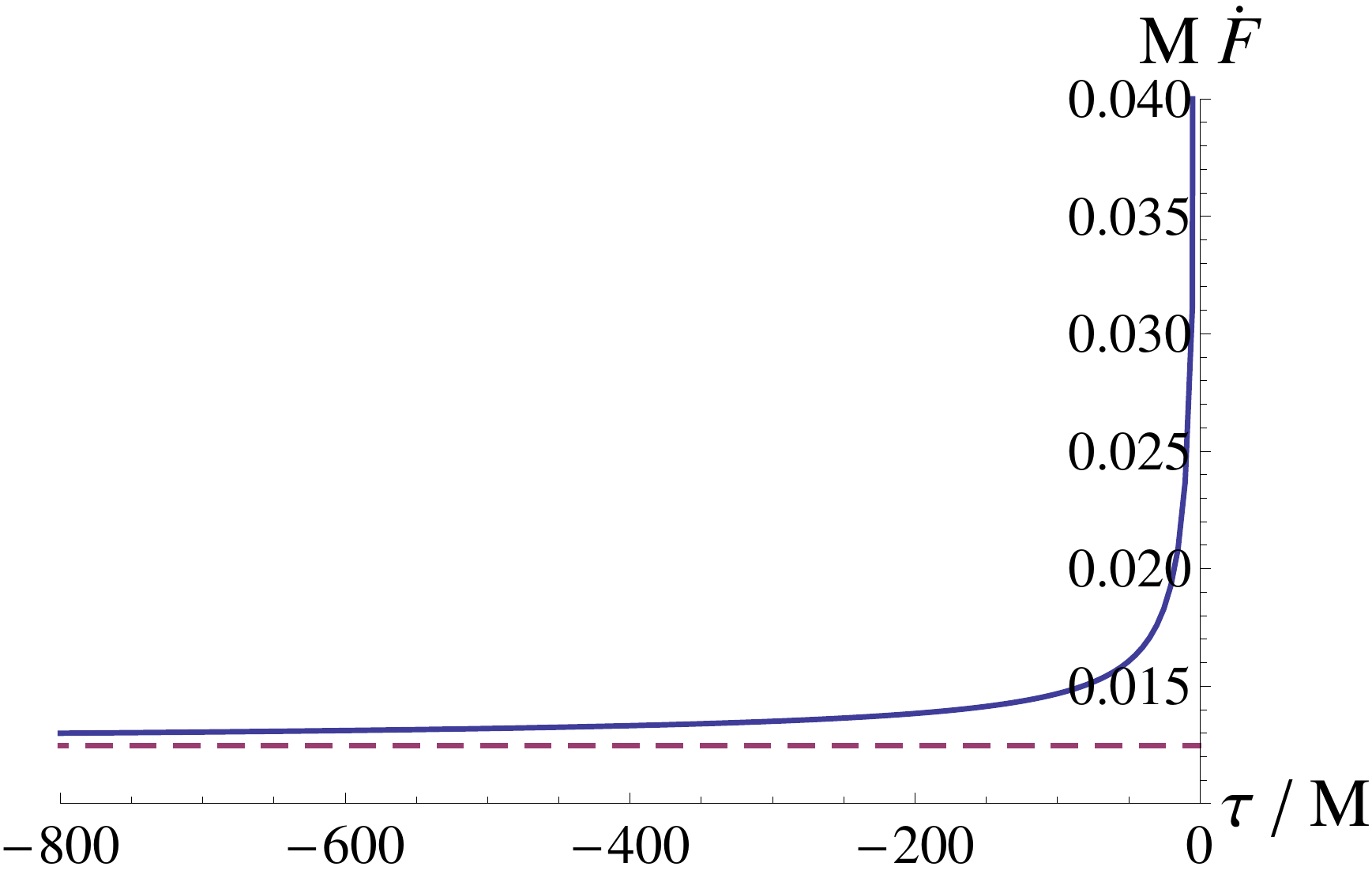}
& \includegraphics[width=0.45\textwidth]{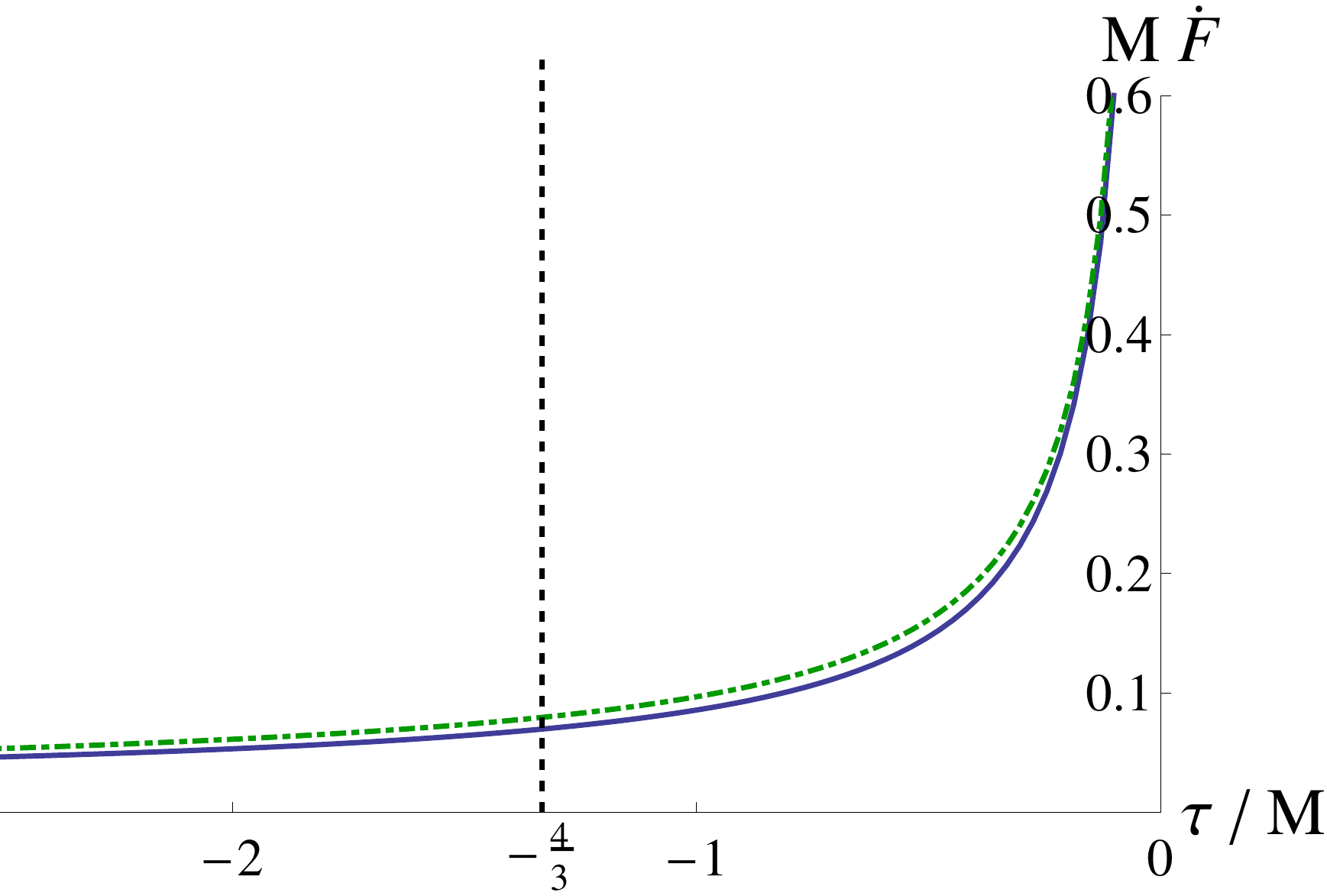}\\
(a) & (b)
\end{tabular}
\caption{(a) The solid (blue) curve shows 
$M \dot{\mathcal{F}}\bigl(1/(4 \pi M)\bigr)$ 
as a function of $\tau/M$ 
for the $E=1$ trajectory in the 
HHI vacuum. 
The dashed (red) line shows the value $1/[4\pi ( \emath^2 -1)]$ 
to which the solid curve asymptotes at $\tau/M \to -\infty$. 
(b) The solid (blue) curve shows a close-up of (a) 
near the horizon-crossing, 
$\tau/M = -4/3$. 
The dash-dotted (green) curve shows the 
$\tau$-dependent terms included in 
the asymptotic 
$\tau\to0$ expression~\eqref{eq:sch-transrate-as-sing}.}
\label{fig:HHt}
\end{figure}

\begin{figure}[p]
\centering
\includegraphics[width=0.6\textwidth]{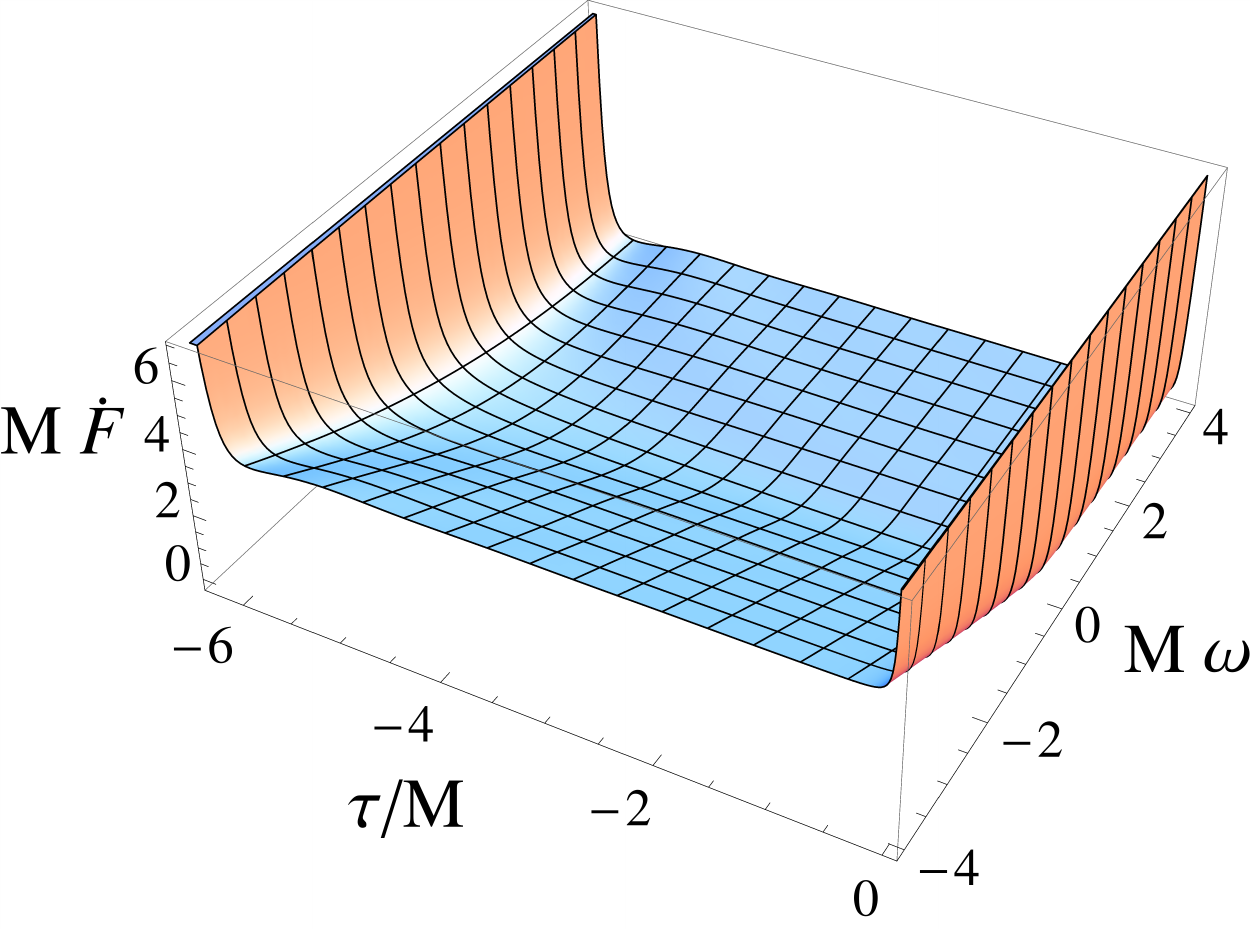}
\caption{The transition rate for a detector on the $E=0$ trajectory 
\eqref{eq:ubar-vbar-ito-varphi} in the HHI vacuum, with the switch-on 
at $\varphi = - 9\pi/10$, where $\tau \approx -3.136M$. 
The white hole singularity is at $\varphi = -\pi$, where $\tau = -\pi
M$. The divergence of the transition rate in the limit of short
detection time and in the limit of approaching the black hole
singularity is evident in the plot.}
\label{fig:wh}
\end{figure}

Numerical evidence also corroborates that the overall magnitude 
of the transition rate increases during the infall and grows 
without bound near the singularity. 
A sample plot for the $E=1$ trajectory in the HHI vacuum is shown in 
Figure~\ref{fig:HHt}. 

Finally, suppose the field is in the HHI vacuum, 
and consider the trajectory that passes 
from the white hole to the black hole through the 
horizon bifurcation point. 
We use the parametrisation \eqref{eq:ubar-vbar-ito-varphi}, 
so that 
\eqref{eq:schw-tau-as-of-phi}
and 
\eqref{eq:schw-r-as-of-phi}
hold with $E=0$. We switch the detector on at $\varphi = - 9\pi/10$, 
close to but well separated from the white 
hole singularity at $\varphi = -\pi$. 
Figure \ref{fig:wh} shows a perspective plot of the transition 
rate as a function of $\omega$ and $\tau$. 
The plot shows clearly both the divergence when 
$\tau$ approaches the switch-on time, 
arising from the last term in~\eqref{eq:Fdot-sharp}, 
and the divergence when the trajectory approaches the black hole singularity. 

To examine the thermal character of the transition rate, 
we have evaluated numerically the quantity 
\begin{align}
T_{\text{as,KMS}}(\omega,\tau) \doteq 
\frac{\omega}{\ln \bigl(\mathcal{\dot{F}}(-\omega,\tau)/
\mathcal{\dot{F}}(\omega,\tau)\bigr)}
\ . 
\end{align}
If $T_{\text{as,KMS}}$ were (approximately) independent of $\omega$ for fixed~$\tau$, 
the transition rate would satisfy (approximately) the KMS condition 
\eqref{eq:KMS-gen} and $T_{\text{as,KMS}}$ 
would be equal to (approximate) KMS temperature, 
which could possibly be $\tau$-dependent. 
Within the parameter range that our numerical experiments have been able to probe, 
we have however found no regimes in which $T_{\text{as,KMS}}$ 
would be (approximately) independent of~$\omega$.

\section{Conclusions\label{sec:conc}}

In this paper we have analysed an UDW detector that is coupled to the
proper time derivative of a scalar field in a $(1+1)$-dimensional
spacetime. Working within first-order perturbation theory, we showed
that although the derivative makes the interaction between the
detector and the field more singular, the singularity is no worse than
that of the non-derivative UDW detector in $(3+1)$ spacetime
dimensions, and issues of switching can be handled by the same
techniques. In particular, even though the transition probability
diverges in the sharp switching limit, the transition rate remains
well defined and allows the detector to address strongly
time-dependent situations.

Our main aim was to show that the derivative-coupling detector
provides a viable tool for probing a $(1+1)$-dimensional massless
field, whose infrared properties create ambiguities for the
conventional UDW detector in time-dependent situations.  We presented
strong evidence that the derivative-coupling detector does remain
well-behaved for the massless field, with and without time-dependence. 
As specific time-dependent examples, we
analysed an inertial detector in a Minkowski spacetime with an
exponentially receding mirror and a detector falling inertially into
the $(1+1)$-dimensional Schwarzschild black hole. In both cases we
recovered the expected thermal results due to the Hawking-Unruh effect
in the appropriate limits. In the receding mirror spacetime we saw the
thermality gradually set in as the mirror's acceleration approaches the
asymptotic late time behaviour, tailored to model the late time
effects of a gravitational collapse. In the $(1+1)$-dimensional
Schwarzschild spacetime we saw thermality gradually lost as the
detector falls, and we saw the transition rate diverge as the detector
approaches the black hole singularity, for both the HHI and Unruh
vacua.

Our results about the time-dependence of the Hawking-Unruh effect
complement those obtained via Bogoliubov coefficient techniques or via
a quasi-temperature approximation to the Wightman function
\cite{Barcelo:2010pj,Barbado:2011dx, Barbado:2012pt,Smerlak:2013sga}.
A key input in our analysis was to characterise the time-dependence of
the response in terms of the instantaneous transition rate, defined by
taking the sharp switching limit, and mathematically well defined in 
our spacetimes even when the time-dependence is strong. 
A~conceptual disadvantage of the instantanous transition rate is however that
it cannot be measured by a single detector, or even by an ensemble of
detectors, but the measurement requires an ensemble of ensembles of detectors
\cite{langlois-thesis,louko-satz-curved}. A~technical disadvantage is
that the instantanous transition rate becomes singular when the
Wightman function has singularities that typically occur with spatial
periodicity~\cite{Martin-Martinez:2014qda}. Further, the limit of
sharp switching and the limit of large energy gap need not
commute~\cite{few-jua-lou}, which becomes an issue when one
attempts to identify characteristics of thermal behavour in the
response of a detector that operates for a genuinely finite interval of
time. While we hence do not advocate the instantaneous transition rate
as a definitive quantifier of time-dependence in the detector's
response, our results strongly suggest that the instantaneous
transition rate conveys a physically expected picture about the onset
and decay of the Hawking-Unruh effect.

\section*{Acknowledgments}

We thank Don Marolf and Adrian Ottewill for discussions and
correspondence, encouraging us to look at the derivative-coupling
detector.  
We also thank 
Chris Fewster, 
Lee Hodgkinson, 
Bei-Lok Hu,
Bernard Kay, 
Eduardo Mart\'in-Mart\'inez, 
Suprit Singh 
and 
Matteo Smerlak 
for helpful discussions and comments. A~special thanks to Don Page
for questions that led to the correction of an error in 
the captions of Figures \ref{fig:HHomega} and~\ref{fig:Unruhomega}.
BAJA was supported by Consejo Nacional de Ciencia y Tecnolog\'ia
(CONACYT), with additional support from Sistema Estatal de Becas de
Veracruz, Mexico. JL was supported in part by STFC (Theory
Consolidated Grant ST/J000388/1).


\appendix

\section{Evaluation of 
$\mathcal{F}_\text{\rm sing}$ \eqref{eq:Fsing0}\label{app:Fsing}}

In this appendix we show that 
$\mathcal{F}_\text{sing}$ \eqref{eq:Fsing0}
can be written as \eqref{eq:Fsing-clean1-appendix}
or~\eqref{eq:Fsing-clean2-appendix}. 

Starting from \eqref{eq:Fsing0} and integrating 
the distributional derivatives by parts, we have 
\begin{align}
\label{eq:Fsing-iparts}
\mathcal{F}_\text{sing}(\omega)
=\int^{\infty}_{-\infty}\,d\tau'\,\int^{\infty}_{-\infty}\,d\tau''\, 
Q_\omega'(\tau') \overline{Q_\omega'(\tau'')} \, 
\mathcal{W}_\text{sing}(\tau',\tau'')
\ , 
\end{align}
where $Q_\omega(\tau) \doteq \emath^{-\ii \omega \tau}
\chi(\tau)$ 
and the prime denotes derivative with respect to the argument. 
Note that the integrand in \eqref{eq:Fsing-iparts} 
is a locally integrable function, containing no distributional parts. 
Using the explicit form of $\mathcal{W}_\text{sing}$~\eqref{eq:Wsing-def}, 
we obtain 
\begin{subequations}
\begin{align}
\label{eq:Fsing-decomp}
\mathcal{F}_\text{sing}(\omega)
&=
\mathcal{F}_{\text{sing},1}(\omega)
+ 
\mathcal{F}_{\text{sing},2}(\omega)
\ , 
\\[1ex]
\mathcal{F}_{\text{sing},1}(\omega)
&= 
- \frac{\ii}{4}
\int^{\infty}_{-\infty}\,d\tau'\,\int^{\infty}_{-\infty}\,d\tau''\, 
Q_\omega'(\tau') \overline{Q_\omega'(\tau'')} \, 
\sgn(\tau'-\tau'')
\ , 
\label{eq:Fsing1}
\\[1ex]
\mathcal{F}_{\text{sing},2}(\omega)
&= 
- \frac{1}{2\pi}
\int^{\infty}_{-\infty}\,d\tau'\,\int^{\infty}_{-\infty}\,d\tau''\, 
Q_\omega'(\tau') \overline{Q_\omega'(\tau'')} \, 
\ln|\tau'-\tau''|
\ . 
\label{eq:Fsing2}
\end{align}
\end{subequations}

For $\mathcal{F}_{\text{sing},1}$, integrating over $\tau''$ in 
\eqref{eq:Fsing1} gives 
\begin{align}
\mathcal{F}_{\text{sing},1}(\omega)
&= 
- \frac{\ii}{2}
\int^{\infty}_{-\infty} d u\,
Q_\omega'(u) \overline{Q_\omega(u)}
\notag
\\[1ex]
&= 
- \frac{\ii}{2}
\int^{\infty}_{-\infty} du\,
[
\chi'(u) - \ii \omega \chi(u) ] 
\, 
\chi(u)
\notag
\\[1ex]
&= 
- \frac{\omega}{2} 
\int_{-\infty}^{\infty} d u \, {[\chi(u)]}^2
\ , 
\label{eq:Fsing1-result}
\end{align}
where we have renamed $\tau'$ as~$u$, used the definition
of~$Q_\omega$, and finally noted that $\int_{-\infty}^{\infty} du \,
\chi'(u) \chi(u) = \frac12 \int_{-\infty}^{\infty} du \,
\frac{d}{du}{[\chi(u)]}^2 = 0$.

For $\mathcal{F}_{\text{sing},2}$, we break the integral in
\eqref{eq:Fsing2} into the subdomains $\tau'>\tau''$ and
$\tau'<\tau''$.  In the subdomain $\tau'>\tau''$ we write $\tau'=u$
and $\tau'' = u-s$, where $u \in \BbbR$ and $0<s<\infty$, and in the
subdomain $\tau'<\tau''$ we write $\tau''=u$ and $\tau' = u-s$, where
again $u \in \BbbR$ and $0<s<\infty$. This gives
\begin{align}
\mathcal{F}_{\text{sing},2}(\omega)
&= 
- \frac{1}{\pi}
\int^{\infty}_{0} d s \ln s
\int_{-\infty}^{\infty} d u 
\Realpart \left[
Q_\omega'(u) \overline{Q_\omega'(u-s)} 
\, 
\right]
\notag
\\[1ex]
&= 
\frac{1}{\pi}
\int^{\infty}_{0} d s \ln s
\int_{-\infty}^{\infty} d u 
\Realpart \left[
Q_\omega(u) \overline{Q_\omega''(u-s)} 
\, 
\right]
\notag
\\[1ex]
&= 
\frac{1}{\pi}
\int^{\infty}_{0} d s \ln s
\, 
\frac{d^2}{ds^2}
\int_{-\infty}^{\infty} d u 
\Realpart \left[
Q_\omega(u) \overline{Q_\omega(u-s)} 
\, 
\right]
\notag
\\[1ex]
&= 
\frac{1}{\pi}
\int^{\infty}_{0} d s \ln s
\, 
\frac{d^2}{ds^2} \! 
\left(
\cos(\omega s) \! 
\int_{-\infty}^{\infty} d u \, 
\chi(u) \chi(u-s) 
\right)
\notag
\\[1ex]
&= 
- \frac{1}{\pi}
\int^{\infty}_{0} \frac{d s}{s}
\, 
\frac{d}{ds} \! 
\left(
\cos(\omega s) \! 
\int_{-\infty}^{\infty} d u \, 
\chi(u) \chi(u-s) 
\right)
\ , 
\label{eq:Fsing2-interm1}
\end{align}
where we have first integrated by parts in~$u$, 
then written the derivatives in $\overline{Q_\omega''(u-s)}$ as 
$s$-derivatives outside the $u$-integral,
then used the definition of~$Q_\omega$, 
and finally integrated by parts in~$s$. 
The substitution term from $s=0$ in the integration by parts vanishes
because $\cos(\omega s)\int_{-\infty}^{\infty} d u \, 
\chi(u) \chi(u-s)$ is even in~$s$, 
and the integral over $s$ 
in the last expression in \eqref{eq:Fsing2-interm1} 
is convergent at small $s$ for the same reason. 

In the last expression in~\eqref{eq:Fsing2-interm1}, writing 
$\chi(u) \chi(u-s) = {[\chi(u)]}^2 - \chi(u) [ \chi(u) - \chi(u-s)]$
gives 
\begin{align}
\mathcal{F}_{\text{sing},2}(\omega)
&= 
\frac{\omega}{\pi}
\left(\int_{-\infty}^{\infty} d u \, {[\chi(u)]}^2\right)
\int_0^\infty ds \, \frac{\sin(\omega s)}{s} 
\notag
\\[1ex]
& \hspace{3ex}
+ 
\frac{1}{\pi}
\int^{\infty}_{0} \frac{d s}{s}
\, 
\frac{d}{ds} \! 
\left(
\cos(\omega s) \! 
\int_{-\infty}^{\infty} d u \, 
\chi(u) [\chi(u) - \chi(u-s)] 
\right)
\notag
\\[1ex]
&= 
\frac{|\omega|}{2}
\int_{-\infty}^{\infty} d u \, {[\chi(u)]}^2
+ 
\frac{1}{\pi}
\int^{\infty}_{0} 
ds \, 
\frac{\cos(\omega s)}{s^2} 
\int_{-\infty}^{\infty} d u \, 
\chi(u) [\chi(u) - \chi(u-s)] 
\ , 
\label{eq:Fsing2-clean1}
\end{align}
where in the first term we have used the identity 
$\int_0^\infty ds \, s^{-1}\sin(\omega s) = \frac12 \pi\sgn\omega$, 
and in the second term we have integrated by parts. 
The integral over $s$ in the second term is convergent at small $s$ because 
$\int_{-\infty}^{\infty} d u \, 
\chi(u) [\chi(u) - \chi(u-s)] $ vanishes at $s=0$ and is even in~$s$. 

Combining 
\eqref{eq:Fsing1-result} 
and~\eqref{eq:Fsing2-clean1}, we obtain 
\begin{align}
\mathcal{F}_\text{sing}(\omega) 
& = 
-\omega \Theta(-\omega) 
\int_{-\infty}^{\infty} d u \, {[\chi(u)]}^2
\notag
\\[1ex]
&\hspace{3ex}
+ 
\frac{1}{\pi}
\int^{\infty}_{0} 
ds \, 
\frac{\cos(\omega s)}{s^2} 
\int_{-\infty}^{\infty} d u \, 
\chi(u) [\chi(u) - \chi(u-s)] 
\ . 
\label{eq:Fsing-clean1-appendix}
\end{align}
An alternative expression is 
\begin{align}
\mathcal{F}_\text{sing}(\omega) 
& = 
- \frac{\omega}{2} 
\int_{-\infty}^{\infty} \! d u \, {[\chi(u)]}^2
+
\frac{1}{\pi} \int_{0}^{\infty} \! \frac{ds}{s^2} \,
\int_{-\infty}^{\infty} \! du \,  
\chi(u) \left[ \chi(u) - \chi(u-s) \right] 
\notag
\\[1ex]
& 
\hspace{3ex}
+ 
\frac{1}{\pi} 
\int_{-\infty}^{\infty} \! du 
\int_{0}^{\infty} \! ds \,
\chi(u) \chi(u-s) 
\, 
\frac{[1 - \cos(\omega s)]}{s^2}
\ , 
\label{eq:Fsing-clean2-appendix}
\end{align}
which may be obtained from 
\eqref{eq:Fsing-clean1-appendix} by writing 
$\cos(\omega s) = 1 - [1 - \cos(\omega s)]$ and using the identity 
\begin{align}
\int_0^\infty
ds \, 
\frac{1-\cos(\omega s)}{s^2}
= \frac{\pi|\omega|}{2}
\ . 
\end{align}

\section{Evaluation of the static detector's transition rate 
in Minkowski (half-)space\label{app:mink-eval}}

In this appendix we verify formulas 
\eqref{eq:mpos-Fdot-combined}
and 
\eqref{eq:mzero-Fdot-combined}
for the transition rate of a static detector in 
Minkowski space and Minkowski half-space. 
We use the Wightman functions found in 
subsection \ref{massFdot} and evaluate 
the transition rate from~\eqref{eq:Fdot-stationary-naive}.

\subsection{$m=0$} 

We consider first the massless field, 
with the Wightman function given by 
\eqref{eq:mzero-Wightman-Mink}
and~\eqref{eq:mzero-Wightman-image}. 

In $\mathcal{M}$, we find from 
\eqref{eq:A-distr}, 
\eqref{eq:Mink-static-traj}
and \eqref{eq:mzero-Wightman-Mink} that
$\mathcal{A}(\tau',\tau'') 
= - 1/\bigl[2 \pi (\tau'-\tau'' - \ii \epsilon)^2\bigr]$. 
Evaluating \eqref{eq:Fdot-stationary-naive} 
as a contour integral gives~\eqref{eq:mzero-Fdot-Minkowski}. 

In $\Mhalfspace$, we find from 
\eqref{eq:mzero-Wightman-image} that the integrand in 
\eqref{eq:Fdot-stationary-naive} contains the additional piece 
\begin{align}
\Delta\mathcal{A}(\tau', \tau'') 
= 
- \frac{\eta}{2 \pi}
\, 
\frac{(\tau'-\tau'' - \ii\epsilon)^2 + 4d^2}{{[(\tau'-\tau'' - \ii\epsilon)^2 - 4d^2]}^2}
\ . 
\label{eq:app-Mink-Amzeroimage}
\end{align}
Evaluating the contribution to \eqref{eq:Fdot-stationary-naive} 
as a contour integral leads to~\eqref{eq:mzero-Fdot-Mtilde}. 

We note that $\Delta\mathcal{A}(\tau', \tau'')$
\eqref{eq:app-Mink-Amzeroimage} has distributional singularities at
$\tau'-\tau''=\pm2d$.  The geometric reason for these singularities is
that the points $\tau'$ and $\tau''$ on the detector's trajectory are
connected by a null ray that is reflected from the mirror.  As we have
seen, the stationary transition rate is well defined despite these
singularities.  Were we however to consider a detector that operates
for a finite duration, the singularities would interfere with the
sharp switching limit manipulations that led to \eqref{eq:Fdot-sharp}
when $\Delta\tau = 2d$.

\subsection{$m>0$}

For the massive field, the Wightman function is given by \eqref{eq:mpos-Wightman-Mink}
and~\eqref{eq:mpos-Wightman-image}. We consider $\mathcal{M}$ and $\Mhalfspace$ in turn. 

\subsubsection{$\mathcal{M}$}

In~$\mathcal{M}$, we find from 
\eqref{eq:A-distr}, 
\eqref{eq:Mink-static-traj}
and \eqref{eq:mpos-Wightman-Mink} that
\begin{align}
\mathcal{A}(\tau',\tau'') 
= \frac{m^2}{2\pi} K_0''\bigl[m\bigl(\epsilon + \ii(\tau'-\tau'')\bigr)\bigr] 
\ , 
\label{eq:app-Mink-Ampos}
\end{align}
where the prime denotes derivative with respect to the argument. 
From 
\eqref{eq:Fdot-stationary-naive}
we then obtain 
\begin{align}
\dot{\mathcal{F}}(\omega)
= \frac{m^2}{2\pi} \int_C 
ds \, \emath^{-\ii \omega s} 
\, 
K_0''(\ii m s) 
\ , 
\label{eq:Fdot-Mink-mpos-one}
\end{align}
where the contour $C$ in the complex $s$ plane follows 
the real axis from $-\infty$ to $+\infty$ 
except that it drops in the lower half-plane near $s=0$, and 
$K_0$ has its principal branch when $s$ is negative imaginary. 
We now assume $\omega\ne-m$: it follows then from the 
asymptotics of $K_0$ at large imaginary argument \cite{dlmf} that
\eqref{eq:Fdot-Mink-mpos-one}
is convergent as an improper Riemann integral. 

From \eqref{eq:Fdot-Mink-mpos-one} we obtain 
\begin{align}
\dot{\mathcal{F}}(\omega)
& = \frac{\omega^2}{2\pi} \int_C 
ds \, \emath^{-\ii \omega s} 
\, 
K_0(\ii m s) 
\notag
\\[1ex]
&= 
\frac{\omega^2}{2} \Imagpart \int_0^\infty
\! ds \, \emath^{-\ii \omega s} 
\, 
H_0^{(2)}(m s)
\ , 
\label{eq:Fdot-Mink-mpos-two}
\end{align}
where we have first integrated by parts twice, 
as allowed by the large $s$ behaviour of the integrand, 
then deformed $C$ to the real $s$ axis, as allowed by the merely 
logarithmic singularity of the integrand at $s=0$, 
and finally used the Bessel function analytic continuation formulas~\cite{dlmf}. 
The integral in \eqref{eq:Fdot-Mink-mpos-two}
was encountered in \cite{Hodgkinson:2013tsa} 
in the context of a non-derivative detector, 
and from equations (5.11) and (5.14) therein we have 
\begin{align}
\dot{\mathcal{F}}(\omega) 
= \frac{\omega^2}{\sqrt{\omega^2-m^2}} 
\, \Theta(-\omega-m)
\ , 
\end{align}
which is the result \eqref{eq:mpos-Fdot-Minkowski}
used in the main text.

\subsubsection{$\Mhalfspace$}

In~$\Mhalfspace$, we find from \eqref{eq:mpos-Wightman-image}
that the integrand in \eqref{eq:Fdot-stationary-naive}
contains the additional piece 
\begin{align}
\Delta\mathcal{A}(\tau, \tau'') 
= 
- \frac{\eta}{2 \pi} 
\frac{d^2}{d\tau'{}^2}
K_0
\left[
m \sqrt{ 4d^2 - (\tau'-\tau'' - \ii \epsilon)^2}
\right] 
\ , 
\label{eq:app-Mink-Amposimage}
\end{align}
where the branch of the square root is as explained in the main text. 
The additional piece in the transition rate
\eqref{eq:Fdot-stationary-naive}
is hence 
\begin{align}
\Delta \dot{\mathcal{F}}(\omega)
&= 
- \frac{\eta}{2\pi} \int_{-\infty}^\infty
\! ds \, \emath^{-\ii \omega s}
\, \frac{d^2}{ds^2}
K_0
\left[
m \sqrt{ 4d^2 - (s - \ii \epsilon)^2}
\right] 
\notag
\\[1ex]
&= 
\frac{\eta\omega^2}{2\pi} \int_{-\infty}^\infty
\! ds \, \emath^{-\ii \omega s}
\, 
K_0
\left[
m \sqrt{ 4d^2 - (s - \ii \epsilon)^2}
\right] 
\notag
\\[1ex]
&= 
\frac{\eta\omega^2}{\pi} \Realpart \int_{0}^\infty
\! ds \, \emath^{-\ii \omega s}
\, K_0
\left[
m \sqrt{ 4d^2 - (s - \ii \epsilon)^2}
\right] 
\ ,  
\label{eq:app-Mink-Fdot-mpos-image}
\end{align}
again assuming $\omega\ne-m$ and integrating by parts twice. 
The integral in \eqref{eq:app-Mink-Fdot-mpos-image} 
was encountered in \cite{Hodgkinson:2013tsa}, 
and from equations (5.15) and (5.25) therein we have 
\begin{align}
\Delta \dot{\mathcal{F}}(\omega)
= \frac{\eta \omega^2 \cos\bigl(2d\sqrt{\omega^2 - m^2}\,\bigr)}{\sqrt{\omega^2-m^2}} 
\, \Theta(-\omega-m)
\ , 
\label{eq:mpos-Fdot-image-appendix}
\end{align}
which leads to the result \eqref{eq:mpos-Fdot-Mtilde} in the main text.

\section{Asymptotic past and future transition rate 
in the receding mirror spacetime\label{app:rec-mirror-calculation}}

In this appendix we find the 
asymptotic past and future forms 
\eqref{pastv0-futurev0} and \eqref{drift:pastv0-futurev0}
 of the transition rate 
of an inertial detector in the 
receding mirror spacetime of Section~\ref{sec:recmir}.

\subsection{Static in the distant past\label{appsub:Static}}

We wish to extract the asymptotic behaviour of 
\eqref{eq:mirror-asstat-Fdot}
as $\tau\to-\infty$ and as $\tau\to\infty$. 


\subsubsection{$\tau\to-\infty$}

Consider~\eqref{eq:mirror-asstat-regfree0-Fdot}. 
Using 
\eqref{eq:pfunc-def} and letting 
$h \doteq
\bigl(1+\emath^{\kappa (d-\tau)}\bigr)^{-1}$, 
we have 
\begin{align}
\dot{\mathcal{F}}_0 (\omega,\tau)
= 
- \omega \Theta(-\omega)
+ \frac{1}{2\pi} \int_0^\infty ds \, 
\cos(\omega s) \! 
\left(  
\frac{1}{X}
+ \frac{1}{s^2} \right) 
\ , 
\label{eq:app:F0past1}
\end{align}
where 
\begin{align}
X = 
- \frac{[1 - h(1 - \emath^{-\kappa s})] 
{\bigl\{\kappa s + \ln[1 - h(1 - \emath^{-\kappa s})]\bigr\}}^2}{\kappa^2 {(1-h)}^2}
\ . 
\label{eq:app:X-def}
\end{align}
The limit $\tau\to-\infty$ is now the limit $h\to0_+$. 

Following the technique used in subsection 5.3 of~\cite{louko-satz-spatial}, 
we make in the integrand of \eqref{eq:app:F0past1}
the re-arrangement 
\begin{align}
\frac{1}{X}
+ \frac{1}{s^2}
= 
\frac{-X - s^2}{s^4}
\left(
1 + \frac{-X - s^2}{s^2}
\right)^{-1}
\ . 
\label{eq:app:F0int-rearr}
\end{align}
Taylor expanding the numerator of \eqref{eq:app:X-def} to quartic order in 
$h(1 - \emath^{-\kappa s})$ shows that the second factor 
in \eqref{eq:app:F0int-rearr} is of the form $1 + O(h)$, 
uniformly in~$s$, and yields for the first factor in \eqref{eq:app:F0int-rearr} 
an estimate that can be applied under the integral over $s$ 
and whose leading term is proportional to~$h$.  We hence have  
\begin{align}
\dot{\mathcal{F}}_0 (\omega,\tau)
= 
- \omega \Theta(-\omega) + O(h)
\ . 
\label{eq:app:F0past0-final}
\end{align}


Consider then~\eqref{eq:mirror-asstat-regfree1-Fdot}. 
Proceeding similarly, we find 
\begin{align}
\dot{\mathcal{F}}_1 (\omega,\tau)
= 
\frac{1}{4 \pi d} + \frac{|\omega|}{2\pi}
\, \bigl[
\cos(2 d \omega) \si(2 d |\omega|)
- 
\sin(2 d |\omega|) 
\Ci (2d |\omega|) 
\bigr]
+ O(h)
\ , 
\label{eq:app:F0past1-final}
\end{align}
where $\si$ and $\Ci$ are the sine and cosine integrals 
in the notation of~\cite{dlmf}. 


Consider finally~\eqref{eq:mirror-asstat-regulator-Fdot}. 
Integrating by parts once reduces the integral to a form 
that can be evaluated exactly in terms of the sine and 
cosine integrals~\cite{dlmf}, with the result 
\begin{align}
\dot{\mathcal{F}}_2 (\omega,\tau)
&= 
\frac{1-h}{2 \pi} 
\biggl\{ 
-\frac{1}{B} 
+ |\omega|
\bigl[\sin (B |\omega|) \Ci (B |\omega|) 
- \cos( B \omega) \si(B |\omega|) \bigr] 
\notag
\\[1ex]
& \hspace{12ex}
+ 2\pi \omega \cos( B \omega)\Theta(-\omega) \biggr\}
\ , 
\label{eq:app:F0pf2-exact}
\end{align}
where $B \doteq 2d - \kappa^{-1}\ln(1-h)$. 
A~small $h$ expansion in \eqref{eq:app:F0pf2-exact}
gives 
\begin{align}
\dot{\mathcal{F}}_2 (\omega,\tau)
&= 
- \frac{1}{4 \pi d} 
+ \frac{|\omega|}{2\pi}
\, 
\bigl[\sin (2d |\omega|) \Ci (2d |\omega|) 
- \cos( 2d \omega) \si(2d |\omega|) \bigr] 
\notag
\\[1ex]
& \hspace{3ex}
+ \omega \cos( B \omega)\Theta(-\omega)
+ O(h)
\ . 
\label{eq:app:F0past2-final}
\end{align}


Combining 
\eqref{eq:app:F0past0-final}, 
\eqref{eq:app:F0past1-final}
and 
\eqref{eq:app:F0past2-final}, we have 
\begin{align}
\dot{\mathcal{F}} (\omega,\tau)
= 
-\omega \, [1-\cos (2 d \omega) ] \, \Theta (-\omega)  + O(\emath^{\kappa \tau})
\hspace{3ex}
\text{as $\tau \to -\infty$}\,. 
\label{eq:app:F0past-finalsum}
\end{align}

\subsubsection{$\tau\to\infty$}

Consider~\eqref{eq:mirror-asstat-regfree0-Fdot}. 
Letting $f\doteq 1/(1+\emath^{\kappa(\tau-d)})$,
and adding and subtracting $\kappa^2 \cos(\omega s) [8\pi\sinh^2(\kappa s/2)]^{-1}$ 
in the integrand, we obtain 
\begin{align}
&\dot{\mathcal{F}}_0 (\omega,\tau)
= 
- \omega \Theta(-\omega)
+ \frac{1}{2\pi} \int_0^\infty ds \, 
\cos(\omega s) \! 
\left(  
\frac{1}{s^2}
- \frac{\kappa^2}{4\sinh^2(\kappa s/2)} \right) 
\notag
\\[1ex]
& \hspace{4ex}
+ \frac{\kappa^2}{2\pi} \int_0^\infty ds \, 
\cos(\omega s) \! 
\left(  
\frac{1}{4\sinh^2(\kappa s/2)} 
- 
\frac{f^2 \, \emath^{\kappa s}}
{[1 + f(\emath^{\kappa s} -1)] 
{\bigl\{\ln[1 + f(\emath^{\kappa s} -1)]\bigr\}}^2}
\right) 
\ . 
\label{eq:app:F0future1}
\end{align}
In the last term in \eqref{eq:app:F0future1}, 
the integrand goes to zero pointwise as $f\to0$, 
and a monotone convergence argument shows that the integral vanishes as $f\to0$. 
The second term plus half of the first term is equal to half of the transition 
rate encountered in subsection \ref{thermalFdot} (with $a\to\kappa$) 
and evaluated to~\eqref{eq:Rindler-transrate}. 
We hence have 
\begin{align}
\dot{\mathcal{F}}_0 (\omega,\tau)
= -\frac{\omega}{2} 
\Theta (-\omega) + \frac{\omega}{2 \, (\emath^{2\pi\omega/\kappa}-1)}
+ o(1) 
\hspace{3ex}
\text{as $f\to0$}\,.
\label{eq:app:F0future0-final}
\end{align}

In~\eqref{eq:mirror-asstat-regfree1-Fdot}, a straightforward monotone convergence argument gives 
$\dot{\mathcal{F}}_1 (\omega,\tau) = o(1)$. 
In~\eqref{eq:mirror-asstat-regulator-Fdot}, \eqref{eq:app:F0pf2-exact} gives 
$\dot{\mathcal{F}}_2 (\omega,\tau) = O(f)$. 

Combining, we have 
\begin{align}
\dot{\mathcal{F}} (\omega,\tau)
= -\frac{\omega}{2} 
\Theta (-\omega) + \frac{\omega}{2 \, (\emath^{2\pi\omega/\kappa}-1)}
+ o(1) 
\hspace{3ex}
\text{as $\tau\to\infty$}\,.
\label{eq:app:F0future-finalsum}
\end{align}

\subsection{Travelling towards the mirror in the distant past\label{appsub:Drifting}}

We wish to extract the asmptotic behaviour of \eqref{eq:mirror-asdrift-Fdot}
as $\tau\to-\infty$ and as $\tau\to\infty$.

\subsubsection{$\tau\to-\infty$}

For~\eqref{eq:mirror-asdrift-regfree0-Fdot}, proceeding as in 
\eqref{eq:app:F0past1}--\eqref{eq:app:F0past0-final}
gives 
\begin{align}
\dot{\mathcal{F}}_0 (\omega,\tau)
&= 
- \omega \Theta(-\omega)
+ O\bigl(\emath^{\emath^\lambda\kappa\tau}\bigr)
\ . 
\label{eq:app:asdrift:F0past0-final}
\end{align}

For~\eqref{eq:mirror-asdrift-regfree1-Fdot}, we have 
\begin{align}
\dot{\mathcal{F}}_1 (\omega,\tau)
= 
\frac{1}{2\pi} \int_0^\infty
\frac{\cos(\omega s) \, ds}{\left(1 + g \emath^{-\kappa s \emath^\lambda} \right)
\left[
s \emath^\lambda - 2\tau \sinh\lambda 
+ \kappa^{-1} \ln \! \left(1 + g \emath^{-\kappa s \emath^\lambda} \right)
\right]^2}
\ , 
\label{eq:app:asdrift:F0past1-1}
\end{align}
where $g = \emath^{\kappa\tau \emath^\lambda}$. 
When $\tau<0$, we may bound the absolute value of $\dot{\mathcal{F}}_1 (\omega,\tau)$ 
by the replacements $\cos(\omega s) \to1$ and $g\to0$ in \eqref{eq:app:asdrift:F0past1-1}, 
and evaluating the integral that ensues gives 
$\dot{\mathcal{F}}_1 (\omega,\tau) = O(\tau^{-1})$. 

For \eqref{eq:mirror-asdrift-regulator-Fdot}, we proceed as 
with~\eqref{eq:app:F0pf2-exact}, obtaining the exact result 
\begin{align}
\dot{\mathcal{F}}_2 (\omega,\tau)
&= 
\frac{(1-h) \, \emath^{2\lambda}}{2 \pi} 
\biggl\{ 
-\frac{1}{C} 
+ |\omega|
\bigl[\sin (C |\omega|) \Ci (C |\omega|) 
- \cos( C \omega) \si(C |\omega|) \bigr] 
\notag
\\[1ex]
& \hspace{16ex}
+ 2\pi \omega \cos( C \omega)\Theta(-\omega) \biggr\}
\ , 
\label{eq:app:drift-F0pf2-exact}
\end{align}
where $h = g/(1+g)$ and 
$C \doteq - (\emath^{2\lambda} -1)\tau - \kappa^{-1} \emath^\lambda \ln(1-h)$. 
As $\tau \to -\infty$, we have $C \to \infty$, 
and using formulas (6.2.17) and (6.12.3) in \cite{dlmf} gives 
\begin{align}
\dot{\mathcal{F}}_2 (\omega,\tau)
= 
\emath^{2\lambda} \, \omega \cos(2 \tau \sinh\lambda \,  \emath^\lambda \omega)
\Theta(-\omega) + O(\tau^{-3})
\ . 
\label{eq:app:asdrift:F0past2-final}
\end{align}

Combining, 
we have 
\begin{align}
\dot{\mathcal{F}} (\omega,\tau)
&= 
-\omega \left[1- \emath^{2\lambda} \cos(2 \tau \sinh\lambda \,  \emath^\lambda \omega) \right] 
\Theta(-\omega) + O(\tau^{-1})
\ . 
\label{eq:app:asdrift:F0past-finalsum}
\end{align}

\subsubsection{$\tau\to\infty$}

For~\eqref{eq:mirror-asdrift-regfree0-Fdot}, proceeding as in \eqref{eq:app:F0future1}
gives 
\begin{align}
\dot{\mathcal{F}}_0 (\omega,\tau)
= 
-\frac{\omega}{2} 
\Theta (-\omega) + \frac{\omega}{2 \, (\emath^{2\pi\emath^{-\lambda}\omega/\kappa}-1)}
+ o(1) 
\hspace{3ex}
\text{as $\tau \rightarrow \infty$}\,.
\label{eq:app:asdrift:F0future0-final}
\end{align}

For~\eqref{eq:mirror-asdrift-regfree1-Fdot}, 
using \eqref{eq:app:asdrift:F0past1-1} and substituting $s = \tau+r$ gives 
\begin{align}
\dot{\mathcal{F}}_1 (\omega,\tau)
= 
\frac{\kappa^2}{2\pi} \int_{-\tau}^\infty
\frac{\cos[\omega(\tau+r)] \, ds}{\left(1 + \emath^{-\kappa r \emath^\lambda} \right)
\left[
\kappa \tau \emath^{-\lambda}
+ \ln \! \left(1 + \emath^{\kappa r \emath^\lambda} \right)
\right]^2}
\ .  
\label{eq:app:asdrift:F0future1-1}
\end{align}
We may assume $\tau >0$. To bound the absolute value of~\eqref{eq:app:asdrift:F0future1-1}, 
we make in the integrand the replacement $\cos[\omega(\tau+r)] \to 1$
and extend the integration to be over the full real axis in~$r$. 
Elementary estimates then show that 
the contribution from $-\infty < r < 0$ is $O(\tau^{-2})$ 
and 
the contribution from from $0 < r < \infty$ is $O(\tau^{-1})$. Hence 
$\dot{\mathcal{F}}_1 (\omega,\tau) = O(\tau^{-1})$. 


For \eqref{eq:mirror-asdrift-regulator-Fdot}, 
\eqref{eq:app:drift-F0pf2-exact} gives 
$\dot{\mathcal{F}}_2 (\omega,\tau)
= O\bigl(\emath^{-\emath^\lambda \kappa\tau}\bigr)$. 


Combining, we have 
\begin{align}
\dot{\mathcal{F}} (\omega,\tau)
= 
-\frac{\omega}{2} 
\Theta (-\omega) + \frac{\omega}{2 \, (\emath^{2\pi\emath^{-\lambda}\omega/\kappa}-1)}
+ o(1) 
\hspace{3ex}
\text{as $\tau \rightarrow \infty$}\,.
\label{eq:app:asdrift:F0future-finalsum}
\end{align}

\section{Near-infinity and near-singularity transition rates 
in the $(1+1)$-dimensional Schwarzschild spacetime\label{app:schw}}

In this appendix we verify the near-infinity and near-singularity
transition rate formulas \eqref{eq:sch-transrate-aspast} and
\eqref{eq:sch-transrate-as-sing} for the inertial detector in the
$(1+1)$-dimensional Schwarzschild spacetime.

\subsection{Near-infinity transition rate} 

We consider the $E\ge1$ trajectories \eqref{eq:schw-allcoords-as-of-chi}
and \eqref{eq:schw-allcoords-as-of-tau-E=1} in Quadrant~I\null. 
We wish to find the
transition rate in the early time limit, assuming that the detector is
switched on in the asymptotic past. 

Using \eqref{eq:Fdot-sharp-aspast} and \eqref{eq:wightmans-in-schw}, 
we find 
\begin{subequations}
\label{eq:app:Fdot-aspast-all-def}
\begin{align}
\dot{\mathcal{F}}_B(\omega,\tau)
& = 
-\omega \Theta(-\omega)
+ 2  \int_{0}^{\infty} 
\! ds \cos(\omega s) 
\left(\mathcal{A}_u(\tau,\tau-s) 
+ 
\mathcal{A}_v(\tau,\tau-s) 
+ \frac{1}{2\pi s^2}\right)
\ , 
\\
\dot{\mathcal{F}}_H(\omega,\tau)
& = 
-\omega \Theta(-\omega)
+ 2  \int_{0}^{\infty} 
\! ds \cos(\omega s) 
\left(\mathcal{A}_{\bar u}(\tau,\tau-s) 
+ 
\mathcal{A}_{\bar v}(\tau,\tau-s) 
+ \frac{1}{2\pi s^2}\right)
\ , 
\\
\dot{\mathcal{F}}_U(\omega,\tau)
& = 
-\omega \Theta(-\omega)
+ 2  \int_{0}^{\infty} 
\! ds \cos(\omega s) 
\left(\mathcal{A}_{\bar u}(\tau,\tau-s) 
+ 
\mathcal{A}_v(\tau,\tau-s) 
+ \frac{1}{2\pi s^2}\right)
\ , 
\end{align}
\end{subequations}
where 
\begin{subequations}
\label{AA-all}
\begin{align}
\mathcal{A}_{{u}}(\tau, \tau') 
& = - \frac{\dot {{u}} (\tau) \dot {{u}} (\tau')}{ 4 \pi {[{u}(\tau)-{u}(\tau')]}^2}
\ , 
\label{Au}
\\
\mathcal{A}_{{v}}(\tau, \tau') 
& = - \frac{\dot {{v}} (\tau) \dot {{v}} (\tau')}{ 4 \pi {[{v}(\tau)-{v}(\tau')]}^2}
\ , 
\label{Av}
\\
\mathcal{A}_{\bar{u}}(\tau, \tau') 
& = - \frac{\dot {\bar{u}} (\tau) \dot {\bar{u}} (\tau')}{ 4 \pi {[\bar{u}(\tau)-\bar{u}(\tau')]}^2}
\ , 
\label{Aubar}
\\
\mathcal{A}_{\bar{v}}(\tau, \tau') 
& = - \frac{\dot {\bar{v}} (\tau) \dot {\bar{v}} (\tau')}{ 4 \pi {[\bar{v}(\tau)-\bar{v}(\tau')]}^2}
\ . 
\label{Avbar}
\end{align}
\end{subequations}
Using \eqref{eq:schw-allcoords-as-of-chi} and~\eqref{eq:schw-allcoords-as-of-tau-E=1}, 
it is straightforward to verify that as $\tau \to -\infty$ with fixed positive~$s$, 
we have 
\begin{subequations}
\begin{align}
\mathcal{A}_{{u}}(\tau, \tau -s) 
& \to - \frac{1}{4\pi s^2}
\ , 
\label{Au-limit}
\\
\mathcal{A}_{{v}}(\tau, \tau-s) 
& \to - \frac{1}{4\pi s^2}
\ , 
\label{Av-limit}
\\
\mathcal{A}_{\bar{u}}(\tau, \tau-s) 
& \to 
- \frac{\emath^{2\lambda}}{4 \pi  {(8 M)}^2 \sinh^2 \bigl(\emath^\lambda s/(8M)\bigr)}
\ , 
\label{Aubar-limit}
\\
\mathcal{A}_{\bar{v}}(\tau, \tau-s) 
& \to 
- \frac{\emath^{-2\lambda}}{4 \pi  {(8 M)}^2 \sinh^2 \bigl(\emath^{-\lambda} s/(8M)\bigr)}
\ , 
\label{Avbar-limit}
\end{align}
\end{subequations}
where $\lambda = \arctanh \bigl(\sqrt{1 - E^{-2}} \, \bigr)$. 
Taking the $\tau \to -\infty$ limit under 
the integrals in~\eqref{eq:app:Fdot-aspast-all-def}, 
justified by the monotone convergence argument given below, 
and proceeding as in subsection~\ref{thermalFdot}, 
leads to the formulas \eqref{eq:sch-transrate-aspast} in the main
text. 

What remains is to provide the monotone convergence argument.
Let $q$ stand
for either $\dot u$ or~$\dot v$, and note from
\eqref{eq:schw-vs-tortoisenull} and \eqref{eq:Schw-geod-eqs} that
\begin{subequations}
\begin{align}
\dot{r} &= - \sqrt{E^2-1 + 2M/r}
\ , 
\\
q &= \frac{1}{E + \eta \sqrt{E^2 - 1 + 2M/r}}
\ , 
\\
\dot{q} &= - \frac{\eta M}{r^2 \, \Bigl( E + \eta \sqrt{E^2 - 1 + 2M/r} \, \Bigr)^2}
\ , 
\end{align}
\end{subequations} 
where $\eta=1$ for $q = \dot v$ and $\eta = -1$ for $q = \dot u$. 
The expressions 
\begin{subequations}
\label{eq:app:monot1}
\begin{align}
& \frac{\int_{\tau-s}^\tau q(\tau') \, d\tau'}{\sqrt{q(\tau) q(\tau - s)}} 
\ , 
\\[1ex]
& \frac{\sinh \! \left(\frac{1}{8M} \int_{\tau-s}^\tau q(\tau') \, d\tau' \right)}{\sqrt{q(\tau) q(\tau - s)}}
\ , 
\end{align}
\end{subequations}
are hence well defined for all
$s>0$ when $\tau$ is sufficiently large and negative. 
For monotone convergence, it suffices to show that 
each of the expressions in
\eqref{eq:app:monot1} is monotone in $\tau$ for all $s>0$ when $\tau$
is sufficiently large and negative. 
Differentiating \eqref{eq:app:monot1} with respect
to~$\tau$, 
it suffices to show that each of the expressions
\begin{subequations}
\label{eq:app:monot2}
\begin{align}
& \int_{\tau-s}^\tau q(\tau') \, d\tau'
- 2[q(\tau) - q(\tau-s)] \left(\frac{\dot q (\tau)}{q(\tau)} 
+ \frac{\dot q (\tau-s)}{q(\tau-s)} \right)^{-1}
\ , 
\label{eq:app:monot2a}
\\[1ex]
& 
\tanh \! \left(\frac{1}{8M} \int_{\tau-s}^\tau q(\tau') \, d\tau'
\right)
- \frac{1}{4M}[q(\tau) - q(\tau-s)] 
\left(\frac{\dot q (\tau)}{q(\tau)} + \frac{\dot q (\tau-s)}{q(\tau-s)} \right)^{-1}
\ , 
\label{eq:app:monot2b}
\end{align}
\end{subequations}
has a fixed sign for all $s>0$ when $\tau$ is sufficiently large and
negative.  
Introducing in \eqref{eq:app:monot2} 
a new integration variable by $p' = 
\sqrt{E^2 - 1 + 2M/r(\tau')}\,$, we see that it suffices to show that 
each of the functions 
\begin{subequations}
\label{eq:app:monot3}
\begin{align}
f_1 (p) & = 
\frac12 \int_p^{p_f} \frac{dp'}{(E + \eta p'){[{p'}^2 - E^2 + 1]}^2}
\notag
\\[1ex]
& \hspace{4ex} 
- \frac{p_f - p}{(E + \eta p){[p_f^2 - E^2 + 1]}^2 + (E + \eta p_f){[p^2 - E^2 + 1]}^2}
\ , 
\label{eq:app:monot3a}
\\[1ex]
f_2 (p) & = 
\tanh \! \left( 
\frac12 \int_p^{p_f} \frac{dp'}{(E + \eta p'){[{p'}^2 - E^2 + 1]}^2} 
\right)
\notag
\\[1ex]
& \hspace{4ex}
- \frac{p_f - p}{(E + \eta p){[p_f^2 - E^2 + 1]}^2 + (E + \eta p_f){[p^2 - E^2 + 1]}^2}
\ , 
\label{eq:app:monot3b}
\end{align}
\end{subequations}
defined on the domain $\sqrt{E^2 -1} < p < p_f$, where 
$p_f \in \bigl( \sqrt{E^2 -1}\, , E\bigr)$ is a parameter, 
has a fixed sign when $p_f$ is sufficiently close to 
$\sqrt{E^2 -1}$. 

Consider~$f_1$. $f_1'$ is
a rational function whose sign can be analysed by elementary methods,
with the outcome that $f_1'$ is negative when $p_f$ is
sufficiently close to $\sqrt{E^2-1}$. 
Hence $f_1$ is positive when $p_f$ is
sufficiently close to $\sqrt{E^2-1}$. 

Consider then~$f_2$. When $p_f$ is sufficiently close
to $\sqrt{E^2-1}$, an elementary analysis shows that the second term
in \eqref{eq:app:monot3b} is negative and strictly increasing,
and there is a $p_1 \in \bigl( \sqrt{E^2 -1}\, , p_f\bigr)$ such that
this term takes the value $-1$ at $p = p_1$. With $p_f$ this close
to $\sqrt{E^2-1}$, it follows that $f_2$ is negative for $p\le
p_1$, whereas for $p_1 < p < p_f$ $f_2$ has the same sign as 
\begin{align}
f_3(p) & = 
\frac12 \int_p^{p_f} \frac{dp'}{(E + \eta p'){[{p'}^2 - E^2 + 1]}^2}
\notag
\\[1ex]
& \hspace{4ex}
- \arctanh \! 
\left
(\frac{p_f - p}{(E + \eta p){[p_f^2 - E^2 + 1]}^2 + (E + \eta
  p_f){[p^2 - E^2 + 1]}^2}
\right)
\ . 
\label{eq:app:monot4b}
\end{align}
$f_3$ can be analysed by the same methods as~$f_1$, with the outcome
that $f_3$ is negative when $p_f$ is  
sufficiently close to $\sqrt{E^2-1}$.  
Collecting, we see that $f_2$ is negative when $p_f$ is
sufficiently close to $\sqrt{E^2-1}$. 

This completes the monotone convergence argument.

\subsection{Near-singularity transition rate}

We consider the trajectories \eqref{eq:schw-allcoords-as-of-chi},
\eqref{eq:schw-allcoords-as-of-tau-E=1},
\eqref{eq:schw-allcoords-as-of-phi}
and~\eqref{eq:ubar-vbar-ito-varphi}, with $E\ge0$, and with the field in the 
HHI and Unruh vacua. The switch-off moment $\tau$ is
assumed to be in Quadrant~II\null. The switch-on-moment $\tau_0$ 
either is finite and in a region 
of the spacetime where the vacuum is regular, or for $E\ge1$ may alternatively 
be pushed to the asymptotic past. 

Let $\tau_{\text{sing}}$ be the value of $\tau$ at the black hole
singularity, and let $\tau_1$ be a constant such that the detector 
is somewhere in Quadrant II at proper time~$\tau_1$. 
In the limit $\tau \to \tau_{\text{sing}}$ with everything else
fixed, we have 
\begin{subequations}
\label{eq:app:Fdot-late-decomp}
\begin{align}
\dot{\mathcal{F}}_H(\omega,\tau)
& = 
G_{\bar u}(\omega,\tau,\tau_1) + G_{\bar v}(\omega,\tau,\tau_1) + O(1)
\ , 
\\
\dot{\mathcal{F}}_U(\omega,\tau,\tau_1)
& = 
G_{\bar u}(\omega,\tau,\tau_1) + G_v(\omega,\tau,\tau_1) + O(1)
\ , 
\end{align}
\end{subequations}
where 
\begin{subequations}
\label{eq:Gthree-div}
\begin{align}
G_v(\omega,\tau,\tau_1)
&= 
2  \int_{\tau_1}^{\tau} 
\! d \tau' \cos[\omega (\tau-\tau')] 
\left(\mathcal{A}_v(\tau,\tau') 
+ \frac{1}{4\pi {(\tau - \tau')}^2}\right)
\ , 
\label{eq:Gthree-div-Gv}
\\
G_{\bar v}(\omega,\tau,\tau_1)
&= 
2  \int_{\tau_1}^{\tau} 
\! d \tau' \cos[\omega (\tau-\tau')] 
\left(\mathcal{A}_{\bar v}(\tau,\tau') 
+ \frac{1}{4\pi {(\tau - \tau')}^2}\right) 
\ , 
\\
G_{\bar u}(\omega,\tau,\tau_1)
&= 
2  \int_{\tau_1}^{\tau} 
\! d \tau' \cos[\omega (\tau-\tau')] 
\left(\mathcal{A}_{\bar u}(\tau,\tau') 
+ \frac{1}{4\pi {(\tau - \tau')}^2}\right)
\ , 
\end{align}
\end{subequations}
and $\mathcal{A}_v$, $\mathcal{A}_{\bar v}$ and 
$\mathcal{A}_{\bar u}$ are given in~\eqref{AA-all}. 
  
Consider first $G_v(\omega,\tau,\tau_1)$, and assume 
$\tau_1<\tau<\tau_{\text{sing}}$. 
Working in the coordinates $(v,r)$, 
well defined in Quadrant~II, 
the equations for the trajectory read 
\begin{subequations}
\label{eq:Schw-v-geod-eqs}
\begin{align}
\dot{r} &= - \sqrt{E^2-1 + 2M/r}
\ , 
\label{eq:Schw-rdot-root-geod-eq}
\\[1ex]
\dot{v} &= \frac{1}{E + \sqrt{E^2 - 1 + 2M/r}}
\ , 
\label{eq:Schw-vdot-root-geod-eq}
\end{align}
\end{subequations}
from which it follows that 
\begin{align}
\ddot{v} &= - \frac{M}{r^2 \, \Bigl( E + \sqrt{E^2 - 1 + 2M/r} \, \Bigr)^2}
\ . 
\label{eq:Schw-vddot-root-geod-eq}
\end{align}
From \eqref{Av} and \eqref{eq:Gthree-div-Gv} we hence have 
\begin{align}
G_v(\omega,\tau,\tau_1)
&= 
\frac{1}{2\pi}
\lim_{\tau' \to \tau}
\left[
\cos[\omega (\tau-\tau')] 
\left( - \frac{\dot {{v}} (\tau) }{{{v}(\tau)-{v}(\tau')}}
+ \frac{1}{{\tau - \tau'}}\right) \right]
+ O(1) 
\notag 
\\[1ex]
&= 
\frac{1}{16 \pi M} \left[ \left(\frac{2M}{r(\tau)}\right)^{3/2}
- E \left(\frac{2M}{r(\tau)}\right)
+\frac{1+E^2}{2} \left(\frac{2M}{r(\tau)}\right)^{1/2}\right]
+ O(1) 
\ , 
\label{eq:app:Gv-latefinal}
\end{align}
where we have first integrated by parts, observing that the new
integral term is $O(1)$ by virtue of near-singularity estimates that
ensue from \eqref{eq:Schw-v-geod-eqs} and
\eqref{eq:Schw-vddot-root-geod-eq}, and then evaluated the limit using
\eqref{eq:Schw-v-geod-eqs} and~\eqref{eq:Schw-vddot-root-geod-eq}.

For $G_{\bar v}(\omega,\tau,\tau_1)$ we may proceed similarly, 
using $\bar v = 4M \exp[v/(4M)]$. 
The differences from $G_v(\omega,\tau,\tau_1)$ turn out to be~$O(1)$, 
so that 
\begin{align}
G_{\bar v}(\omega,\tau,\tau_1)
&= 
\frac{1}{16 \pi M} \left[ \left(\frac{2M}{r(\tau)}\right)^{3/2}
- E \left(\frac{2M}{r(\tau)}\right)
+\frac{1+E^2}{2} \left(\frac{2M}{r(\tau)}\right)^{1/2}\right]
+ O(1) 
\ . 
\label{eq:app:Gvbar-latefinal}
\end{align}
For $G_{\bar u}(\omega,\tau,\tau_1)$ the analysis is as for 
$G_{\bar v}(\omega,\tau,\tau_1)$ but with $E \to -E$, with the result 
\begin{align}
G_{\bar u}(\omega,\tau,\tau_1)
&= 
\frac{1}{16 \pi M} \left[ \left(\frac{2M}{r(\tau)}\right)^{3/2}
+ E \left(\frac{2M}{r(\tau)}\right)
+\frac{1+E^2}{2} \left(\frac{2M}{r(\tau)}\right)^{1/2}\right]
+ O(1) 
\ . 
\label{eq:app:Gubar-latefinal}
\end{align}

Combining \eqref{eq:app:Fdot-late-decomp}, 
\eqref{eq:app:Gv-latefinal}, 
\eqref{eq:app:Gvbar-latefinal} 
and 
\eqref{eq:app:Gubar-latefinal}
yields~\eqref{eq:sch-transrate-as-sing}.


\begin{thebibliography}{99}

\bibitem{birrell-davies}
N.~D. Birrell
and
P.~C.~W. Davies, 
{\it Quantum Fields in Curved Space\/}  
(Cambridge University Press, 1982).

\bibitem{wald-smallbook}
R.~M. Wald, 
{\it Quantum field theory in curved spacetime and black hole thermodynamics\/}
(University of Chicago Press, Chicago, 1994).

\bibitem{unruh}
W.~G.~Unruh,
``Notes on black hole evaporation,''
Phys.\ Rev.\  D {\bf 14}, 870 (1976).

\bibitem{hawking}
S.~W. Hawking,
``Particle creation by black holes,''
Commun.\ Math.\ Phys.\  {\bf 43}, 199 (1975)
[Erratum-ibid.\  {\bf 46}, 206 (1976)].

\bibitem{gibb-haw:dS} 
G.~W. Gibbons and S.~W. Hawking,
``Cosmological event horizons, thermodynamics, and particle creation,''
Phys.\ Rev.\  D {\bf 15}, 2738 (1977).

\bibitem{dewitt}
B.~S. DeWitt,
``Quantum gravity: the new synthesis'', 
in {\it General Relativity: an Einstein centenary survey}, 
edited by S.~W.~Hawking and W.~Israel (Cambridge University Press, Cambridge, 1979)
680. 

\bibitem{Takagi:1986kn}
S.~Takagi,
``Vacuum noise and stress induced by uniform acceleration: Hawking-Unruh
effect in Rindler manifold of arbitrary dimension,''
Prog.\ Theor.\ Phys.\ Suppl.\  {\bf 88}, 1 (1986).

\bibitem{Crispino:2007eb} 
L.~C.~B.~Crispino, A.~Higuchi and G.~E.~A.~Matsas,
``The Unruh effect and its applications,''
Rev.\ Mod.\ Phys.\  {\bf 80}, 787 (2008)
[arXiv:0710.5373].

\bibitem{Davies:2002bg}
P.~C.~W.~Davies and A.~C.~Ottewill,
``Detection of negative energy: 4-dimensional examples,''
Phys.\ Rev.\ D {\bf 65}, 104014 (2002) 
[arXiv:gr-qc/0203003].

\bibitem{schlicht}
S.~Schlicht,
``Considerations on the Unruh effect: Causality and regularization,''
Class.\ Quant.\ Grav.\  {\bf 21}, 4647 (2004)
[arXiv:gr-qc/0306022].

\bibitem{Langlois:2005nf} 
P.~Langlois,
``Causal particle detectors and topology,''
Annals Phys.\  {\bf 321}, 2027 (2006)
[arXiv:gr-qc/0510049].

\bibitem{langlois-thesis}
P.~Langlois,
``Imprints of spacetime topology in the Hawking-Unruh effect'',
PhD Thesis, University of Nottingham (2005)
[arXiv:gr-qc/0510127]. 

\bibitem{louko-satz-spatial}
J.~Louko and A.~Satz,
``How often does the Unruh-DeWitt detector click? 
Regularisation by a spatial profile,''
Class.\ Quant.\ Grav.\  {\bf 23}, 6321 (2006)
[arXiv:gr-qc/0606067].

\bibitem{satz}
A.~Satz,
``Then again, how often does the Unruh-DeWitt detector click if we switch it carefully?,''
Class.\ Quant.\ Grav.\  {\bf 24}, 1719 (2007)
[arXiv:gr-qc/0611067].

\bibitem{louko-satz-curved}
J.~Louko
and
A.~Satz,
``Transition rate of the Unruh-DeWitt detector in curved spacetime'',
Class.\ Quant.\ Grav.\  {\bf 25}, 055012 (2008)
[arXiv:0710.5671 [gr-qc]].

\bibitem{hodgkinson-louko}
L.~Hodgkinson
and
J.~Louko,
``How often does the Unruh-DeWitt detector click beyond four dimensions?'', 
{\it J. Math. Phys. \bf{53}}, 082301 (2012) 
[arXiv:1109.4377v3 [gr-qc]]. 

\bibitem{hodgkinson-louko-btz}
L.~Hodgkinson and J.~Louko,
``Static, stationary and inertial Unruh-DeWitt detectors on the BTZ black hole,''
Phys.\ Rev.\ D {\bf 86}, 064031 (2012)
[arXiv:1206.2055 [gr-qc]].

\bibitem{Barbado:2012fy} 
L.~C.~Barbado and M.~Visser,
``Unruh-DeWitt detector event rate for trajectories with time-dependent acceleration,''
Phys.\ Rev.\ D {\bf 86}, 084011 (2012)
[arXiv:1207.5525 [gr-qc]].

\bibitem{Hodgkinson:2013tsa} 
L.~Hodgkinson,
``Particle detectors in curved spacetime quantum field theory,''
PhD Thesis, University of Nottingham (2013)  
[arXiv:1309.7281v2 [gr-qc]].

\bibitem{Hodgkinson:2014}
L.~Hodgkinson, J.~Louko and A.~C. Ottewill, 
``Static detectors and circular-geodesic detectors 
on the Schwarzschild black hole,''
Phys.\ Rev.\ D {\bf 89}, 104002 (2014)
[arXiv:1401.2667 [gr-qc]]. 

\bibitem{ng:schw-ads-stationary}
K.~K.~Ng, 
L.~Hodgkinson, 
J.~Louko, 
R.~B.~Mann
and 
E.~Mart\'in-Mart\'inez, 
``Unruh-DeWitt detector response along static 
and circular-geodesic trajectories for Schwarzschild-AdS black holes,''
Phys.\ Rev.\ D {\bf 90}, 064003 (2014)
[arXiv:1406.2688 [quant-ph]].

\bibitem{Raval:1995mb} 
A.~Raval, B.~L.~Hu and J.~Anglin,
``Stochastic theory of accelerated detectors in a quantum field,''
Phys.\ Rev.\ D {\bf 53}, 7003 (1996)
[arXiv:gr-qc/9510002].

\bibitem{Lin:2006jw} 
S.~-Y.~Lin and B.~L.~Hu,
``Backreaction and the Unruh effect: 
New insights from exact solutions of uniformly accelerated detectors,''
Phys.\ Rev.\ D {\bf 76}, 064008 (2007)
[gr-qc/0611062].

\bibitem{Ostapchuk:2011ud} 
D.~C.~M.~Ostapchuk, S.~-Y.~Lin, R.~B.~Mann and B.~L.~Hu,
``Entanglement Dynamics between Inertial and Non-uniformly Accelerated Detectors,''
JHEP {\bf 1207}, 072 (2012)
[arXiv:1108.3377 [gr-qc]].

\bibitem{Brown:2012pw} 
E.~G.~Brown, E.~Mart\'in-Mart\'inez, N.~C.~Menicucci and R.~B.~Mann,
``Detectors for probing relativistic quantum physics beyond perturbation theory,''
Phys.\ Rev.\ D {\bf 87}, 084062 (2013)
[arXiv:1212.1973 [quant-ph]].

\bibitem{Bruschi:2012rx} 
D.~E.~Bruschi, A.~R.~Lee and I.~Fuentes,
``Time evolution techniques for detectors in relativistic quantum information,''
J.\ Phys.\ A {\bf 46}, 165303 (2013)
[arXiv:1212.2110 [quant-ph]].

\bibitem{Hu:2012jr} 
B.~L.~Hu, S.~-Y.~Lin and J.~Louko,
``Relativistic quantum information in detectors-field interactions,''
Class.\ Quant.\ Grav.\  {\bf 29}, 224005 (2012)
[arXiv:1205.1328 [quant-ph]].

\bibitem{Davies:1976hi}
P.~C.~W.~Davies and S.~A.~Fulling,
``Radiation from a moving mirror in two-dimensional space-time conformal anomaly,''
Proc.\ Roy.\ Soc.\ Lond.\ A {\bf 348}, 393 (1976).

\bibitem{Davies:1977yv}
P.~C.~W.~Davies and S.~A.~Fulling,
``Radiation from moving mirrors and from black holes,''
Proc.\ Roy.\ Soc.\ Lond.\ A {\bf 356}, 237 (1977). 

\bibitem{decanini-folacci}
Y.~D\'ecanini and A.~Folacci,
``Hadamard renormalization of the stress-energy tensor 
for a quantized scalar field in a general spacetime 
of arbitrary dimension,''
Phys.\ Rev.\ D {\bf 78}, 044025 (2008)
[arXiv:gr-qc/0512118].

\bibitem{fulling-ruijsenaars}
S.~A. Fulling
and 
S.~N.~M. Ruijsenaars, 
``Temperature, periodicity and horizons,''
Phys.\ Rept.\ {\bf 152}, 135
(1987). 

\bibitem{Kay:2000fi} 
B.~S.~Kay,
``Application of linear hyperbolic PDE to linear quantum fields 
in curved space-times: Especially black holes, time machines and a new semilocal vacuum concept,''
Journ\'ees \'Equations aux d\'eriv\'ees partielles, 
Nantes, 5 au 9 juin 2000, GDR 1151 (CNRS), IX-1
[arXiv:gr-qc/0103056]. 

\bibitem{Raine:1991kc} 
D.~J.~Raine, D.~W.~Sciama and P.~G.~Grove,
``Does an accelerated oscillator radiate?''
Proc.\ Roy.\ Soc.\ A {\bf 435}, 205
(1991).

\bibitem{Wang:2013lex} 
Q.~Wang and W.~G.~Unruh,
``Motion of a mirror under infinitely fluctuating quantum vacuum stress,''
Phys.\ Rev.\ D {\bf 89}, 085009 (2014)
[arXiv:1312.4591 [gr-qc]].

\bibitem{Kubo:1957mj} 
  R.~Kubo,
  ``Statistical mechanical theory of irreversible processes. 1. 
 General theory and simple applications in magnetic and conduction problems,''
  J.\ Phys.\ Soc.\ Jap.\  {\bf 12}, 570 (1957).

\bibitem{Martin:1959jp} 
  P.~C.~Martin and J.~S.~Schwinger,
 ``Theory of many particle systems. 1.,''
  Phys.\ Rev.\  {\bf 115}, 1342 (1959).

\bibitem{Martin-Martinez:2014qda} 
E.~Mart\'in-Mart\'inez and J.~Louko,
``Particle detectors and the zero mode of a quantum field,''
Phys.\ Rev.\ D {\bf 90}, 024015 (2014)
[arXiv:1404.5621 [quant-ph]].

\bibitem{Anderson:2014jua} 
  P.~Anderson, R.~Balbinot, A.~Fabbri and R.~Parentani,
  ``Gray-body factor and infrared divergences in BEC acoustic black holes,''
  arXiv:1404.3224 [gr-qc].

\bibitem{braunstein-et-al} 
S.~L. Braunstein, ``Black hole entropy as entropy of
entanglement, or it's curtains for the equivalence principle,''
arXiv:0907.1190v1 [quant-ph]; 
S.~L. Braunstein, S.~Pirandola and K.~\.Zyczkowski, 
``Better Late than Never: Information
Retrieval from Black Holes,''
Phys.\ Rev.\ Lett.\  {\bf 110}, 101301 (2013)
[arXiv:0907.1190v3 [quant-ph]]. 

\bibitem{Mathur:2009hf} 
  S.~D.~Mathur,
  ``The information paradox: a pedagogical introduction,''
  Class.\ Quant.\ Grav.\  {\bf 26}, 224001 (2009)
  [arXiv:0909.1038 [hep-th]].

\bibitem{Almheiri:2012rt} 
  A.~Almheiri, D.~Marolf, J.~Polchinski and J.~Sully,
  ``Black Holes: Complementarity or Firewalls?,''
  JHEP {\bf 1302}, 062 (2013)
  [arXiv:1207.3123 [hep-th]].

\bibitem{Hotta:2013clt} 
M.~Hotta, J.~Matsumoto and K.~Funo,
``Black hole firewalls require huge energy of measurement,''
Phys.\ Rev.\ D {\bf 89}, 124023 (2014)
[arXiv:1306.5057 [quant-ph]].

\bibitem{Almheiri:2013wka} 
  A.~Almheiri and J.~Sully,
  ``An Uneventful Horizon in Two Dimensions,''
  JHEP {\bf 1402}, 108 (2014)
  [arXiv:1307.8149 [hep-th]].

\bibitem{Fewster:1999gj} 
C.~J.~Fewster,
``A general worldline quantum inequality,''
Class.\ Quant.\ Grav.\  {\bf 17}, 1897 (2000)
[arXiv:gr-qc/9910060].

\bibitem{junker} 
W.~Junker and E.~Schrohe,
  ``Adiabatic vacuum states on general space-time manifolds: Definition,
  construction, and physical properties,''
  Ann.\ Henri Poincar\'e 
  {\bf 3}, 1113 
  (2002)
  [arXiv:math-ph/0109010].

\bibitem{hormander-vol1}
L.~H\"ormander, 
\textit{The Analysis of Linear Partial Differential Operators~I
(Distribution Theory and Fourier Analysis)\/}, 
2nd Edition 
(Springer, Berlin, 1990), 
Theorem 8.2.4. 

\bibitem{hormander-paper1}
L.~H\"ormander, 
``Fourier Integral Operators.~I'', 
Acta Mathematica \textbf{127}, 
79
(1971), 
Theorem 2.5.11'. 
Reprinted in: 
J.~Br\"uning 
and 
V.~W. Guillemin (Editors), 
\textit{Fourier Integral Operators\/}
(Springer, Berlin, 1994). 

\bibitem{dlmf}
NIST Digital Library of Mathematical Functions. 
{\tt http://dlmf.nist.gov/}, 
Release 1.0.6 of 2013-05-06. 

\bibitem{grad-ryzh}
I.~S.~Gradshteyn 
and 
I.~M.~Ryzhik,
{\it Table of Integrals, Series, and Products\/},
7th edition 
(Academic Press, New York, 2007).

\bibitem{Carlitz:1986nh} 
R.~D.~Carlitz and R.~S.~Willey,
``Reflections on moving mirrors,''
Phys.\ Rev.\ D {\bf 36}, 2327 (1987).

\bibitem{Good:2013lca} 
M.~R.~R.~Good, P.~R.~Anderson and C.~R.~Evans,
``Time dependence of particle creation from accelerating mirrors,''
Phys.\ Rev.\ D {\bf 88}, 025023 (2013)
[arXiv:1303.6756 [gr-qc]].

\bibitem{Misner:1974qy}
C.~W.~Misner, K.~S.~Thorne and J.~A.~Wheeler,
{\it Gravitation\/}
(Freeman, San Francisco, 1973). 

\bibitem{Boulware-scalar} 
D.~G.~Boulware,
``Quantum field theory in Schwarzschild and Rindler spaces,''
Phys.\ Rev.\ D {\bf 11}, 1404 (1975).

\bibitem{Hartle:1976tp}
J.~B.~Hartle and S.~W.~Hawking,
``Path integral derivation of black hole radiance,''
Phys.\ Rev.\  D {\bf 13}, 2188 (1976).
  
\bibitem{Israel:1976ur}
W.~Israel,
``Thermofield dynamics of black holes,''
Phys.\ Lett.\  A {\bf 57}, 107 (1976).

\bibitem{Dappiaggi:2009fx} 
C.~Dappiaggi, V.~Moretti and N.~Pinamonti,
``Rigorous construction and Hadamard property 
of the Unruh state in Schwarzschild spacetime,''
Adv.\ Theor.\ Math.\ Phys.\  {\bf 15}, 355 (2011)
[arXiv:0907.1034 [gr-qc]].

\bibitem{Barcelo:2010pj}
C.~Barcelo, S.~Liberati, S.~Sonego and M.~Visser,
``Minimal conditions for the existence of a Hawking-like flux,''
Phys.\ Rev.\ D {\bf 83} (2011) 041501
[arXiv:1011.5593 [gr-qc]].

\bibitem{Barbado:2011dx}
L.~C.~Barbado, C.~Barcelo and L.~J.~Garay,
``Hawking radiation as perceived by different observers,''
Class.\ Quant.\ Grav.\  {\bf 28} (2011) 125021
[arXiv:1101.4382 [gr-qc]].

\bibitem{Barbado:2012pt}
L.~C.~Barbado, C.~Barcelo and L.~J.~Garay,
``Hawking radiation as perceived by different observers:
An analytic expression for the effective-temperature function,''
Class.\ Quant.\ Grav.\  {\bf 29} (2012) 075013
[arXiv:1201.3820 [gr-qc]].

\bibitem{Smerlak:2013sga} 
M.~Smerlak and S.~Singh,
``New perspectives on Hawking radiation,''
Phys.\ Rev.\ D {\bf 88}, 104023 (2013)
[arXiv:1304.2858 [gr-qc]].

\bibitem{few-jua-lou} 
C.~J. Fewster, 
B.~A. Ju\'arez-Aubry 
and 
J.~Louko, 
in preparation (2014). 






\end{thebibliography}
\end{document}